\documentclass[]{jingdong}
\usepackage[toc,page,header]{appendix}
\usepackage{latexsym}
\usepackage[T1]{fontenc}
\usepackage[utf8]{inputenc}
\usepackage{microtype}
\usepackage{inconsolata}
\usepackage{graphicx}
\usepackage{hyperref}       
\usepackage{url}            
\usepackage{booktabs}       
\usepackage{amsfonts}       
\usepackage{nicefrac}       
\usepackage{stackengine}
\usepackage{microtype}      
\usepackage{colortbl}
\usepackage{xcolor}
\usepackage{amsmath}
\usepackage{amssymb}
\usepackage{amsthm}
\usepackage{mathrsfs}
\usepackage{pifont}
\usepackage{MnSymbol}
\usepackage{balance}
\usepackage{enumitem}
\usepackage{listings}
\usepackage{xcolor}
\usepackage{natbib}
\usepackage{multicol}

\AtBeginDocument{%
  \providecommand\BibTeX{{%
    \normalfont B\kern-0.5em{\scshape i\kern-0.25em b}\kern-0.8em\TeX}}}

\makeatletter
\DeclareRobustCommand\onedot{\futurelet\@let@token\@onedot}
\def\@onedot{\ifx\@let@token.\else.\null\fi}

\usepackage{setspace}
\usepackage{mathtools}

\usepackage{multirow,booktabs}
\usepackage{subcaption}

\newcommand{\owo}[1]{\textsc{OAgents}}

\definecolor{lightgreen}{RGB}{144, 238, 144} 
\definecolor{lightred}{RGB}{255, 105, 97}

\newtcolorbox{promptbox}[2][Prompt]{
colback=black!5!white,
arc=5pt, 
boxrule=0.5pt,
fonttitle=\bfseries,
title=#1, 
before upper={\small}, fontupper=\fontfamily{ptm}\selectfont,
colframe=#2, 
}
\definecolor{ogreen}{RGB}{34, 139, 34}

\usepackage{color}
\usepackage{multirow}
\usepackage{booktabs}
\usepackage{wrapfig}
\usepackage{soul}
\usepackage{colortbl}
\usepackage{bbding}
\usepackage[ruled,linesnumbered]{algorithm2e}
\definecolor{mygray}{gray}{.9}
\definecolor{mypink}{rgb}{.99,.5,.5}
\definecolor{mycyan}{cmyk}{.3,0,0,0}

\definecolor{headerpink}{HTML}{FDECEC}   
\definecolor{headerblue}{HTML}{E8F2FF}   
\definecolor{rowyellow}{HTML}{FFFBE6}    
\definecolor{posgreen}{HTML}{1A801A}     
\definecolor{negred}{HTML}{D10000}       

\usepackage{cleveref}
\theoremstyle{plain}

\theoremstyle{definition}

\theoremstyle{remark}
\usepackage{subcaption}

\usepackage[textsize=tiny]{todonotes}
\usepackage{fontawesome}

\newcommand{\bestall}[1]{\textbf{#1}}
\newcommand{\bestsec}[1]{\underline{#1}}

\title{JoyAI-Image: Awaking Spatial Intelligence in Unified Multimodal Understanding and Generation}

\author[*]{Lin Song}
\author[*]{Wenbo Li}
\author[*]{Guoqing Ma}
\author{Wei Tang}
\author{Bo Wang}
\author{Yuan Zhang}
\author{Yijun Yang}
\author{Yicheng Xiao}
\author{Jianhui~Liu}
\author{Yanbing Zhang}
\author{Guohui Zhang}
\author{Wenhu Zhang}
\author{Hang Xu}
\author{Nan Jiang}
\author{Xin Han}
\author{Haoze Sun}
\author{Maoquan Zhang}
\author[\dagger]{Haoyang~Huang}
\author{Nan Duan}

\affiliation{Joy Future Academy, JD}

\contribution[*]{Equal contribution.}
\contribution[\dagger]{Corresponding author: Haoyang Huang (\email{huanghaoyang.ocean@jd.com}).}
\contribution[\ddagger]{See \hyperref[sec:contributions]{Contributors} section for the full contributor list.}

\abstract{

We present \textbf{JoyAI-Image}, a unified multimodal foundation model for visual understanding, text-to-image generation, and instruction-guided image editing. JoyAI-Image couples a spatially enhanced Multimodal Large Language Model (MLLM) with a Multimodal Diffusion Transformer (MMDiT), allowing perception and generation to interact through a shared multimodal interface. Around this architecture, we build a scalable training recipe that combines unified instruction tuning, long-text rendering supervision, spatially grounded data, and both general and spatial editing signals. This design gives the model broad multimodal capability while strengthening geometry-aware reasoning and controllable visual synthesis. Experiments across understanding, generation, long-text rendering, and editing benchmarks show that JoyAI-Image achieves state-of-the-art or highly competitive performance. More importantly, the bidirectional loop between enhanced understanding, controllable spatial editing, and novel-view-assisted reasoning enables the model to move beyond general visual competence toward stronger spatial intelligence. These results suggest a promising path for unified visual models in downstream applications such as vision-language-action systems and world models.

}

\checkdata[Code]{\url{https://github.com/jd-opensource/JoyAI-Image}}

\begin{document}
\maketitle

\begin{figure}[!ht]
\centering
\vspace{-2mm}
\includegraphics[width=0.60\linewidth]{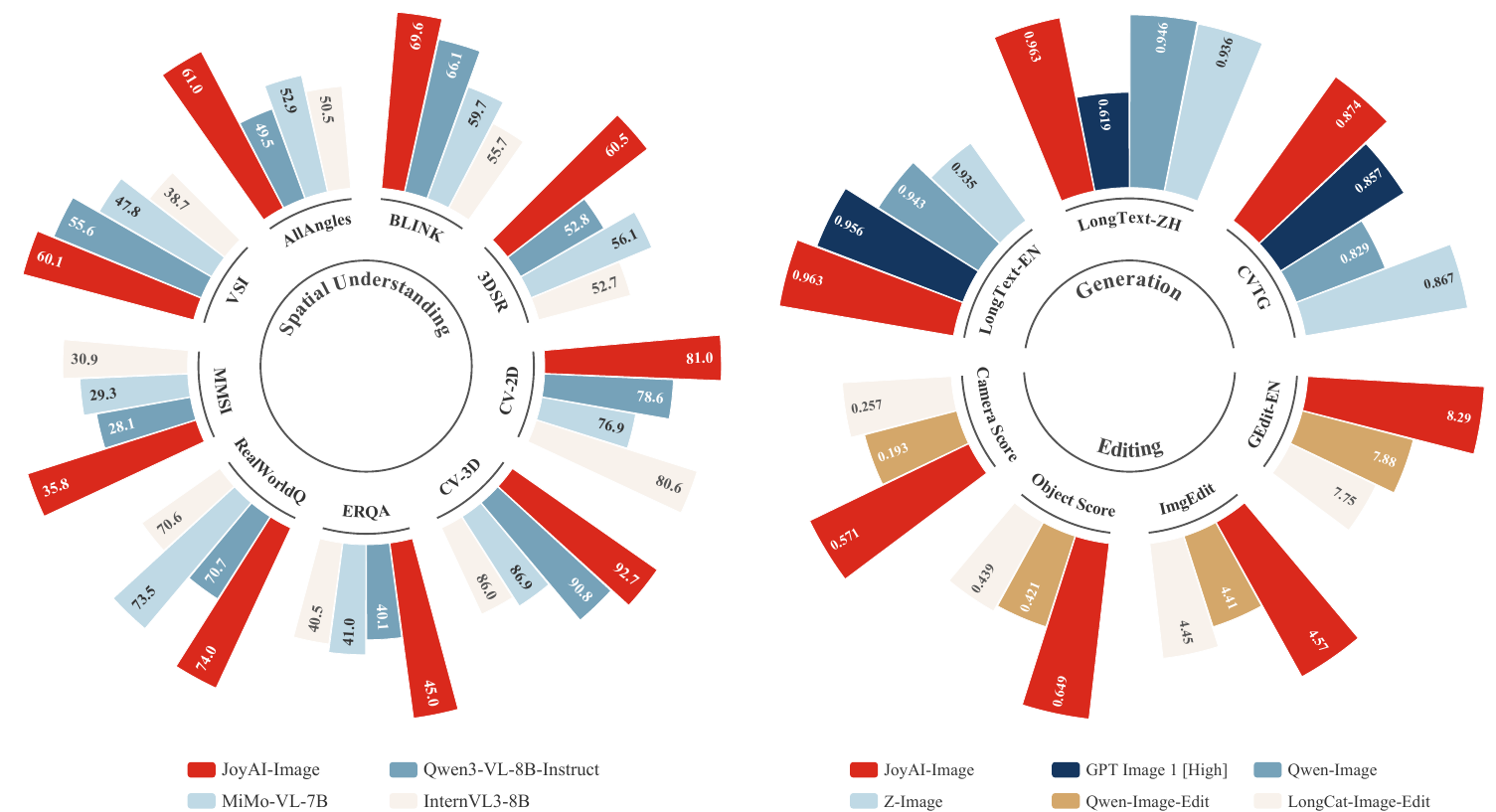}
\vspace{-1mm}
\caption{Overall performance of JoyAI-Image across understanding, generation, and editing.}
\label{fig:teaser}
\vspace{-3mm}
\end{figure}

\begin{figure}[htbp]
\begin{center}
   \includegraphics[width=0.96\linewidth]{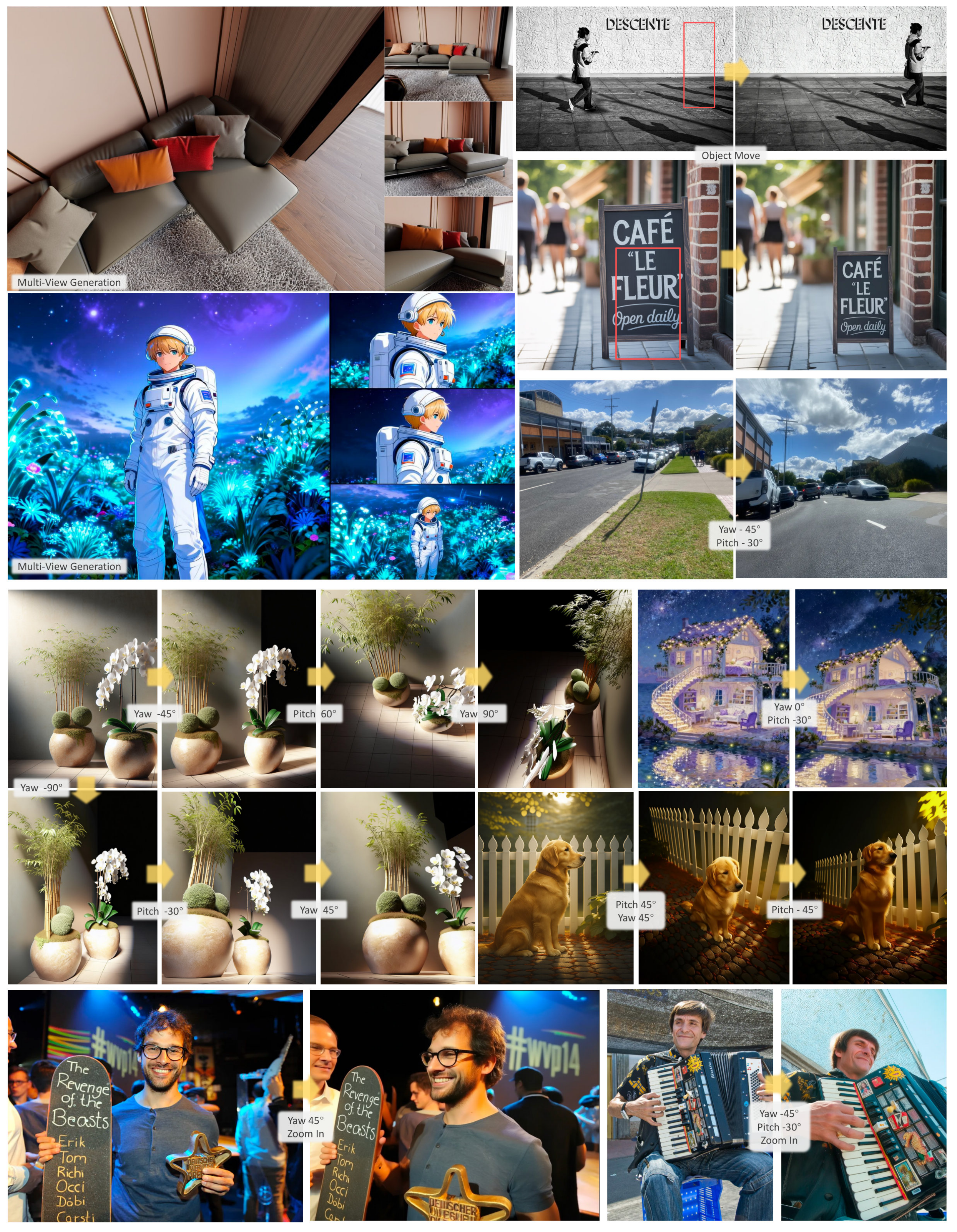}
   \caption{Showcase of JoyAI-Image’s spatial reasoning and editing capabilities, including multi-view generation, geometry-aware transformations, and precise, location-specific object editing.}
   \label{fig:spatial_cases}
\end{center}
\end{figure}

\begin{figure}[htbp]
\begin{center}
   \includegraphics[width=0.96\linewidth]{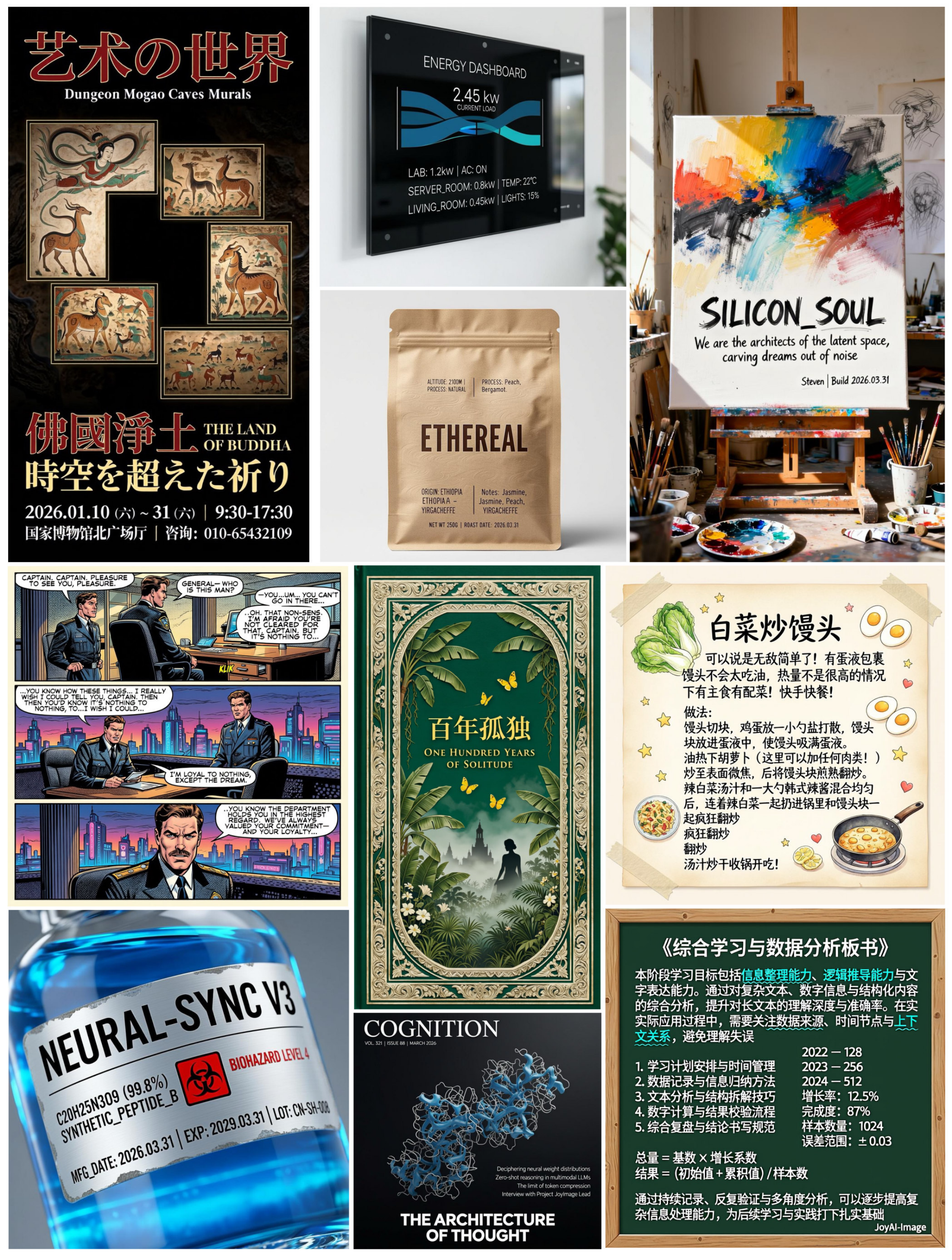}
   \caption{Showcase of JoyAI-Image’s advanced text rendering capabilities across diverse and challenging scenarios, including multi-panel comics, dense multi-line text, multilingual content, long-form layout composition, real-world scene text, and handwritten styles.}
   \label{fig:text_render_cases}
\end{center}
\end{figure}

\clearpage

\section{Introduction}

Recent advances in Multimodal Large Language Models (MLLMs)~\cite{qwen2.5vl,qwen3vl2025,gemini2.5,coreteam2025mimovltechnicalreport} and diffusion models~\cite{chen2024textdiffuser,flux2024,stablediffusion} have accelerated the development of unified models that jointly support image understanding, generation, and editing. This trend reflects a shift from task-specific pipelines toward general-purpose visual intelligence, where a single model is expected to interpret visual content, synthesize new images, and perform instruction-guided modifications. A key benefit of this unification is the possibility of tighter coordination across tasks, allowing understanding, generation, and editing to mutually benefit from better architecture design, data construction, and training strategies. Recent systems~\cite{qwenimage,xiao2025mindomni,LongCat-Image,nanopro,seedream2025seedream,deng2025bagel}, have demonstrated the potential of this paradigm through large-scale data curation, staged training, and scalable diffusion architectures.

Despite recent progress, current unified models still face two important limitations. First, although visual understanding, generation, and editing are increasingly integrated into a single framework, their interaction remains weak in practice. Visual understanding is not fully exploited to guide grounded generation and editing, while generative transformations are rarely used to provide useful feedback for perception and reasoning. Second, these models still lack strong spatial intelligence for the physical world. Real-world scenes are fundamentally shaped by object layout, relative geometry, viewpoint changes, and cross-view consistency, yet existing systems remain limited in fine-grained spatial understanding and geometrically precise manipulation. As a result, these weaknesses not only constrain controllable generation and editing, but also prevent unified visual models from further extending toward broader spatial intelligence, with important implications for applications such as visual-language-action systems~\cite{zitkovich2023rt,openvla} and world models~\cite{bruce2024genie,cen2025worldvla}.

In this work, we present \textbf{JoyAI-Image}, a unified multimodal foundation framework for understanding, generation, and editing, designed to improve overall visual performance by systematically strengthening spatial intelligence. JoyAI-Image combines a spatially enhanced MLLM with a Multimodal Diffusion Transformer (MMDiT) for high-fidelity image synthesis~\cite{flux-2-2025,qwenimage,team2025zimage,openai2025gptimage,ultrapixel,wu2025omnigen2} and instruction-based editing~\cite{brooks2023instructpix2pix,zhang2025icedit,yu2025anyedit,zhao2024ultraedit,camedit}. The MLLM serves not only as the core engine for scene understanding and instruction parsing, but also as the main interface for generative tasks, providing semantically rich and spatially grounded conditioning signals for downstream generation and editing. In this way, JoyAI-Image goes beyond a loose combination of perception and generation modules, and instead builds a unified, understanding-driven visual framework with stronger cross-task coupling.

A central principle of JoyAI-Image is to \textit{awaken spatial intelligence} throughout the unified training and reasoning process. Rather than treating spatial capability as an isolated module or a late-stage extension, we inject spatially grounded data construction, task design, and supervision into the full pipeline, so that spatial awareness develops jointly with understanding, generation, and editing. This design also establishes a bidirectional collaborative paradigm. On the one hand, stronger spatial understanding improves generation and editing through better scene parsing, relational grounding, and instruction decomposition. On the other hand, generative transformations, such as geometrically meaningful edits and novel-view expansion, provide complementary visual evidence for spatial understanding and downstream reasoning. In this way, JoyAI-Image strengthens both task collaboration and spatial capability within a unified model.


To realize this goal, JoyAI-Image is trained within a unified instruction-following framework that harmonizes understanding, generation, and editing objectives through a multi-stage curriculum. Our training regime leverages a multi-faceted data suite that spans ubiquitous visual tasks to specialized spatial operations. Specifically, it integrates general-purpose understanding with fine-grained spatial reasoning, high-fidelity synthesis with long-text typography, and versatile content editing ranging from general attribute modification to precise spatial manipulation. By balancing broad-domain robustness with pinpoint spatial control, JoyAI-Image delivers a versatile suite of capabilities, encompassing spatial understanding, typography-enhanced generative synthesis, general and spatial editing, and view-assisted reasoning.

\noindent The key contributions of JoyAI-Image can be summarized as follows:
\begin{itemize}

    \item \textbf{A Strong Unified Multimodal Foundation.} We present JoyAI-Image, a unified framework for image understanding, text-to-image generation, and instruction-based editing via a shared MLLM/MMDiT interface. As shown in Figure~\ref{fig:teaser}, it achieves strong results across broad visual tasks, especially in spatial understanding, long-text rendering, multi-view generation, and controllable editing.

    \item \textbf{A Practical Data and Training Recipe.} We build a scalable learning pipeline with detailed data construction and multi-stage optimization strategies and provide a practical recipe for training unified multimodal understanding-and-generation models with strong general-purpose capability. 

    \item \textbf{Awakening Spatial Intelligence in a Unified Model.} Beyond strong general-purpose performance, JoyAI-Image strengthens spatial understanding, controllable spatial editing, and novel-view-assisted reasoning through a bidirectional loop between understanding and generation, laying a practical foundation for broader spatial intelligence with implications for robotic systems and world models.

\end{itemize}

\begin{figure}[t]
\begin{center}
   \includegraphics[width=1\linewidth]{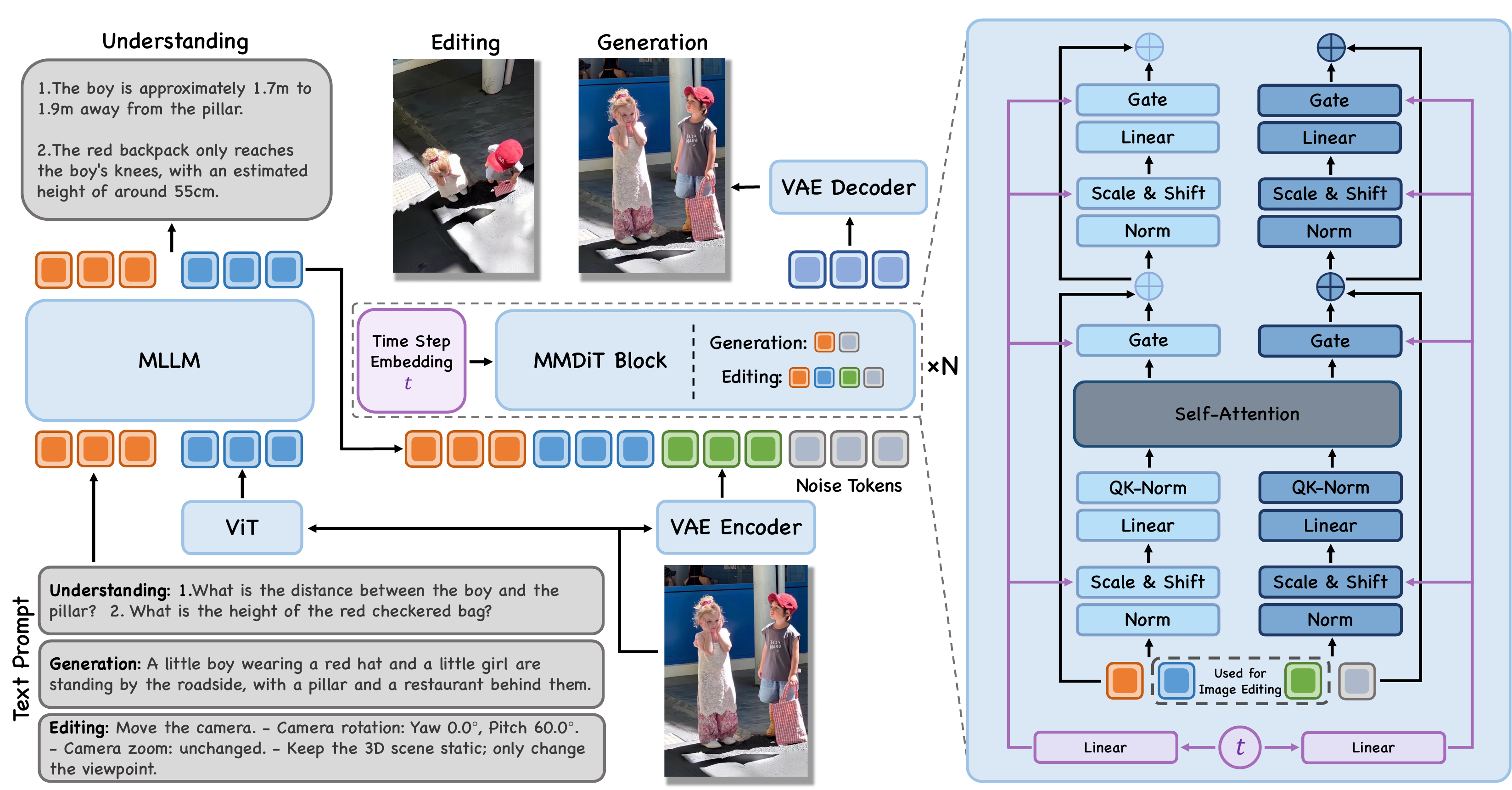}
   \caption{Overall architecture of JoyAI-Image, a unified foundation model for multimodal understanding, generation, and editing, integrating a Multimodal Large Language Model (MLLM), a Variational Autoencoder (VAE), and a dual-stream Multimodal Diffusion Transformer (MMDiT). For image understanding, the MLLM jointly encodes visual and textual inputs to enable semantic comprehension and reasoning. For image generation, the MLLM converts textual prompts into latent guidance, which the MMDiT transforms into images through iterative denoising. For image editing, the MLLM interprets both the user instruction and the source image, while the MMDiT synthesizes the final output by integrating MLLM-interleaved priors with VAE-encoded image features and noise tokens.}
   \label{fig:main_arch}
\end{center}
\end{figure}

\section{Model}

As illustrated in Figure~\ref{fig:main_arch}, JoyAI-Image is a unified framework for image understanding and generation, combining a spatially enhanced Multimodal Large Language Model (MLLM) with a Variational Autoencoder (VAE) and a Multimodal Diffusion Transformer (MMDiT)~\cite{esser2024scaling}. This paradigm facilitates a seamless transition from standalone scene comprehension to high-fidelity image synthesis and instruction-based editing. The operational workflow follows a principled three-stage pipeline:
\begin{itemize}
\item \textbf{Multimodal Understanding:} Serving as the "cognitive brain" of the architecture, the MLLM assumes a dual role. Primarily, it functions as a standalone understanding engine capable of general scene parsing and intricate spatial reasoning. Subsequently, it acts as an intent-explanation mediator, where it interprets interleaved instructions and reference signals that guide the downstream generative process. 
\item \textbf{Latent Encoding:} A Variational AutoEncoder (VAE) bridges the pixel-level data and the latent manifold. This stage ensures efficient spatio-temporal compression, mapping raw visual inputs into a compact representation space suitable for robust diffusion modeling.
\item \textbf{Conditional Generation:} The MMDiT serves as the core generative engine, modeling the conditional distribution between noise and latents. Through its dual-stream architecture, the MMDiT facilitates deep cross-modal fusion, effectively consuming the MLLM-derived priors to support both high-fidelity generation and fine-grained, multimodal-conditioned editing.
\end{itemize}

The architecture follows a progressive training paradigm: we first fine-tune the MLLM for robust visual-spatial understanding, then train the MMDiT from scratch for high-fidelity generation using MLLM-derived priors, and finally optimize the framework for precise, instruction-based editing.

\subsection{Multimodal Large Language Model}



We employ an MLLM as the primary interaction interface for parsing user inputs and facilitating cross-modal alignment. By utilizing the pre-trained knowledge and reasoning architecture of the MLLM, JoyAI-Image establishes a structured foundation for holistic scene comprehension and intent parsing, providing the necessary semantic priors for both image synthesis and instruction-based editing. Specifically, our comprehension module is built upon Qwen3-VL-8B-Instruct~\cite{qwen3vl2025}. To achieve precise geometric awareness and multi-view structural consistency, we further fortify its spatial reasoning through a dedicated data engine and specialized training (see Section~\ref{sec:spatial_und}). This enhancement is critical for tasks requiring high spatial fidelity, such as viewpoint-controllable synthesis and geometry-preserving manipulation.

The MLLM operates in two distinct functional modes based on the task objective:
\begin{itemize}
\item \textbf{Standalone Understanding:} For pure understanding tasks (\textit{e.g.}, image captioning or spatial reasoning), the MLLM functions as a generative language model, directly decoding its internal representations into human-readable text.
\item \textbf{Generative Conditioning:} For synthesis and editing, the MLLM processes input queries via task-specific workflows to guide the subsequent diffusion process:
\begin{itemize}
\item \textit{Text-to-Image Generation:} The MLLM parses text into structured semantic representations.
\item \textit{Instruction-based Editing:} The model processes interleaved inputs, namely the original image and the instruction, to resolve the mapping between linguistic modifiers and specific visual attributes.
\end{itemize}
To integrate these cognitive insights into the generative pipeline, we extract the hidden states from the final layer of the MLLM backbone for synthesis and editing tasks. These high-dimensional features serve as the primary conditioning signal, encapsulating high-level semantic-spatial cues to guide the MMDiT.
\end{itemize}

\subsection{Variational Auto-Encoder and Multimodal Diffusion Transformer}


To facilitate efficient and high-fidelity synthesis, we employ Wan-2.1-VAE~\cite{wan2025wan} as our latent compressor. It leverages causal 3D convolutions for superior spatio-temporal compression, preserving fine-grained structures and high-frequency details (\textit{e.g.}, small text rendering) during reconstruction.
The generative core of JoyAI-Image is a 16B-parameter MMDiT, which jointly models the multimodal representations from the MLLM and the latent representations from the VAE. This dual-stream architecture facilitates deep cross-modal fusion, supporting both denoising-based generation and complex multimodal-conditioned editing. We optimize the backbone efficiency by replacing the MSRoPE used in Qwen-Image~\cite{qwenimage} with a standard MRoPE, aligning the model's rotary positional embeddings more effectively with our structural conditioning objectives. The detailed architectural hyperparameters are summarized in Table~\ref{tab:joyai-image-arch}.

\begin{table}[h]
\centering
\caption{Multimodal Diffusion Transformer Hyperparameters of JoyAI-Image.}
\label{tab:joyai-image-arch}
\small
\renewcommand{\arraystretch}{1.0}
\setlength{\tabcolsep}{8pt}
\begin{tabular}{lc @{\hspace{5em}} lc}
\toprule
\textbf{Parameter} & \textbf{Value} & \textbf{Parameter} & \textbf{Value} \\
\midrule
Input/Output Dim & 16 & Attention Heads & 32 \\
Patch Size & $1 \times 2 \times 2$ & Modulation Type & WanX \\
Number of Layers & 40 & Position Embedding & MRoPE \\
Hidden Dim & 4096 & RoPE Base $\theta$ & 10,000 \\
Text Hidden Dim & 4096 & RoPE Dim List & [16, 56, 56] \\
\bottomrule
\end{tabular}
\end{table}
\begin{figure}[htbp]
\begin{center}
   \includegraphics[width=1\linewidth]{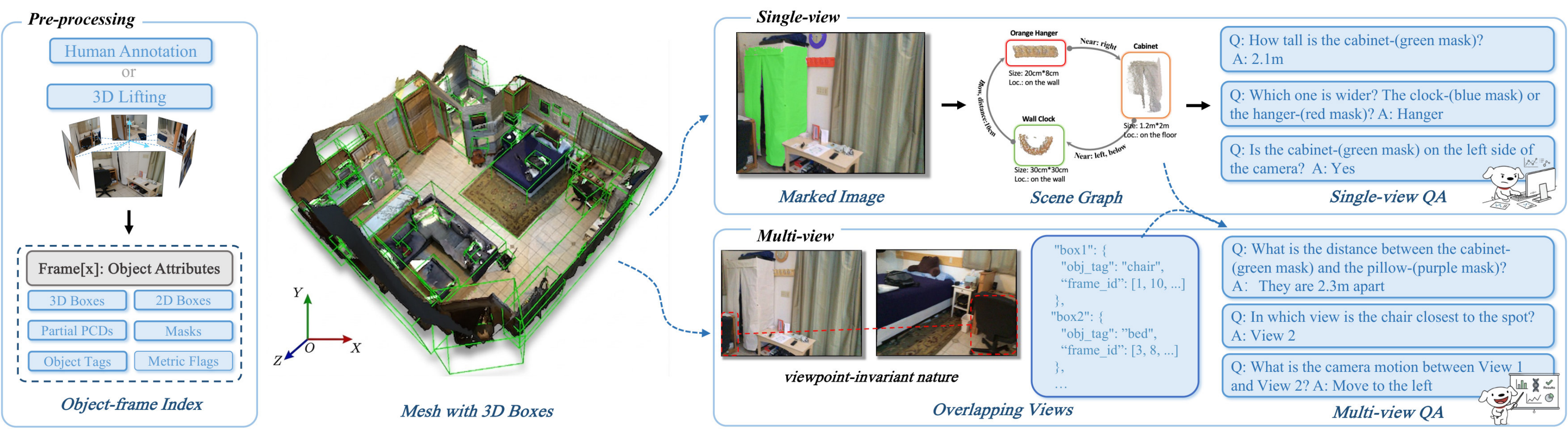}
   \caption{System overview of the OpenSpatial engine. Driven by a box-centric paradigm, the OpenSpatial engine features an automated or semi-automated workflow to generate high-quality and diversified spatial understanding data.}
   \label{fig:pipeline3d}
\end{center}
\end{figure}

\section{Advanced Spatial Understanding}
\label{sec:spatial_und}

\subsection{Data}

\subsubsection{Automated Spatial Data Synthesis}
\label{sec:data_synthesis}

\begin{figure}[htbp]
\begin{center}
   \includegraphics[width=0.82\linewidth]{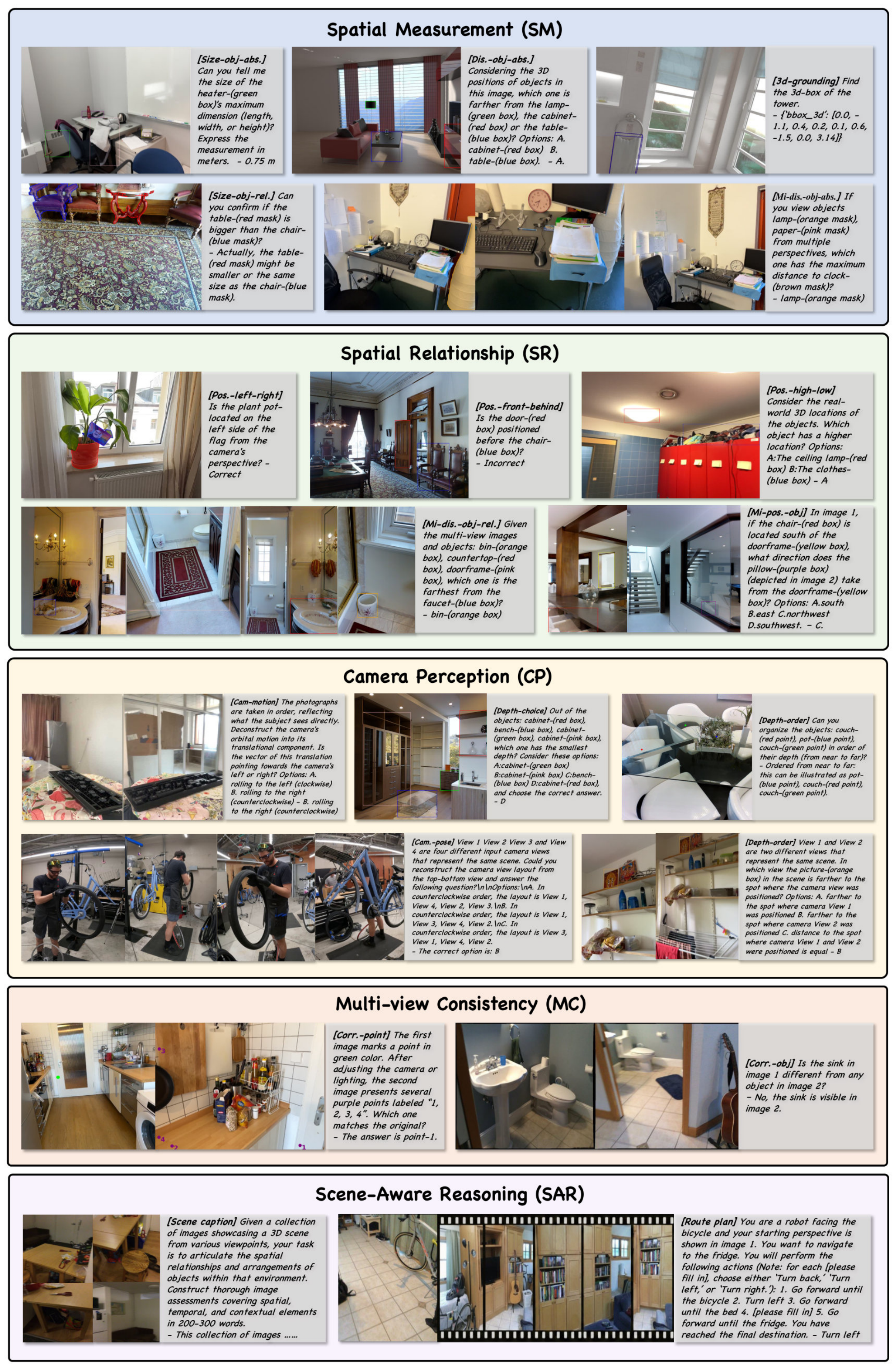}
   \caption{Overview of the OpenSpatial dataset. OpenSpatial consists of high-quality spatial understanding samples organized into a five-category taxonomy: Spatial Measurement (SM), Spatial Relationship (SR), Camera Perception (CP), Multi-view Consistency (MC), and Scene-Aware Reasoning (SAR). QA pairs have been condensed for clarity.} 
   \label{fig:LogiSpatial}
\end{center}
\end{figure}

To bridge the gap between 2D semantic understanding and 3D spatial intelligence, we introduce \textbf{OpenSpatial} (Figure~\ref{fig:pipeline3d}), an automated data engine designed to synthesize spatially-grounded QA pairs from a unified, 3D box-centric representation. A key strength of OpenSpatial is its ability to scale beyond labor-intensive 3D scans by leveraging a robust 3D lifting mechanism, which transforms unconstrained, in-the-wild web videos into high-fidelity training data. Leveraging this engine, we curate \textbf{OpenSpatial-3M}, a comprehensive training suite comprising 3 million entries. This dataset spans five foundational capabilities, including Spatial Measurement (SM), Spatial Relationship (SR), Camera Perception (CP), Multi-view Consistency (MC), and Scene-Aware Reasoning (SAR), as illustrated in Figure~\ref{fig:LogiSpatial}, across 19 diverse sub-tasks, establishing an extensible cornerstone for general-purpose spatial understanding.

The data engine ingests a variety of sources, encompassing high-precision 3D indoor scans (\textit{e.g.}, ScanNet~\cite{scannet}, Matterport3D~\cite{matterport3d}, ARKitScenes~\cite{arkitscenes}, ScanNet++~\cite{scannet++}, and Hypersim~\cite{hypersim}) in addition to the aforementioned web-scale video sequences. To maintain a consistent geometric foundation, all ingested assets are normalized within a canonical coordinate system.

The technical workflow of OpenSpatial begins with the acquisition of scene-level 3D oriented bounding boxes (OBBs), obtained either through manual curation or the 3D lifting procedure. These scene-level primitives are subsequently distilled into frame-level object attributes via a rigorous pipeline of projection, visibility filtering, and mask refinement. This yields a unified object-frame index, a shared representation that synchronizes 3D/2D boxes, instance masks, partial point clouds, and metric metadata. There are two downstream branches:
\begin{itemize}
\item \textbf{Single-view QA:} Extracts fine-grained queries from per-frame scene graphs, grounding language in the 2D plane through explicit visual anchors.
\item \textbf{Multi-view QA:} Capitalizes on the viewpoint-invariant nature of 3D boxes to synchronize objects across overlapping frames. This shared geometric index enables the synthesis of cross-view queries that require consistent spatial reasoning despite significant perspective shifts.
\end{itemize}

At the core of our strategy is the 3D box-centric design, which serves as a robust geometric anchor for all annotations. Unlike traditional 2D-based methods, we utilize 3D OBBs to encapsulate absolute metric scale, centroids, and orientations. For datasets lacking native 3D labels, our lifting mechanism propagates 2D instance masks into 3D space via depth-map integration. To guarantee spatial fidelity, we enforce a multi-view cycle-consistency constraint: a candidate 3D box is validated only if its projections consistently align with observed instance masks across multiple viewpoints. 

\begin{figure}[htbp]
\begin{center}
   \includegraphics[width=\linewidth]{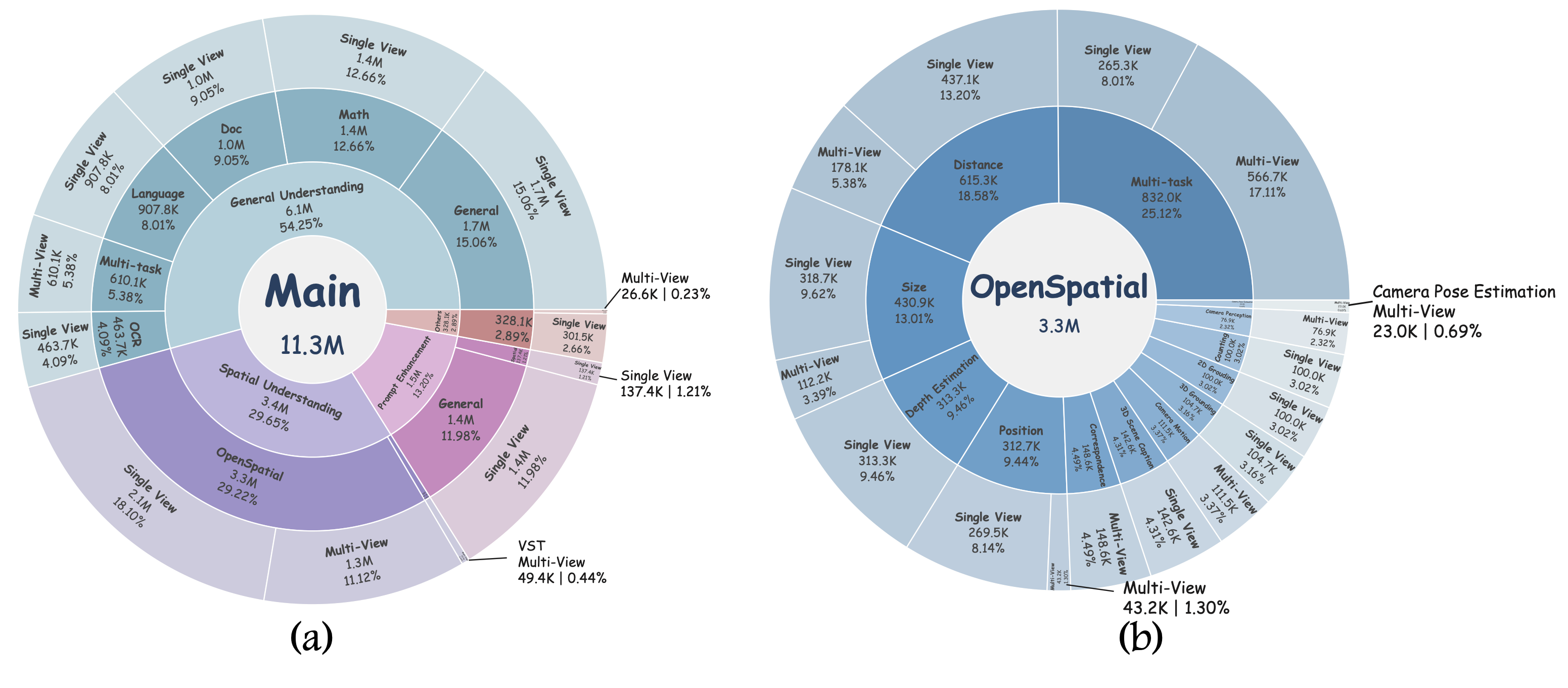}
   \caption{Overview of the data recipe for enhancing spatial understanding.} 
   \label{fig:spatial_data}
\end{center}
\end{figure}

\subsubsection{Dataset Overview \& Statistics}
\label{sec:data-overview}

Our training corpus is organized into four categories: General Understanding, Spatial Understanding, Prompt Enhancement, and Others. 
In total, the corpus contains approximately \textit{11.3M} samples. 
As shown in Figure~\ref{fig:spatial_data}, the data distribution is intentionally non-uniform; therefore, we adopt per-dataset sampling ratios rather than uniform mixing to mitigate the substantial scale imbalance across sources.

\noindent\textbf{General Understanding. }
This is the largest portion of the corpus, comprising about \textit{6.1M} samples (54.25\%). 
It serves as the foundation for preserving broad multi-modal competence, including document understanding, language understanding, OCR, multi-task instruction following, mathematical reasoning, and general visual question answering. Concretely, this category includes large-scale General VQA data (1.7M), Math data (1.4M), Doc/Chart data (1.0M), Language data (907.8K), Multi-task data (610.1K), and OCR data (463.7K). 
Most samples in this category are single-view, making it the main anchor for retaining strong general-purpose visual-language capabilities.


\noindent\textbf{Spatial Understanding. }
This is the core subset for spatial intelligence, comprising about \textit{3.4M} samples (29.65\%). It mainly contains two sources:
\begin{itemize}
  \item \textbf{OpenSpatial} (3.3M): Our principal spatial supervision source, covering a diverse set of fine-grained skills such as distance, size, depth estimation, position, correspondence, 3D scene captioning, camera motion, as well as smaller subsets for 3D grounding, orientation, and multi-view camera pose estimation. This source contains both single-view and multi-view supervision.
  \item \textbf{VST Subset} (49.4K): A compact but important multi-view subset centered on camera motion, providing explicit supervision for dynamic viewpoint changes.
\end{itemize}
Overall, the spatial branch provides comprehensive coverage across single-view, multi-view, and video data, establishing a holistic foundation for 3D/4D spatial reasoning.


\noindent\textbf{Prompt Enhancement.}
In support of downstream image generation and editing tasks, we incorporate two prompt rewriting sources designed to improve instruction density and robustness:
\begin{itemize}
\item \textbf{Instruction Rewriting} (1.4M, 11.98\%): This subset transforms concise, low-entropy descriptions into detailed and stylistically diverse instructions. By leveraging a systematic rewriting pipeline, we expand descriptive granularity while strictly preserving original semantics, enabling the model to interpret complex generative prompts with high fidelity.
\item \textbf{Spatial Editing} (137.4K, 1.21\%): A suite that maps spatial instructions to their corresponding visual transitions. Given a specific prompt, the module normalizes the instruction format to infer the resulting visual content of the target perspective. By explicitly characterizing transformations relative to the original image, it ensures the model captures the precise geometric and semantic changes. 
\end{itemize}


\vspace{0.05in}\noindent\textbf{Others.} (328.1K, 2.89\%): A curated collection of in-house multi-modal understanding sampled from JingDong. These data provide complementary long-tail supervision and improve distributional diversity.

\subsection{Training}
\label{sec:training-recipe}

We perform spatial-specialized supervised fine-tuning (SFT) on Qwen3-VL-8B-Instruct~\cite{qwen3} using a full-parameter training setup. 
To handle the inherent length heterogeneity of our multimodal spatial corpus, where short grounding queries coexist with long multi-turn dialogues, we adopt a dynamic sequence packing strategy that greedily bins multiple short sequences into a single training slot. Packed sub-sequences are kept causally independent via Flash Attention's variable-length interface. 
This design substantially reduces padding waste and decouples throughput from worst-case sequence length. 
To preserve the pre-trained visual representations while allowing the language backbone to adapt more aggressively, we employ a decoupled learning rate schedule that assigns a smaller update rate to the vision encoder. 


In addition to standard cross-entropy SFT, we incorporate an \textbf{online knowledge distillation} objective to retain the original capabilities of the pre-trained model. Specifically, a frozen teacher model provides soft supervision via a KL divergence penalty on intermediate hidden-state representations~\cite{ma2025x2i}. The final training objective is defined as:
\begin{equation}
    \mathcal{L} = \mathcal{L}_{\text{SFT}} + \lambda \cdot \mathcal{L}_{\text{KL}},
\end{equation}
where $\mathcal{L}_{\text{KL}}$ denotes the layer-averaged KL divergence computed over the response tokens. Crucially, this regularization is selectively applied only to general-purpose datasets, while being omitted for spatial understanding tasks. Since the base model's intrinsic spatial capabilities are often limited, imposing KL constraints on such data would inadvertently hinder the acquisition of new spatial knowledge. Conversely, for general domains where the original training recipe is inaccessible and our fine-tuning samples are sparse, the $\mathcal{L}_{\text{KL}}$ term serves as a vital anchor to prevent catastrophic forgetting and maintain the model's foundational knowledge. The detailed training hyperparameters are described in Table~\ref{tab:sft_stage1_core}.


\begin{table}[htbp]
\centering
\caption{Supervised fine-tuning hyperparameters.}
\label{tab:sft_stage1_core}
\small
\renewcommand{\arraystretch}{1.1} 
\setlength{\tabcolsep}{10pt}      
\begin{tabular}{lc @{\hspace{5em}} lc} 
\toprule
\textbf{Parameter} & \textbf{Value} & \textbf{Parameter} & \textbf{Value} \\
\midrule
Training Strategy & FSDP2 & ViT Learning Rate & $5\times10^{-6}$ \\
Total Batch Size & 128 & LR Scheduler & Cosine \\
Micro Batch Size & 1 & Training Epochs & 1 \\
Mixed Precision & BF16 & Max Length & 8192 \\
Learning Rate & $5\times10^{-5}$ & KL Weight ($\lambda$) & 10 \\
\bottomrule
\end{tabular}
\end{table}

\subsection{Evaluation}
We conduct an extensive evaluation based on VLMEvalKit~\cite{duan2024vlmevalkit} to comprehensively assess the performance of our spatial-specialized vision-language model. We introduce Gemini-2.5-Flash~\cite{gemini2.5} as the judger for open-ended questions. 
The benchmarks evaluated in tables are categorized into three tiers:

\noindent\textbf{Level 1: 2D Semantic Perception. }
Benchmarks such as MMBench~\cite{mmbench} and MMStar~\cite{mmstar} serve as the foundation, assessing general visual question answer and cross-modal alignment. 
While MMStar is specifically curated to eliminate language bias, OCRB~\cite{ocrbench} evaluates fine-grained text recognition capabilities. 
MathVista~\cite{lu2023mathvista} evaluate the mathematical reasoning capabilities of foundation models within diverse visual contexts.
These benchmarks verify that our model retains decent general-purpose performance.

\noindent\textbf{Level 2: 3D Spatial Understanding. }
This level focuses on the physical world. 
For example, BLINK~\cite{blink} and CV-3D~\cite{cambrian} assess low-level geometric cues like depth, surface normals, and relative size. In contrast, 3DSR~\cite{3dsrbench} and MMSI~\cite{mmsibench} shift the focus toward spatial logic, requiring the model to resolve complex linguistic relations (\textit{e.g.}, ``between,'' ``behind'') within a 3D coordinate frame. 
RealWorldQA~\cite{grok15v} further tests these capabilities in autonomous driving and outdoor scenarios.

\noindent\textbf{Level 3: 4D Spatio-Temporal Reasoning. }
The most challenging tier involves VSI-Bench~\cite{vsibench} and AllAnglesBench~\cite{yeh2025seeing}. 
Unlike static 3D benchmarks, these require temporal coherence and view-invariant reasoning. 
AllAnglesBench specifically challenges the model to maintain a consistent spatial mental map across extreme viewpoint shifts, while VSI-Bench evaluates the tracking of spatial identities over time.

Table~\ref{tab:main} presents a comprehensive comparison between our model and state-of-the-art proprietary and open-source VLMs across 13 benchmarks. Notably, our model achieves a new state-of-the-art in spatial understanding, reaching an average score of 64.4. This represents a substantial 5.3 point improvement over the base model, even matching the performance of the proprietary Gemini-2.5-Pro. While establishing this dominance in 3D and 4D spatial tasks, our model simultaneously maintains competitive results on general multimodal benchmarks like MMBench and MMStar, demonstrating its versatility without compromising foundational capabilities.

\begin{table}[htbp]
\centering
\caption{Quantitative comparison with state-of-the-art VLMs on 9 spatial benchmarks and 4 general benchmarks. \textbf{Bold} and \underline{underlined} indicate the best and second-best results. The row with a \colorbox[HTML]{E1F5FE}{blue} background denotes our method.}
\setlength{\tabcolsep}{2pt}
\renewcommand{\arraystretch}{1.2}
\resizebox{\textwidth}{!}{
\begin{tabular}{l|c|ccccccccc|cccc}
\toprule
\multirow{2}{*}{\textbf{Benchmarks}} &
\multirow{2}{*}{
    \begin{tabular}[c]{@{}c@{}} 
        \textbf{Spa.} \\ \textbf{Avg.} 
    \end{tabular}
} &
\multicolumn{9}{c|}{\textbf{Spatial Understanding}} &
\multicolumn{4}{c}{\textbf{General Understanding}} \\
\cline{3-15}
& &
VSI & AllAngles &
BLINK & 3DSR\_C & CV-2D & CV-3D & ERQA & RealWorldQA & MMSI &
MMB\_CN & MathVista & MMStar & OCRB \\
\midrule
\multicolumn{15}{c}{\textbf{\textit{Proprietary Models}}} \\ \midrule
Gemini-2.5-Pro~\cite{gemini2.5} & 64.4 &48.4 & 61.3 & 70.6 & 57.6 & 80.4 & 91.3 & 55.8 & 77.3 & 36.9 & 90.2 & 84.9 & 79.1 & 86.6 \\
GPT-4o~\cite{gpt4}  & 57.7 &34.0 & 52.4 & 65.9 & 44.3 & 75.8 & 83.0 & 57.0 & 76.2 & 30.3 & 83.9 & 63.8 & 65.1 & 80.6 \\
\midrule
\multicolumn{15}{c}{\textbf{\textit{Open-Source General/Spatial VLMs}}} \\ \midrule
InternVL2.5-4B~\cite{internvl2.5}& 50.6 & 28.3 & 45.1 & 50.8 &44.0  & 77.1 & 76.4 & 41.0 & 64.2 & 28.5 & 77.6 & 61.6 & 58.5 & 82.6  \\
InternVL2.5-8B~\cite{internvl2.5}& 54.6 & 39.3 & 48.9 & 54.9 &51.0  &  78.6& 79.9 & 40.8 &  69.4& 28.6 & 81.3 & 63.4 & 62.6 & 82.1   \\
InternVL3-2B~\cite{internvl3}  & 50.6 &30.4 & 48.6 & 52.8 & 46.4 & 71.9 & 77.3 & 36.2 & 65.5 & 25.9 & 77.1 & 58.3 & 61.5 & 83.7 \\
InternVL3-8B~\cite{internvl3} & 56.2 &38.7 & 50.5 & 55.7 & 52.7 & \underline{80.6} & 86.0 & 40.5 & 70.6 & 30.9 & 81.9 & 70.8 & 68.2 & \underline{88.0} \\
SpaceR-7B~\cite{spacer}  & 53.3 &44.4 & 49.8 & 54.3 & 47.5 & 73.9 & 76.2 & 40.5 & 64.2 & 29.4 & 80.3 & 65.8 & 61.6 & 85.9 \\
MiMo-VL-7B~\cite{coreteam2025mimovltechnicalreport}  & 58.2 &47.8 & \underline{52.9} & 59.7 & \underline{56.1} & 76.9 & 86.9 & 41.0 & \underline{73.5} & 29.3 & 80.9 & \textbf{81.3} & \underline{71.1} & 84.5 \\
Qwen2.5-VL-3B-Instruct~\cite{qwen2.5vl}& 47.9 & 32.0 & 42.8 &49.0  & 45.2 & 66.1 & 64.8 & 40.8 & 65.2 & 25.0 & 76.9 & 62.1 & 56.6 & 82.6  \\
Qwen2.5-VL-7B-Instruct~\cite{qwen2.5vl} & 52.8 &36.0 & 50.1 & 55.3 & 49.0 & 75.6 & 73.8 & 41.0 & 68.1 & 26.5 & 82.3 & 68.6 & 70.9 & 87.9 \\
VST-7B-SFT~\cite{vst}  & \underline{60.2} &55.3 & 49.5 & 62.1 & 53.3 & 77.9 & \textbf{94.8} & \underline{43.8} & 71.5 & \underline{33.3} & 80.4 & 65.7 & 63.1 & 86.3 \\
Qwen3-VL-4B-Instruct~\cite{qwen3vl2025}& 58.8 & 53.6 & 49.1 &62.6  &52.5  &79.5  & 92.3 & 40.2 & 71.4 & 28.0 & 82.5 & 70.6 & 67.5 & \underline{88.0}  \\
Qwen3-VL-8B-Instruct~\cite{qwen3vl2025} & 59.1 &\underline{55.6} & 49.5 & \underline{66.1} & 52.8 & 78.6 & 90.8 & 40.1 & 70.7 & 28.1 & \underline{83.3} &\underline{75.0}  & 70.1 & \textbf{90.3} \\
\midrule
\multicolumn{15}{c}{\textbf{\textit{Our Method}}} \\ \midrule
\rowcolor[HTML]{E1F5FE} \textbf{JoyAI-Image-Und (Ours)}& \textbf{64.4}&\textbf{60.1}&\textbf{61.0}&\textbf{69.6}&\textbf{60.5}&\textbf{81.0}&\underline{92.7}&\textbf{45.0}&\textbf{74.0}&\textbf{35.8}&\textbf{83.7}&74.4&\textbf{71.3}&87.9 \\
\rowcolor[HTML]{E1F5FE} 
\textbf{$\Delta$ (Ours vs. Base)} & \textbf{+5.3} & \textbf{+4.5} & \textbf{+11.5} & \textbf{+3.5} & \textbf{+7.7} & \textbf{+2.4} & \textbf{+1.9} & \textbf{+4.9} & \textbf{+3.3} & \textbf{+7.7} & \textbf{+0.4} & -0.6 & \textbf{+1.2} & -2.4 \\
\bottomrule
\end{tabular}
}
\label{tab:main}
\end{table}

\section{Text-to-Image: JoyAI-Image}

\subsection{Data Pipeline}

Our data pipeline is designed as a progressive, multi-stage system that jointly optimizes data quality and distributional coverage. The pipeline consists of five core modules: (1)~a \textbf{Data Filtering} module that applies increasingly stringent quality criteria across training stages; (2)~a \textbf{Captioning} module that generates multi-level textual descriptions using a vision-language model; (3)~a \textbf{Rebalancing} module that leverages a large-scale semantic taxonomy to correct long-tail distributional biases; (4)~an \textbf{Annotating} module that employs human experts to establish fine-grained quality standards for quality-tuning data; and (5)~a \textbf{Multi-view Generation} module that curates a million-scale Blender-rendered multi-view corpus with geometric annotation to support viewpoint-controllable generation. We describe each module in detail below.

\subsubsection{Data Filtering}
\label{sec:filtering}

We construct our training data from a large-scale, diverse collection of billions of images sourced from professional photography platforms, web-crawled repositories, and curated internal collections. To ensure data quality throughout the iterative development of JoyAI-Image, we design a multi-stage filtering pipeline that progressively raises the admission threshold as training advances from low-resolution stages to high-resolution stages. The filtering criteria are organized along several complementary dimensions.

In the initial stage, we apply a set of fundamental filters to remove clearly unsuitable samples from the raw data pool, including broken file detection, minimum resolution enforcement (progressively raised from $\min(\text{width}, \text{height}) > 128$ at 208p to $> 512$ at 1024p), deduplication based on MD5 hash and semantic similarity, and NSFW content exclusion. Beyond these basic filters, we deploy specialized operators targeting image quality, aesthetic appeal, and text-image alignment.
Beyond the standard filters, we highlight two filters that prove particularly effective in our pipeline.

\paragraph{In-House Image Quality Assessment (IQA).}
Off-the-shelf image quality metrics typically evaluate a single perceptual dimension and are insufficient for the diverse failure modes in web-crawled data. We develop an in-house IQA operator that jointly considers \textit{statistical image properties} and \textit{learned perceptual quality indicators}, fused through cascaded decision logic. On the statistical side, four low-level indicators are computed from pixel values: \textit{Brightness}, \textit{Entropy}, \textit{Saturation}, and \textit{Sharpness Variance}, which collectively detect overexposure, underexposure, uniform regions, color degradation, blur, and over-sharpening artifacts. On the perceptual side, three learned models provide complementary assessments: \textit{NIQE}~\cite{mittal2012making}, which measures deviation from natural scene statistics; \textit{CLIP-IQA}~\cite{wang2023exploring}, which leverages CLIP's semantic understanding for human-perception-aligned scoring; and \textit{MUSIQ}~\cite{ke2021musiq}, a multi-scale Transformer capturing quality features across spatial scales.

The final decision follows a cascaded protocol: images with extremely low brightness are skipped; samples passing \textit{both} statistical and perceptual thresholds are accepted directly; borderline cases may be recovered through relaxed perceptual thresholds conditioned on sufficient saturation; and images with very high NIQE scores or severely abnormal statistics are unconditionally rejected. Human verification on a representative subset confirms 90\% accuracy in alignment with human judgment.

\paragraph{Caption-Based Content Filter.}
Rather than relying on image-level classifiers, we perform systematic keyword and pattern matching on VLM-generated captions to identify undesirable content, including composite layouts (\textit{e.g.}, collages, split-screens), prominent watermarks or logos, and low-information screenshots or memes. This approach leverages the already-deployed VLM captioner at no additional cost, is orders of magnitude faster than image classification at billion-scale, and can be updated instantaneously by modifying keyword lists. Empirically, it captures the vast majority of problematic samples found by image-based methods while additionally detecting subtle cases that visual classifiers frequently miss.

The filtering pipeline is configured differently for each training stage. Table~\ref{tab:filtering_config} summarizes key threshold configurations across the three main pre-training stages. In later stages, we additionally employ Artimuse~\cite{cao2025artimuse}, a vision-language model-based aesthetic evaluator, retaining only images scoring above 60 in continual training and above 65 in the annotating stage.

\begin{table}[h]
\centering
\scriptsize
\caption{Filtering configurations across pre-training stages. Thresholds are progressively tightened to elevate data quality as training advances to higher resolutions.}
\label{tab:filtering_config}
\begin{tabular}{lccc}
\toprule
\textbf{Criterion} & \textbf{Stage 1 (208p)} & \textbf{Stage 2 (512p)} & \textbf{Stage 3 (1024p)} \\
\midrule
Min Resolution & $\min(h,w)>128$ & $\min(h,w)>256$ & $\min(h,w)>512$ \\
Aesthetic Score & $\geq 3.0$ & $\geq 4.6$ & $\geq 4.6$ \\
IQA Filter & N/A & Retention $\sim$34\% & Retention $\sim$20\% \\
Dense OCR Resolution & N/A & $>512$ & $>768$ \\
Border Detection & N/A & $\checkmark$ & $\checkmark$ \\
Rebalance & N/A & N/A & $\checkmark$ \\
\bottomrule
\end{tabular}
\end{table}

\subsubsection{Captioning}
\label{sec:captioning}

High-quality textual descriptions are critical for training text-to-image models, as they directly determine the upper bound of text-image alignment. We address this through two complementary strategies: (1) a multi-level captioning system that generates diverse textual representations at varying granularities, and (2) an OCR-aware captioning pipeline that ensures faithful text rendering in generated images. Both strategies use Qwen3-VL-8B-Instruct~\cite{qwen3vl2025} as the unified captioning backbone, and all captions are generated in both Chinese and English to support multilingual generation. During training, caption types and language variants are sampled with predefined probabilities, ensuring exposure to diverse textual formats.

\paragraph{Caption Types.}
We generate four types of textual descriptions, each targeting a distinct granularity:
\textbf{Short Captions} are concise one-to-two-sentence descriptions that capture the most salient aspects of the image, mimicking the distribution of real user prompts.
\textbf{Long Captions} are paragraph-length descriptions that comprehensively depict subjects, objects, spatial relationships, background elements, lighting, artistic style, and atmosphere, establishing dense visual-textual mappings for fine-grained generation learning.
\textbf{Extended Long Captions} go further in granularity, providing meticulous descriptions of textures, materials, spatial layouts, and subtle visual details to improve visual fidelity.
\textbf{Structured Captions} are JSON-formatted annotations describing the image along predefined dimensions (subject appearance, background, artistic style, compositional attributes, and visible text), improving per-dimension controllability and enabling flexible data composition during training.

\begin{figure}[htbp]
    \centering
    \includegraphics[width=0.9\textwidth]{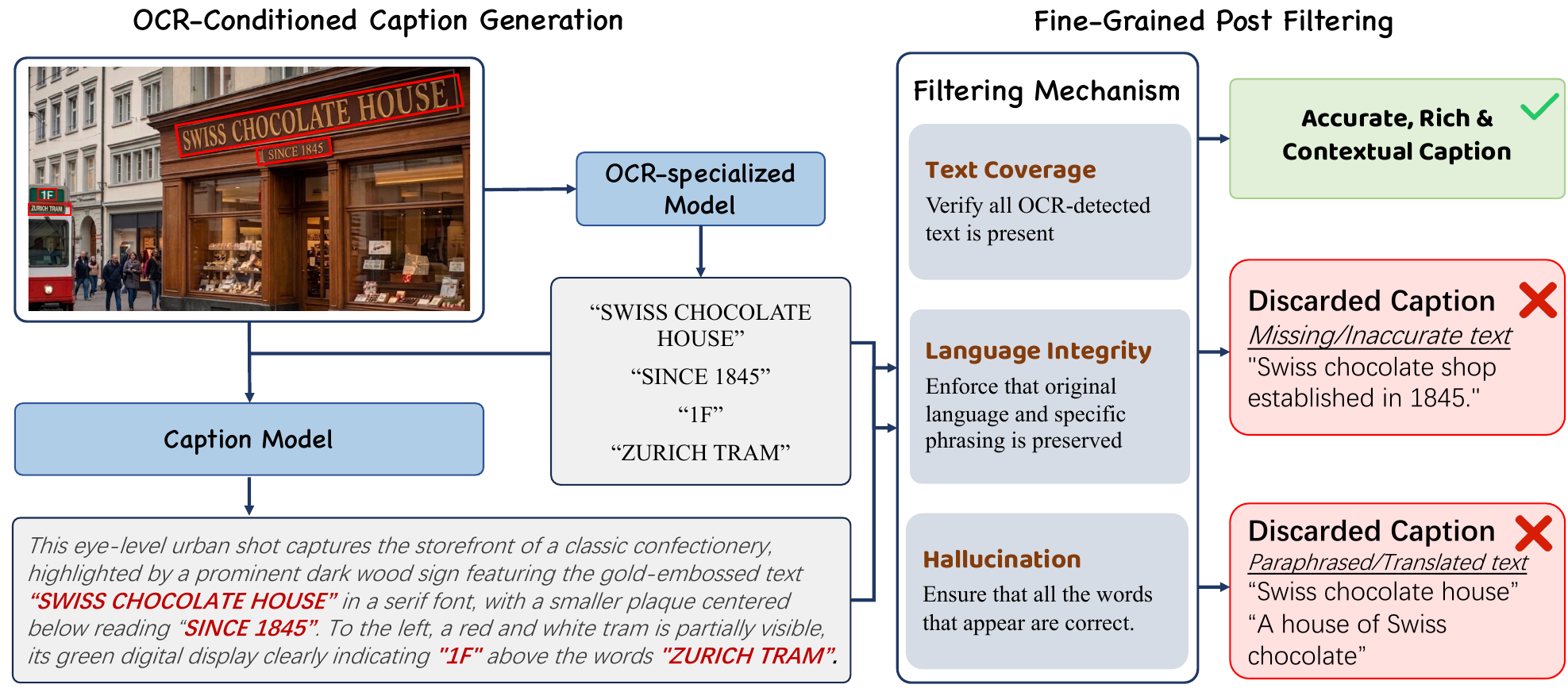}
    \caption{
    Overview of the OCR-conditioned captioning pipeline.
    Given an input image, an OCR-specialized model first extracts textual tokens present in the scene.
    These OCR tokens are then fused with visual features and fed into a captioning model to generate a text-grounded description.
    To ensure fidelity, a fine-grained post-filtering module is applied, and only captions that satisfy all constraints are retained as accurate and contextually grounded outputs.
    }
    \label{fig:ocr_aware_captioning}
\end{figure}

\paragraph{OCR-Aware Captioning.}
Accurate text rendering is a critical capability for image generation models.
We find that incorporating explicit OCR signals into captions is essential for achieving text-faithful generation.
As shown in Figure~\ref{fig:ocr_aware_captioning}, we adopt a two-stage OCR-conditioned framework: an OCR-specialized model~\cite{cui2026paddleocr} first extracts all textual tokens from the image, and an MLLM then generates captions conditioned on both visual features and recognized text.
This design effectively mitigates missing or hallucinated text, particularly in dense and multilingual scenarios.
To further ensure fidelity, we introduce a fine-grained post-filtering mechanism that enforces (i) full coverage of OCR-detected text, (ii) consistency between OCR outputs and caption content, and (iii) preservation of the original language without translation.

To complement this OCR-aware annotation pipeline, we further develop a text-rendering system that generates high-quality training data with controllable typography.
An LLM first converts prompts into structured layout blueprints, while removing typography-related tokens and applying negative prompting to suppress unintended text generation.
Content is then organized under poster-style grid constraints and rendered via HTML/CSS using a Chromium engine, achieving web-grade typographic fidelity.
To improve robustness on rare Chinese characters, we perform frequency-based filtering and morphological decomposition (radical and stroke), followed by balanced sampling to ensure diverse coverage.
Finally, a VLM is used to ground text regions, and the extracted text is incorporated into a recaptioning prompt to produce spatially aware, text-faithful captions while suppressing irrelevant artifacts.

Together, the combination of multi-level captions and OCR-aware captioning provides comprehensive textual supervision that enables JoyAI-Image to learn both holistic scene understanding and fine-grained attribute control.

\subsubsection{Rebalancing}
\label{sec:rebalancing}

Large-scale web-crawled datasets exhibit severe long-tail distributions, where a small number of common concepts dominate while the majority appear infrequently, degrading generation quality on rare concepts and styles. We design a rebalancing pipeline that constructs a ``Caption $\rightarrow$ Embedding $\rightarrow$ Tag $\rightarrow$ Retrieval $\rightarrow$ Rebalance'' processing flow, enabling semantically-aware category balancing at scale.

\paragraph{Taxonomy and Tag Tree.}
Our rebalancing system is built upon a hierarchical classification tree derived from established visual taxonomies~\cite{zhang2025bamboo, wang2024v3detchallenge2024vast}. We select the finest-grained level containing approximately 285K leaf-node categories as our tag vocabulary.

\paragraph{Embedding-Based Tagging.}
We compute embeddings for both the 285K tag vocabulary and each sample's caption, then retain the top-$K$ ($K{=}1000$) tags by cosine similarity as initial candidates, which achieves 86\% human verification accuracy in embedding similarity matching tests, and enabling efficient large-scale tagging without per-sample classification inference.

\paragraph{Adaptive Diversity Sampling.}
A naive top-$K$ selection often yields semantically redundant tags, as related concepts cluster in embedding space. We design an adaptive diversity sampling algorithm that leverages the taxonomy hierarchy: candidate tags are aggregated at intermediate tree levels, a ``parent node'' level is dynamically selected based on the target tag count, and only the highest-scoring child within each parent's subtree is retained. This ensures the final tag set covers diverse semantic dimensions rather than concentrating on a single concept cluster.

\paragraph{Rebalance Strategy.}
The tagged dataset exhibits extreme skew: roughly 2\% of categories account for over 30\% of total frequency (individual counts exceeding 10M), while approximately 50\% of categories appear fewer than 100K times. We address this through stratified sampling: \textit{tail categories} ($<$100K) are fully retained; \textit{head categories} ($\geq$100K) are downsampled via an inverse-logarithmic schedule $N_{\text{sample}} \propto 2/\log(\text{Count})$ with segment-specific base rates for different frequency ranges. Additionally, categories linked to weak model capabilities that are identified through systematic evaluation receive a 20 to 50\% sampling boost. Sampling proceeds from lowest to highest frequency with running deduplication to avoid double-counting.

\subsubsection{Annotating}
\label{sec:annotating}

While automated filtering and scoring provide scalable quality control for the bulk of training data, the later stages of training demand a higher standard of curation that automated methods alone cannot achieve. To this end, we design a human-in-the-loop annotation pipeline that establishes rigorous quality benchmarks for selecting SFT data.

\paragraph{Multi-Dimensional Quality Scoring.}
Human annotators evaluate each candidate image along three orthogonal dimensions, each scored on a discrete scale of \{5, 4, 3, 0\}. \textbf{Aesthetics} (weight: 50\%) evaluates overall visual appeal, including sensory comfort, compositional tension, lighting intent, and color sophistication. \textbf{Information Density} (weight: 30\%) measures the quantity and diversity of learnable visual content; high-scoring images exhibit rich ``subject + object + environment + interaction'' compositions with abundant texture detail, while visually sparse images are penalized despite being technically well-captured. \textbf{Style Purity} (weight: 20\%) assesses the consistency and distinctiveness of the image's visual style, including whether photographic grain, oil painting brushstrokes, ink wash diffusion, or digital illustration precision, without cross-style contamination or characteristic ``AI-generated'' artifacts. Table~\ref{tab:scoring_rubric} provides the full dimension-specific rubric.

\begin{table}[htbp]
    \centering
    \scriptsize
    \caption{Dimension-specific scoring rubric for human quality annotation. Each image receives an independent score per dimension.}
    \label{tab:scoring_rubric}
    \setlength{\tabcolsep}{4pt}
    \begin{tabular}{l >{\raggedright\arraybackslash}p{4.05cm} >{\raggedright\arraybackslash}p{4.05cm} >{\raggedright\arraybackslash}p{4.05cm}}
    \toprule
    \textbf{Score} & \textbf{Aesthetics (50\%)} & \textbf{Information Density (30\%)} & \textbf{Style Purity (20\%)} \\
    \midrule
    5 (Stunning)
        & Cinematic or fine-art caliber; elicits an immediate ``wow.'' Masterful lighting, composition, and color harmony.
        & Dense, multi-layered scene with rich subject, object, and environment interactions and abundant texture variety.
        & Unmistakable, committed style (\textit{e.g.}, film grain, oil impasto, precise vector art) with zero cross-style contamination. \\
    \addlinespace[3pt]
    4 (Outstanding)
        & Professional stock or post-processed quality; intentional stylistic choices and meticulous lighting.
        & Multiple distinct visual elements with clear spatial relationships and moderate texture diversity.
        & Consistent identifiable style with only minor ambiguities that do not undermine the overall impression. \\
    \addlinespace[3pt]
    3 (Good)
        & Well-composed and properly exposed, but without distinctive artistic merit.
        & Adequate content with a clear subject and basic context, yet lacking compositional complexity.
        & Recognizable style category, but executed generically without distinctive character. \\
    \addlinespace[3pt]
    0 (Eliminated)
        & Flat casual snapshots, visible defects (distortion, watermarks, incorrect lighting), or unusable content (blur, blank).
        & Visually sparse (\textit{e.g.}, single object on plain background) or chaotic scenes with no coherent structure.
        & Unclassifiable style, heavy cross-style mixing, or overt AI-generated appearance. \\
    \bottomrule
    \end{tabular}
\end{table}

\paragraph{Quality Control.}
Annotators are calibrated with target score distributions: scores 4 to 5 should account for approximately 10 to 30\% of samples, score 3 for 30 to 50\%, and score 0 for 30 to 40\%; score inflation is actively monitored. Each batch is seeded with 5\% pre-labeled sentinel samples whose ground-truth scores are known; annotators falling below 90\% accuracy on sentinels are flagged for retraining. Daily random audits sample 5\% of the previous day's accepted images, and if the low-quality pass rate exceeds 5\%, the entire batch is returned for re-annotation.

\paragraph{Iterative Refinement.}
Annotation standards are continuously refined via quality audit feedback. Common calibration issues, such as inflated scores for dark-toned artistic photography, under-appreciation of traditional art forms (\textit{e.g.}, Chinese ink painting), or confusion between intentional stylistic blur and photographic deficiency, are addressed through targeted training sessions incorporating exemplars from professional photography competitions, cinematic stills, and high-end commercial photography.

\subsubsection{Multi-view Generation}
To support controllable text-to-image generation under explicit viewpoint constraints, we curate a multi-view generation corpus at approximately 1M-scale. The dataset focuses on both object-centric and scene-centric settings with a single dominant subject, including object-focused cases and interior-design scenarios, and is designed for multi-view collage generation. This design aligns training data with practical inference requirements, where users may request either one target view or a coherent set of views under shared semantics.

\paragraph{Rendering Pipeline.}
We render multi-view images using Blender 4.5. Each image consists of one designated main view and several supporting sub-views, all captured by cameras oriented toward the target object's center. Camera intrinsics are sampled from a commonly used range, and the camera-to-object distance is uniformly sampled from a range scaled relative to the object's bounding extent, ensuring consistent framing across varying object sizes. To discard infeasible camera placements, we cast rays from the central region of each camera's image plane toward the target object and retain only views where a sufficient proportion of rays arrive unobstructed.

\paragraph{Dense and Structured Captioning.}
We adopt a dual-prompt annotation pipeline with Gemini-3-Flash~\cite{google2025gemini3blog} as the labeling model. A dense-caption prompt produces fluent, detail-rich natural language descriptions that emphasize collage layout, main-view content, and complementary evidence from supporting views. In parallel, a structured prompt outputs schema-constrained JSON annotations with fixed fields for subject summarization, rotation modeling, main-view anchoring, per-view details, and layout specification, as summarized in Table~\ref{tab:mv_structure}. This combination improves both linguistic richness and geometric controllability, while providing reliable machine-parsable supervision for downstream data composition and training.

\begin{table}[htbp]
  \centering
  \scriptsize
  \caption{Fields and descriptions in structured captions for multi-view collage data.}
  \label{tab:mv_structure}
  \begin{tabular}{l>{\raggedright\arraybackslash}p{12.7cm}}
  \toprule
  \textbf{Name} & \textbf{Description} \\
  \midrule
  Subject Summary     & A concise global summary of the central object/scene shared across sub-images, including common visual attributes, lighting, and rendering style. \\
  Rotation Logic      & Description of cross-view horizontal rotation progression and viewpoint transition pattern relative to the anchored main view. \\
  Main View Index     & Index of the selected main view that best represents the subject, based on completeness, composition balance, and informativeness. \\
  View Details        & Per-sub-image annotations following reading order (left-to-right, top-to-bottom), including role (Main/Other), angle description, and independent content notes. \\
  Layout Description  &  Description of collage organization, including grid arrangement, number of valid sub-images, blank slots (if any), and overall layout/background cues. \\
  \bottomrule
  \end{tabular}
\end{table}

\subsection{JoyAI-Image Model}
\subsubsection{Pre-Training}
We adopt a flow matching objective~\cite{lipman2022flow}, where the model learns to predict the velocity field along a linear interpolation path between noise and clean data. 
Given an image $x$, we obtain its latent representation $z_1$ via WanVAE~\cite{wan2025wan} and sample Gaussian noise $z_0 \sim \mathcal{N}(0, I)$. 
We construct $
z_t = t z_1 + (1 - t) z_0,$
and train the model to predict the velocity field $z_1 - z_0$:
\[
\mathcal{L} = \mathbb{E}_{t, z_0, z_1, y} 
\left[ \left\| f_\theta(z_t, y, t) - (z_1 - z_0) \right\|^2 \right],
\]
where $y$ denotes the text condition.


In the pre-training stage, we train the MMDiT-based text-to-image model with a progressive multi-resolution schedule. Specifically, we organize the pre-training process into three stages, including low-resolution training (208P), mid-resolution training (512P), and high-resolution training (1024P). The main purpose of this stage is to establish the model’s basic image generation capability under text conditioning, rather than to directly optimize generation quality at the target resolution. To further support downstream spatial editing capabilities, we additionally introduce multi-view data during the high-resolution (1024P) stage. Specifically, for a given prompt, the model is trained to generate images from different viewpoints, which encourages consistent 3D-aware understanding and improves controllability across views.  As shown in Figure~\ref{fig:spatial_cases}, the pretrained models is able to generate multi-view images with single prompt.


\subsubsection{Continue Training}
After the large-scale pre-training stage, the model exhibits strong generalization ability but also inherits the high variance of data. To further improve generation quality, we perform a continue training stage with a narrowed data distribution. Specifically, we construct a high-quality subset from the original training corpus through strict filtering and reweighting. This subset emphasizes visually appealing, compositionally coherent, and semantically accurate samples, effectively removing noisy or low-quality modes. Compared to pre-training, this stage focuses on reducing distribution entropy and guiding the model toward a more concentrated high-fidelity region. This distribution narrowing process enables the model to stabilize generation behaviors, suppress artifacts, and significantly improve visual consistency, while preserving the core semantic coverage learned during pre-training.

\subsubsection{Supervised Fine-Tuning}
Building upon the continued training stage, we further perform supervised fine-tuning (SFT) to enhance two key capabilities: \textbf{complex text rendering} and \textbf{multi-view generation}. To this end, we construct a task-oriented high-quality dataset with thousands of samples and fine-grained human annotations. Compared to the previous stage, the SFT data is more focused and specifically designed to strengthen these two aspects. First, for text rendering, we emphasize complex layout scenarios, including bilingual text (\textit{e.g.}, Chinese and English), dense text regions, and structured typography. We curate data with diverse font styles, spatial arrangements, and long text to improve rendering accuracy and readability.
Second, for multi-view generation, we introduce data that contains consistent subjects across different viewpoints. These samples are designed to improve the model’s ability to maintain identity, structure, and spatial consistency under viewpoint changes.
\subsubsection{Reinforcement Learning}
We mainly follow Flow-GRPO~\cite{liu2025flow} as our reinforcement learning framework for text-to-image generation. To construct diverse and informative online training data, we first collect a large pool of diverse prompts covering a wide range of scenes, subjects, styles, and compositions. For each prompt $c$, we sample a group of images $\{x_0^i\}_{i=1}^G$ together with their reverse-time trajectories $\{(x_T^i, x_{T-1}^i, \dots, x_0^i)\}_{i=1}^G$, and perform group-relative policy optimization based on their rewards.

Following Flow-GRPO, the group-wise normalized advantage is computed as
\begin{equation}
\hat{A}_t^i =
\frac{
R(x_0^i, c) - \mathrm{mean}\left(\{R(x_0^i, c)\}_{i=1}^G\right)
}{
\mathrm{std}\left(\{R(x_0^i, c)\}_{i=1}^G\right)
}.
\end{equation}
The training objective is defined as
\begin{equation}
\begin{aligned}
J_{\mathrm{Flow\mbox{-}GRPO}}(\theta)
&=
\mathbb{E}_{\substack{c \sim \mathcal{C}\; ,\; \{x^i\}_{i=1}^G \sim \pi_{\theta_{\mathrm{old}}}(\cdot|c)}} \\
&\quad\left[
\frac{1}{G}
\sum_{i=1}^G
\frac{1}{T}
\sum_{t=0}^{T-1}
\min\!\left(
 r_t^i(\theta)\hat{A}_t^i,\,
\mathrm{clip}\!\left(r_t^i(\theta), 1-\epsilon, 1+\epsilon\right)\hat{A}_t^i
\right)
-
\beta D_{\mathrm{KL}}(\pi_\theta \,\|\, \pi_{\mathrm{ref}})
\right].
\end{aligned}
\end{equation}

In practice, we employ multiple reward models to provide complementary supervision signals, including an aesthetic reward model for visual quality and a text-image alignment reward model for semantic consistency with the prompt. This multi-reward design provides a more balanced optimization target and improves both perceptual quality and prompt faithfulness.

\subsection{Evaluation}
\subsubsection{Quantitative Results}
\label{sec:image_evaluation}
We conduct a systematic evaluation of JoyAI-Image on representative image generation benchmarks, focusing on long-text rendering, instruction following, and stylistic quality. Overall, the results show that JoyAI-Image achieves highly competitive performance across different settings. In particular, it demonstrates clear advantages on text generation tasks, where it delivers consistently strong bilingual performance in both English and Chinese. 

\paragraph{LongText-Bench.}
To further evaluate long-text rendering capability, we conduct experiments on the LongText-Bench benchmark, which measures the accuracy of generating long texts in both English and Chinese. As shown in Table~\ref{tab:longtext}, JoyAI-Image achieves 0.963 on LongText-Bench-EN and 0.963 on LongText-Bench-ZH, outperforming existing methods. Compared to prior models that exhibit performance gaps across languages, our model maintains consistently high accuracy in both English and Chinese, indicating stable long-text rendering capability across different language settings.

\paragraph{CVTG-2k.}
On the CVTG-2K benchmark shown in Table~\ref{tab:t2i_merged}, our model demonstrates strong text rendering capability, particularly in terms of word accuracy and structural consistency. As shown in Table~\ref{tab:t2i_merged}, JoyAI-Image achieves the highest Word Accuracy of 0.8739, outperforming prior methods such as Z-Image (0.8671) and GPT Image 1 (0.8569), indicating more precise character-level generation. In addition, our model attains a competitive normalized edit distance (NED) score of 0.9369, which measures normalized edit distance between rendered text and ground truth, reflecting strong robustness to spelling and structural errors. These results highlight that our model not only improves exact text correctness (word accuracy) but also maintains high overall string-level fidelity, demonstrating its effectiveness in complex visual text generation scenarios.

\paragraph{OneIG.}
On the OneIG benchmark shown in Table~\ref{tab:t2i_merged}, JoyAI-Image demonstrates competitive bilingual generation performance in both English and Chinese settings. Specifically, our model achieves 0.542 on the English split, ranking second among all compared methods, and obtains 0.521 on the Chinese split, remaining competitive with strong existing systems. 

\paragraph{DPG.}
On the DPG benchmark shown in Table~\ref{tab:t2i_merged}, JoyAI-Image achieves an overall score of 88.05, demonstrating strong general text-to-image generation capability. This result shows that our model maintains high overall generation quality across diverse prompts and evaluation settings.

\paragraph{CoreBench.}
To further evaluate compositional and reasoning capabilities, we conduct experiments on the T2I-CoReBench benchmark, which contains a diverse set of composition and reasoning tasks. As shown in Table~\ref{tab:corebench}, JoyAI-Image achieves an overall score of 68.7. Notably, our model obtains the best performance on the composition split with a mean score of 94.2, outperforming all compared methods across the four composition-related dimensions. On the reasoning split, JoyAI-Image reaches a mean score of 55.9, which is also highly competitive and ranks second overall. These results indicate that our model not only excels at compositional fidelity but also maintains strong reasoning ability in complex text-to-image generation tasks.


\begin{table}[t]
\centering
\caption{Quantitative evaluation results on LongText-Bench~\cite{geng2025xomni}.}
\label{tab:longtext}
\resizebox{0.7\textwidth}{!}{
\begin{tabular}{l c c}
\toprule
Model & LongText-Bench-EN$\uparrow$ & LongText-Bench-ZH$\uparrow$ \\
\midrule
Janus-Pro~\cite{chen2025januspro} & 0.019 & 0.006 \\
BLIP3-o~\cite{chen2025blip3} & 0.021 & 0.018 \\
HiDream-I1-Full~\cite{cai2025hidream} & 0.543 & 0.024 \\
Kolors 2.0~\cite{kolors} & 0.258 & 0.329 \\
FLUX.1 [Dev]~\cite{labs2025flux} & 0.607 & 0.005 \\
OmniGen2~\cite{wu2025omnigen2} & 0.561 & 0.059 \\
BAGEL~\cite{deng2025emerging} & 0.373 & 0.310 \\
GPT Image 1 [High]~\cite{openai2025gptimage} & \underline{0.956} & 0.619 \\
X-Omni~\cite{geng2025xomni} & 0.900 & 0.814 \\
Seedream 3.0~\cite{gao2025seedream30} & 0.896 & 0.878 \\
Z-Image-Turbo~\cite{team2025zimage} & 0.917 & 0.926 \\
Z-Image~\cite{team2025zimage} & 0.935 & 0.936 \\
Qwen-Image~\cite{wu2025qwenimage} & 0.943 & \underline{0.946} \\
\rowcolor[HTML]{E1F5FE}
JoyAI-Image & \textbf{0.963} & \textbf{0.963} \\
\bottomrule
\end{tabular}
}
\end{table}

\begin{table*}[t]
\centering
\setlength{\tabcolsep}{6pt}
\caption{Quantitative evaluation results on representative general T2I benchmarks. 
For OneIG \cite{chang2025oneig} and DPG \cite{hu2024ella}, we report overall scores.}
\label{tab:t2i_merged}
\resizebox{0.8\textwidth}{!}{
\begin{tabular}{l cc ccc c}
\toprule
\multirow{2}{*}{\textbf{Model}} 
& \multicolumn{2}{c}{\textbf{OneIG}}
& \multicolumn{3}{c}{\textbf{CVTG-2K}}
& \textbf{DPG} \\
\cmidrule(lr){2-3} \cmidrule(lr){4-6}
& \textbf{EN} & \textbf{ZH} & \textbf{NED} & \textbf{CLIPScore} & \textbf{Word Acc.} & \textbf{Overall} \\
\midrule
Seedream 3.0~\cite{gao2025seedream30} & 0.530 & \underline{0.528} & 0.8537 & 0.7821 & 0.5924 & \underline{88.27} \\
GPT Image 1 [High]~\cite{openai2025gptimage} & 0.533 & 0.474 & \textbf{0.9478} & \underline{0.7982} & 0.8569 & 85.15 \\
Z-Image~\cite{team2025zimage} & \textbf{0.546} & 0.535 & 0.9367 & 0.7969 & \underline{0.8671} & 88.14 \\
Qwen-Image~\cite{wu2025qwenimage} & 0.539 & \textbf{0.548} & 0.9116 & \textbf{0.8017} & 0.8288 & \textbf{88.32} \\
\rowcolor[HTML]{E1F5FE}
JoyAI-Image & \underline{0.542} & 0.521 & \underline{0.9369} & 0.7990 & \textbf{0.8739} & 88.05 \\
\bottomrule
\end{tabular}
}
\end{table*}
\begin{table*}[t]
\centering
\scriptsize
\setlength{\tabcolsep}{4pt}
\caption{Quantitative evaluation results on T2I-CoReBench \cite{li2026easier}. Best and second-best results are marked in \textbf{bold} and \underline{underline} only for the aggregate metrics (Composition Mean, Reasoning Mean, and Overall). Qwen-Image denotes the Qwen-Image-2512 variant, and the results are referenced from \cite{li2026easier}.}
\label{tab:corebench}
\resizebox{\textwidth}{!}{
\begin{tabular}{l ccccc ccccccccc c}
\toprule
\multirow{2}{*}{\textbf{Model}} & \multicolumn{5}{c}{\textbf{Composition}} & \multicolumn{9}{c}{\textbf{Reasoning}} & \multirow{2}{*}{\textbf{Overall $\uparrow$}} \\
\cmidrule(lr){2-6} \cmidrule(lr){7-15}
& \textbf{MI} & \textbf{MA} & \textbf{MR} & \textbf{TR} & \textbf{Mean}
& \textbf{LR} & \textbf{BR} & \textbf{HR} & \textbf{PR} & \textbf{GR} & \textbf{AR} & \textbf{CR} & \textbf{RR} & \textbf{Mean} & \\
\midrule
Janus-Pro-1B~\cite{chen2025januspro} & 51.0 & 54.5 & 33.8 & 2.9 & 35.5 & 12.9 & 18.1 & 24.7 & 13.4 & 7.1 & 15.1 & 6.7 & 6.4 & 13.0 & 20.5 \\
PixArt-$\alpha$~\cite{chen2023pixart} & 40.2 & 42.2 & 14.2 & 3.3 & 25.0 & 11.6 & 11.6 & 21.1 & 30.4 & 22.6 & 44.4 & 26.7 & 20.9 & 23.7 & 24.1 \\
Janus-Pro-7B~\cite{chen2025januspro} & 54.4 & 59.3 & 40.9 & 7.5 & 40.5 & 19.8 & 20.9 & 34.6 & 22.4 & 11.5 & 30.4 & 8.7 & 9.8 & 19.8 & 26.7 \\
PixArt-$\Sigma$~\cite{chen2024pixart} & 47.2 & 49.7 & 23.8 & 2.8 & 30.9 & 14.7 & 18.3 & 26.7 & 39.2 & 25.7 & 44.9 & 33.9 & 24.3 & 28.5 & 29.3 \\
SD3 Medium~\cite{esser2024scaling} & 59.1 & 57.9 & 35.4 & 9.5 & 40.4 & 22.1 & 21.1 & 35.3 & 51.0 & 37.4 & 47.3 & 35.0 & 27.1 & 34.5 & 36.5 \\
FLUX.1 [Dev]~\cite{labs2025flux} & 58.6 & 60.3 & 44.1 & 31.1 & 48.6 & 24.8 & 23.0 & 36.0 & 61.8 & 42.4 & 57.2 & 36.3 & 30.3 & 39.0 & 42.2 \\
HiDream-I1-Full~\cite{cai2025hidream} & 62.5 & 62.0 & 42.9 & 33.9 & 50.3 & 34.2 & 24.5 & 40.9 & 53.2 & 34.2 & 50.3 & 46.1 & 31.7 & 39.4 & 43.0 \\
Seedream 3.0~\cite{gao2025seedream30} & 79.9 & 78.0 & 63.7 & 47.6 & 67.3 & 36.8 & 33.6 & 50.3 & 75.1 & 54.9 & 61.7 & 59.1 & 31.2 & 50.3 & 56.0 \\
Z-Image-Turbo~\cite{team2025zimage} & 79.5 & 72.2 & 62.7 & 83.9 & 74.6 & 36.9 & 28.8 & 48.7 & 74.0 & 56.2 & 55.8 & 52.0 & 26.0 & 47.3 & 56.4 \\
Qwen-Image$^{*}$~\cite{wu2025qwenimage} & 88.5 & 82.5 & 71.9 & 91.9 & \underline{83.7} & 42.7 & 34.6 & 53.6 & 82.0 & 62.0 & 57.2 & 60.3 & 21.7 & 51.7 & 62.4 \\
GPT Image 1 [High]~\cite{openai2025gptimage} & 84.1 & 75.9 & 72.7 & 86.4 & 79.8 & 59.0 & 54.8 & 65.6 & 87.3 & 76.5 & 82.0 & 70.9 & 56.1 & \textbf{69.0} & \textbf{72.6} \\
\rowcolor[HTML]{E1F5FE}
JoyAI-Image & 91.1 & 96.2 & 91.8 & 97.6 & \textbf{94.2} & 54.8 & 38.9 & 47.9 & 91.6 & 70.1 & 73.6 & 52.7 & 17.8 & \underline{55.9} & \underline{68.7} \\
\bottomrule
\end{tabular}
}
\vspace{1mm}
\end{table*}

\subsubsection{Qualitative Results}

\begin{figure}[htbp]
    \centering
    \includegraphics[width=0.74\linewidth]{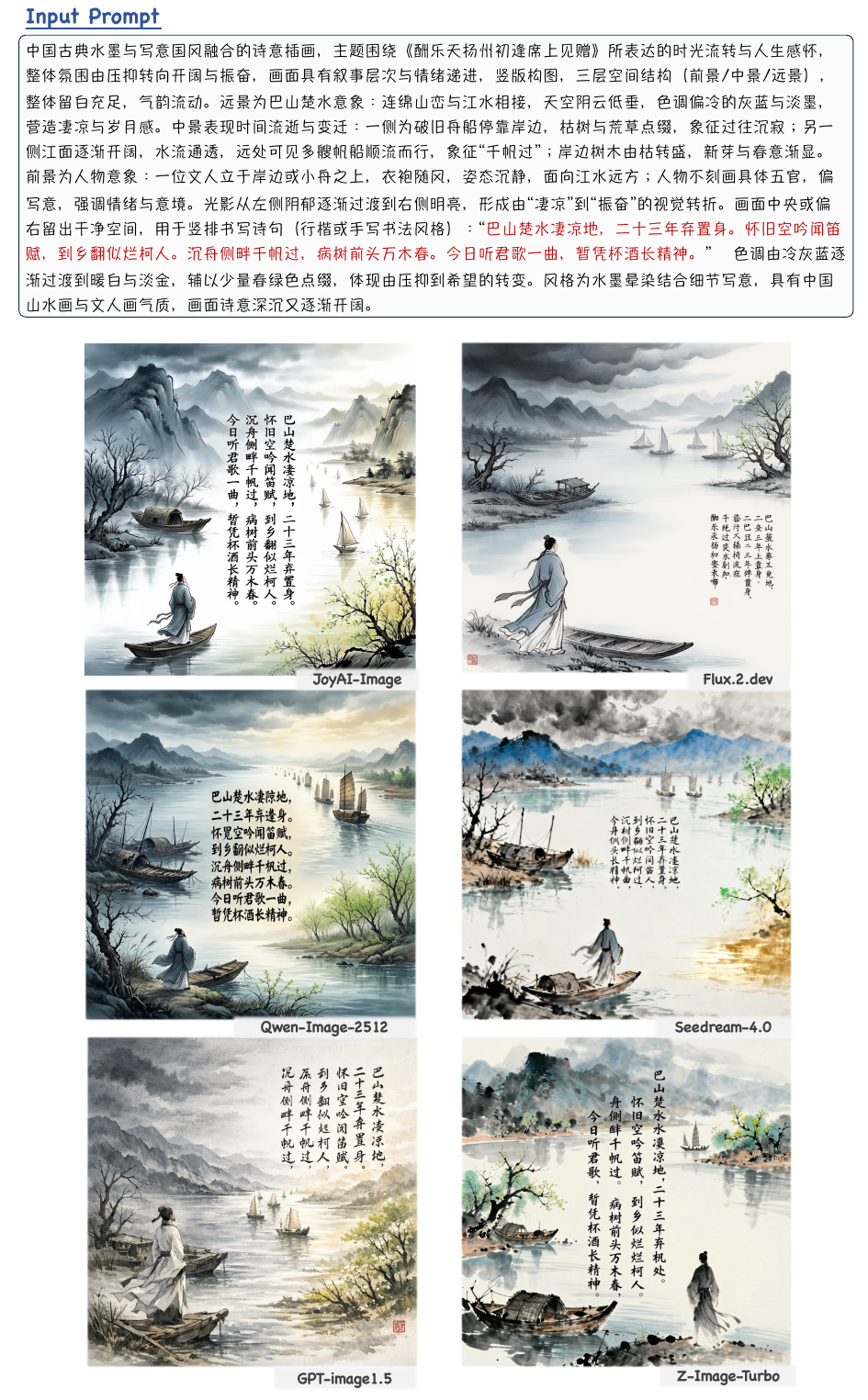}
    \caption{Comparison of different T2I models on a challenging Chinese text-rendering prompt. Among the compared models, JoyAI-Image produces the most faithful and complete rendering of the Chinese text, while also maintaining strong visual coherence.}
    \label{fig:t2i-case1}
\end{figure}

\begin{figure}[htbp]
    \centering
    \includegraphics[width=0.70\linewidth]{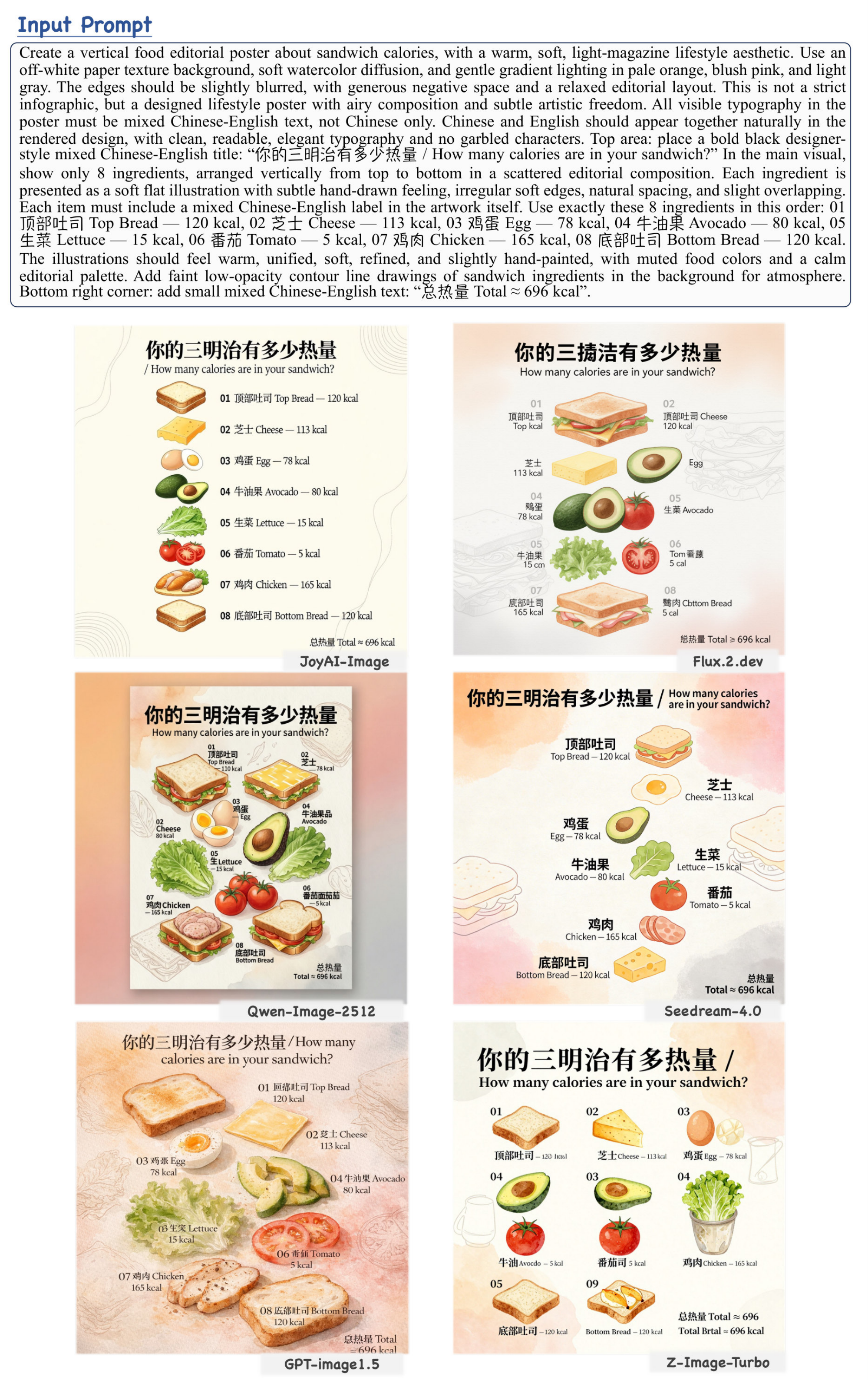}
    \caption{Comparison of different T2I models on a complex bilingual Chinese-English layout prompt. JoyAI-Image demonstrates stronger text rendering ability, producing more accurate bilingual content, clearer typography, and better overall layout fidelity.    }
    \label{fig:t2i-case2}
\end{figure}

\begin{figure}[htbp]
    \centering
    \includegraphics[width=0.76\linewidth]{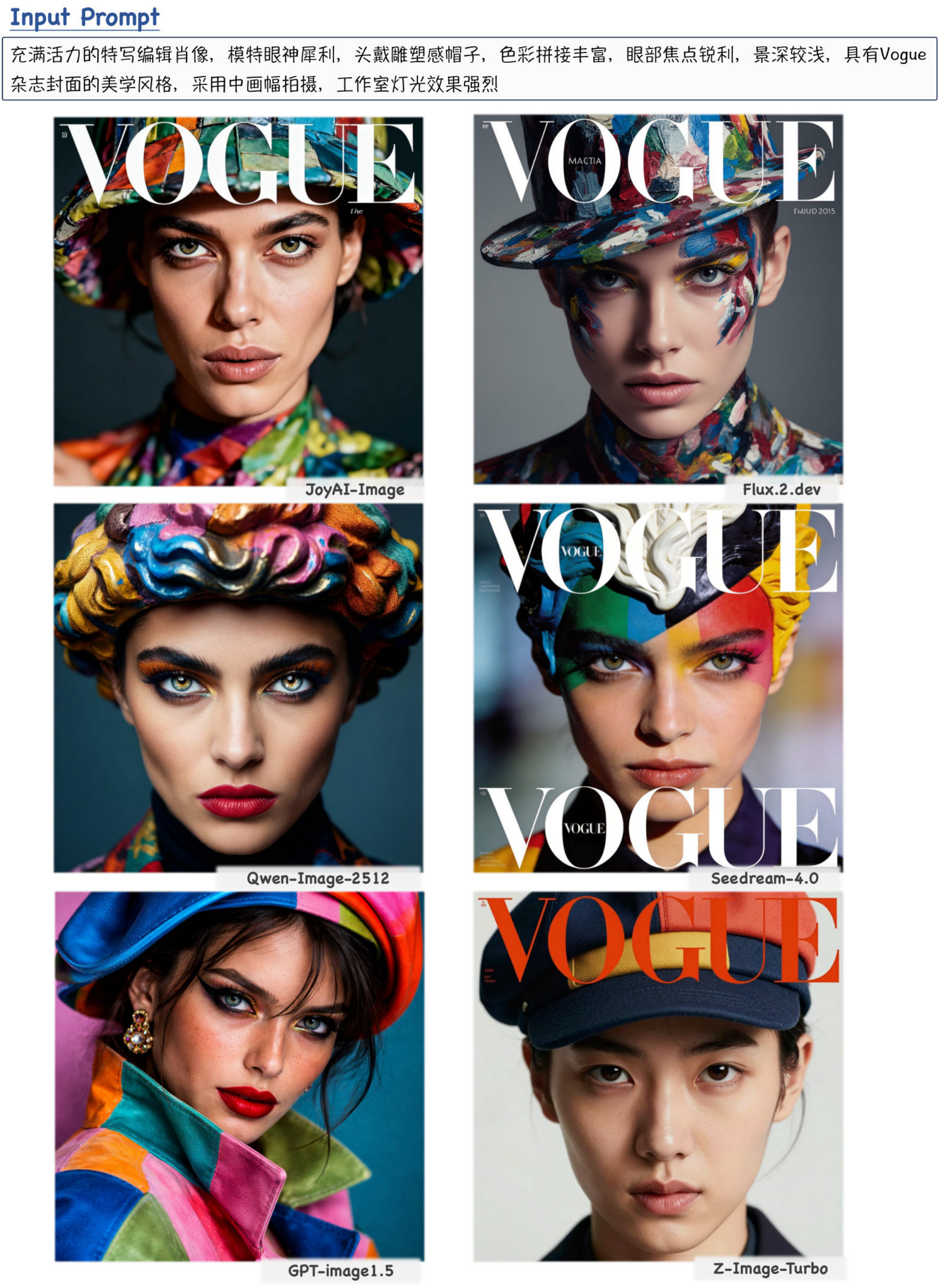}
    \caption{Comparison of different T2I models on a stylized fashion editorial prompt. JoyAI-Image achieves stronger aesthetic appeal, with a more polished composition, richer color harmony, and a high-fashion visual style.}
    \label{fig:t2i-case3}
\end{figure}

We provide qualitative comparisons to illustrate the effectiveness of JoyAI-Image across diverse and challenging generation scenarios. As shown in Figures~\ref{fig:t2i-case1} through~\ref{fig:t2i-case3}, we evaluate three representative settings: (1) long-form Chinese text rendering in artistic scenes, (2) complex bilingual layout generation with structured typography, and (3) stylized high-fidelity visual synthesis in editorial contexts. 

Compared to existing T2I models, JoyAI-Image consistently produces more accurate and complete text rendering, better preserves multilingual consistency, and maintains stronger visual coherence and layout fidelity. In addition, it demonstrates improved aesthetic quality in stylized scenarios, highlighting its ability to jointly model text, layout, and visual appearance.

\section{Image Editing: JoyAI-Image-Edit}

This section presents \textbf{JoyAI-Image-Edit}, our image editing system designed for diverse, instruction-following editing scenarios. Different from pure image generation, image editing requires the model to make targeted modifications while preserving visual fidelity, content identity, and non-target regions. To support this goal, we build the training pipeline around a heterogeneous data system that combines broad-coverage general editing data with specialized supervision for challenging settings such as spatial manipulation, text-centric editing, and multi-image composition. 


\subsection{Image Editing Data}

\begin{figure*}[htbp]
    \centering
    \includegraphics[width=0.95\textwidth]{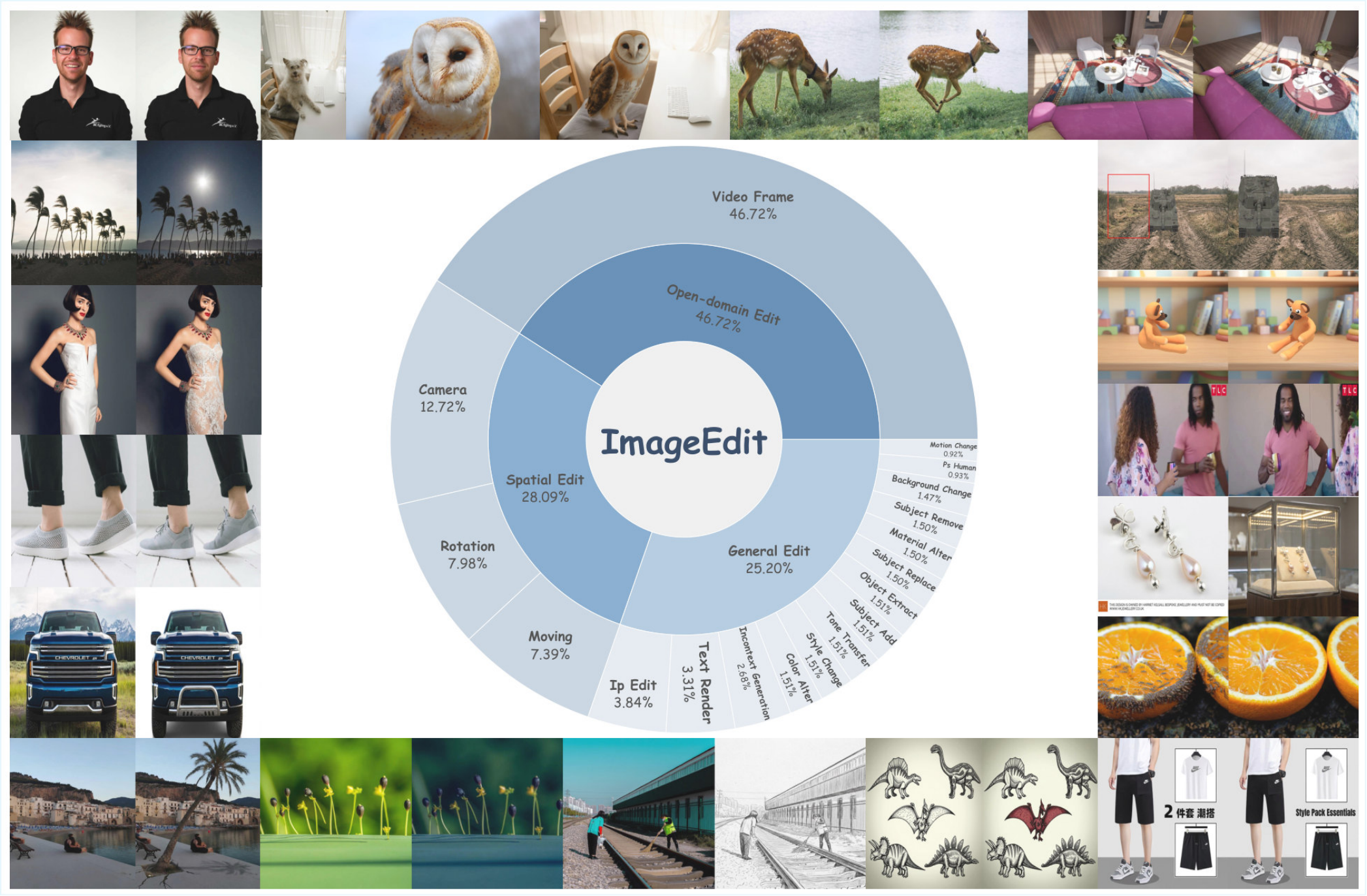}
    \caption{Overview of the training data distribution for editing task.}
    \label{fig:edit_traindata}
\end{figure*}

\subsubsection{Data Distribution}
Our image editing corpus is organized as a {capability-oriented data mixture}. At a high level, it contains three complementary parts: {open-domain editing}, which account for nearly half of the corpus and expose the model to broad, naturally occurring visual changes; {spatial editing}, which occupy a substantial portion and explicitly emphasize geometry-aware transformations such as camera variation, rotation, and object movement; and {specialized general editing}, which cover high-value but challenging scenarios such as text rendering, IP-preserving edits, in-context generation, and fine-grained local modifications. 

This distribution is intentionally designed to reflect the overall goal of JoyAI-Image-Edit. The large open-domain portion builds robust editing semantics and content-preserving behavior under diverse real-world conditions. The strong spatial branch injects supervision for layout, viewpoint, and geometric consistency, aligning the editing engine with our broader effort to strengthen spatial intelligence in a unified framework. The remaining specialist data focuses on controllability-critical long-tail cases, enabling the model to handle precise local edits, typography-sensitive modifications, and reference-conditioned composition. In this sense, our data design is not centered on any single dataset family, but on balancing {broad-domain robustness}, {spatially grounded control}, and {practical long-tail coverage} within one unified editing engine.

\subsubsection{General Editing Data Engine}
Our general editing corpus is designed to support a diverse set of fine-grained image editing tasks, including IP-preserving edits, text rendering, in-context generation, human portrait retouching, motion-aware modification, and a broad family of localized appearance and content transformations such as color alteration, style change, tone transfer, subject addition, removal, and replacement, object extraction, material alteration, and background change. To cover this long-tail task space in a scalable and controllable manner, we construct the corpus from four complementary data streams: open-source editing datasets, expert-model distilled data, in-house text editing data, and multi-image editing data. These sources provide different but mutually reinforcing supervision signals, allowing the model to learn robust general editing semantics while improving faithfulness, controllability, and performance on challenging specialist tasks.

\paragraph{Open-source editing data.}
We incorporate several high-quality open-source editing datasets~\cite{qian2025picobanana400,wang2025gptedit,wu2025omnigen2} as a strong initialization source for general instruction-following behavior. These resources provide diverse source, instruction, and target triplets spanning common edit types such as object insertion and removal, local replacement, attribute modification, style transfer, and background transformation. They offer broad task coverage and relatively mature instruction-image alignment, making them well suited for learning general edit semantics and stable content-preserving behavior at scale. In practice, all samples are normalized into a unified triplet format to ensure consistent downstream training across heterogeneous sources.

\paragraph{Expert-model distilled data.}
To further improve coverage on fine-grained and difficult editing cases, we construct an expert-model distilled dataset using strong image editing models as data generators. This branch is particularly useful for tasks that are either underrepresented or insufficiently clean in existing public resources, such as IP editing, in-context generation, portrait retouching, subtle subject manipulation, and other high-precision local transformations. We use expert models to produce candidate edits under carefully designed task templates, and then apply multimodal verification and filtering to retain samples with strong instruction faithfulness, clear source-target differences, and good visual naturalness. Compared with raw open-source data, this branch provides stronger supervision for long-tail editing behaviors and more controllable edit patterns.

\paragraph{Text editing data.}
We build a dedicated text editing dataset to strengthen the model's ability on text-centric image editing tasks, including text replacement, insertion, removal, and text rendering. These tasks require not only semantic correctness, but also accurate preservation of layout, typography, spacing, and local visual coherence. To balance realism and controllability, we combine real-world text-rich images with a rendering-based data pipeline. The real-world portion provides natural text appearances and complex backgrounds, while the synthetic branch enables scalable construction of layout-aware text modifications under controlled conditions. By combining the two, this data stream provides targeted supervision for fine-grained text edits that are difficult to learn from generic editing corpora alone.

\paragraph{Multi-image editing data.}
To support reference-driven and compositional editing, we construct a multi-image editing dataset using multiple reference images per sample. This branch focuses on tasks that require the model to jointly reason over several visual inputs, such as identity-preserving editing, subject composition, attribute transfer, IP-consistent generation, and structured content borrowing across images. The data includes both curated multi-image resources and high-quality synthesized examples, with dedicated quality control to reduce reference confusion and compositional artifacts. This branch complements single-image editing data by providing supervision for long-context visual conditioning and more complex editing scenarios that depend on multi-reference consistency.

\subsubsection{Open-domain Editing Data Engine}
Open-domain editing data is designed to equip the model with broad instruction-following ability under diverse real-world visual conditions. Rather than focusing on a narrow set of predefined operations, this branch emphasizes naturally occurring content changes, semantic diversity, and robust preservation of non-target regions, thereby serving as the foundation for general-purpose image editing. 

To improve realism and broaden the coverage of naturally occurring edits in open-domain editing data, we construct the entire corpus from video-derived editing pairs that capture visual transformations difficult to obtain from static-image editing data alone. Specifically, we first segment videos into semantically coherent shots and uniformly sample frames within each shot, followed by sharpness-based filtering to remove frames affected by motion blur or defocus. Candidate frame pairs are then constructed from adjacent or short-interval frames within the same shot, such that the overall scene content remains largely consistent while still exhibiting clear local or temporal changes. Based on these frame pairs, we use a multimodal language model to perform difference-based annotation, where the model is instructed to describe only the observable visual changes and rewrite them into executable natural-language editing instructions, with explicit constraints to avoid hallucinating non-existent edits. The resulting annotations are stored in a unified structured schema for the open-domain corpus, forming complete video-derived image editing triplets. Compared with synthetic editing pairs, this video-based open-domain data source provides more realistic supervision for edits involving human motion, object displacement, pose variation, illumination shifts, and subtle scene updates, thereby enriching the model's ability to handle temporally grounded and physically plausible transformations~\cite{qu2025vincie}.

\begin{figure*}[htbp]
    \centering
    \includegraphics[width=0.99\textwidth]{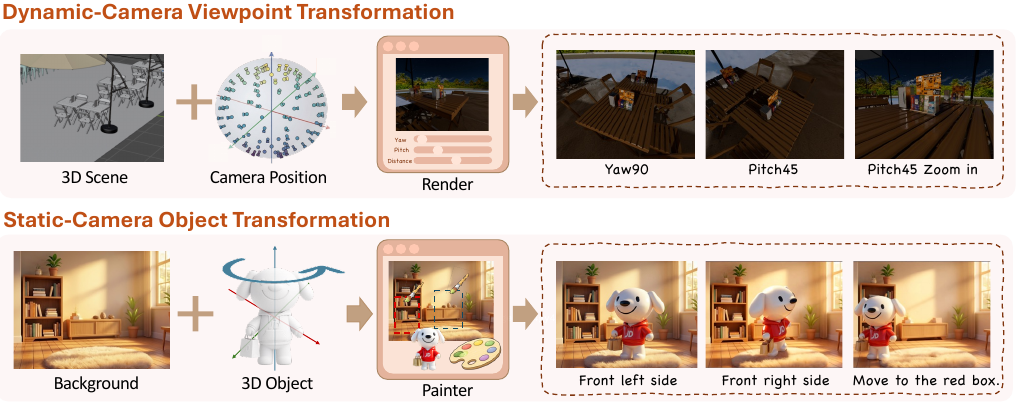}
    \caption{\textbf{Overview of the spatial editing data generation pipeline}. We leverage Blender~\cite{blender} to synthesize both objects and scenes, while preprocessing 3D assets using the segmentation model~\cite{sam3} and the VLM~\cite{gemini2.5}. The object-level engine constructs two inpainting-based data branches to generate object transformations, including rotation, translation, and scaling. The camera-level engine produces viewpoint transformation data by sampling different camera poses, resulting in variations in yaw, pitch, and zoom.
  }
    \label{fig:se_datapipe}
\end{figure*}

\subsubsection{Spatial Editing Data Engine}
Spatial editing imposes a stronger requirement on training data than general appearance editing: the target transformation must be not only visually plausible, but also geometrically unambiguous and instruction-consistent. In practice, such supervision is difficult to obtain from naturally occurring image pairs, since real-world data rarely provides paired examples with explicit object motion, viewpoint change, or controllable spatial intent. To address this limitation, we build a spatial edit data engine~\cite{SpatialEdit-Bench}, a scalable 3D-driven data generation framework that produces paired editing samples with explicit spatial transformations. As illustrated in Figure~\ref{fig:se_datapipe}, the engine is organized into two complementary branches: {Static-Camera Object Transformation}, which focuses on local object-level manipulation under a fixed camera, and {Dynamic-Camera Viewpoint Transformation}, which focuses on global viewpoint control by varying camera poses while preserving the underlying scene structure. Together, these two branches provide unified supervision for geometry-aware spatial editing.

\paragraph{Asset preparation and preprocessing.}
The engine is built upon a curated collection of 3D objects and 3D scenes. Before rendering, we apply an asset preprocessing stage to ensure that the generated samples are semantically meaningful and visually reliable. For object assets, we first canonicalize object orientation and camera setup in Blender so that each asset admits a consistent nominal frontal view. We then use a vision-language model to verify recognizability and remove assets with ambiguous geometry or invalid canonical poses. Multi-view renderings are further checked with SAM-based segmentation to ensure that the foreground object can be reliably localized and remains sufficiently visible under subsequent transformations. For scene assets, we identify visually salient target objects that can serve as anchors for camera control, and discard scenes that yield unstable focus targets, severe occlusion, or visually degenerate renderings. This preprocessing stage is critical for turning raw 3D assets into a stable source of controllable spatial supervision.

\paragraph{Static-Camera Object Transformation.}
The first branch of the engine targets {object-level spatial editing} under a fixed camera. Starting from a canonical rendering of a foreground object, we synthesize spatial edits by applying controlled transformations to the object while keeping the camera and the overall scene layout unchanged. The resulting edits cover representative object-level operations, including {translation}, {scaling}, and {rotation}. To improve realism, the transformed object is composited into semantically compatible backgrounds, and inpainting-based construction is used to maintain local visual coherence after object movement or resizing. This branch produces training pairs in which the edit intent is spatially localized and the non-target context remains stable, making it particularly suitable for learning identity-preserving object manipulation. Compared with generic editing data, these samples provide substantially cleaner supervision for disentangling foreground transformation from background preservation.

\begin{figure*}[t]
\centering
\includegraphics[width=0.8\textwidth]{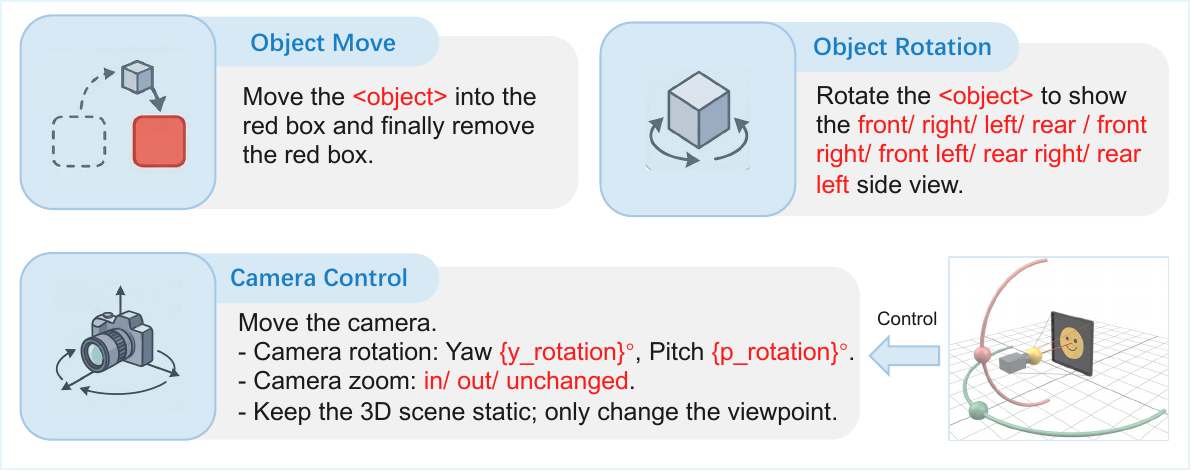}
\caption{Prompt template for spatial editing task. The <red> parts are the user input infomation.}
\label{fig:se_prompt}
\end{figure*}

\paragraph{Dynamic-Camera Viewpoint Transformation.}
The second branch targets {viewpoint-level spatial editing} by explicitly varying camera poses in a fixed 3D scene. For each selected focus object, we parameterize camera motion with three degrees of freedom: {yaw}, {pitch}, and {distance}. Blender is then used to systematically sample camera poses around the focus object while keeping scene geometry and camera intrinsics otherwise controlled. This process generates source-target image pairs with globally consistent scene content but different viewing conditions, covering common viewpoint transformations such as horizontal orbiting, vertical tilting, and zooming in or out. Unlike local object transformation, this branch supervises the model to perform scene-level geometric reconfiguration while preserving object identity, spatial relations, and framing consistency. It therefore serves as a dedicated data source for training global camera-aware editing behaviors that are difficult to express using conventional 2D image editing corpora.

\paragraph{Instruction interface and unified spatial supervision.}
Beyond image synthesis, the engine converts structured spatial transformations into a unified instruction interface for training, as illustrated in Figure~\ref{fig:se_prompt}. Rather than treating different spatial edits as separate supervision formats, we represent them with a shared prompt template that standardizes user intent across transformation types. This interface supports both language-based instructions and image-grounded interaction signals, such as region indicators or interface-like visual guidance, allowing the training data to better reflect practical editing scenarios. By unifying all generated samples under the same instruction format and quality standard, the Spatial Edit Data Engine provides a coherent supervision space for geometry-aware editing, and serves as the core data foundation for learning precise spatial control in JoyAI-Image-Edit.

\subsubsection{Data Curation and Quality Control}

We apply a unified curation pipeline across all data branches to ensure that the final corpus provides valid and learnable editing supervision. At the image level, we remove samples with low visual fidelity, severe artifacts, excessive blur, or weak source-target correspondence. At the supervision level, we filter out pairs with ambiguous edit intent, negligible visual change, or inconsistent instruction-image alignment. We further apply branch-specific checks for specialized data, such as visibility and geometric validity for spatial edits, readability and layout stability for text edits, and reference consistency for multi-image editing. This multi-stage filtering process reduces noisy supervision and improves the precision of edit signals throughout the training corpus.

In addition to automatic filtering, we use model-based quality assessment and targeted manual inspection to monitor data quality across heterogeneous sources. This includes checking semantic faithfulness, content preservation, and visual plausibility of edited results, as well as identifying recurring failure modes such as instruction drift, identity inconsistency, or implausible compositions. By combining scalable automatic screening with focused human verification, we maintain a high-quality dataset while preserving sufficient diversity across editing tasks and visual domains.

\subsubsection{Instruction Refinement and Unified Data Representation}

Since our training data is collected from diverse sources, including open-source datasets, video-derived pairs, text editing data, multi-image samples, and 3D-driven spatial editing data, we convert all samples into a unified representation for downstream training. Each sample is standardized into a common format consisting of the source image, optional reference inputs, a natural-language editing instruction, the target image, and optional structured metadata when needed. This shared representation allows different data branches to be mixed seamlessly while preserving task-specific supervision signals such as spatial transformation parameters or text region annotations.

To improve instruction quality, we further refine noisy, underspecified, or overly templated descriptions into more explicit and executable editing prompts. The refinement process aims to better align the instruction with the actual source-target difference, while reducing ambiguity and improving consistency across data sources. For structured tasks such as spatial editing, the same framework also supports normalized prompt templates and visual interaction cues, enabling a unified instruction interface that covers both language-driven and interface-driven editing scenarios. This design improves the compatibility of heterogeneous data and strengthens instruction-following during training.

\subsubsection{Data Balancing and Training Mixture}

After curation and normalization, we organize the final corpus with a balanced training mixture to avoid over-representation of dominant edit types. In practice, naturally abundant data sources, such as generic single-image editing pairs, can easily overwhelm more specialized but important tasks, including spatial editing, text editing, and multi-image editing. We therefore balance the dataset across multiple axes, including task category, edit granularity, input condition, language, and semantic domain, so that the model receives sufficiently diverse supervision throughout different training process.

\subsection{JoyAI-Image-Edit Model}

As show in Figure~\ref{fig:main_arch}, JoyAI-Image-Edit is built upon a {Multi-Modal Diffusion Transformer (MMDiT)} architecture, which jointly models text conditions, source-image conditions, and noisy latent inputs within a unified denoising framework. Given a source image and an editing instruction, the model predicts the edited target by preserving non-target content while applying the requested modifications. Compared with pure text-to-image generation, image editing places a stronger emphasis on conditional faithfulness, local controllability, and structure preservation. To support these requirements, the model is trained to fuse textual intent with visual evidence from the source image, so that the generated result remains semantically aligned with the instruction and visually consistent with the original content. For spatial editing, camera and geometric transformations are formulated within the unified language instruction template, so that the model learns to perform viewpoint and spatial edits directly from natural-language prompts while preserving scene coherence and object identity.

\subsubsection{Pre-Training}
In the pre-training stage, we initialize JoyAI-Image-Edit with large-scale image generation and reconstruction priors, and then expose the model to coarse-grained editing supervision derived from broad image-text and video-related data. The primary goal of this stage is not to achieve final editing quality, but to establish the model's basic capability to perceive the source image, understand the difference between source and target content, and associate natural-language instructions with corresponding visual operations. In particular, video-derived pairs provide abundant supervision on temporally grounded changes, allowing the model to observe how two highly related images differ and to connect such differences with executable editing intent. As a result, pre-training equips the model with the fundamental ability to interpret instructions as transformations over an input image, laying the foundation for subsequent controllable editing.

\subsubsection{Continue Training}
In the continue training stage, we further adapt the model to high-quality image editing with a curated mixture of general editing data and specialized data branches, including spatial editing, text-centric editing, and multi-image editing. The objective of this stage is to turn the coarse editing capability acquired during pre-training into a more practical and controllable editing behavior. Concretely, continue training improves instruction following, strengthens content preservation, enhances identity and structure consistency, and increases the model's sensitivity to fine-grained user intent. At the same time, this stage places greater emphasis on visual aesthetics and overall image quality, so that edited results remain natural, coherent, and visually pleasing rather than merely satisfying the literal instruction. In this sense, continue training serves as the main stage for aligning editability and image quality.

\subsubsection{Supervised Fine-Tuning}
After continuing training, we optionally apply {supervised fine-tuning (SFT)} to further strengthen specific editing dimensions and improve robustness on user-facing scenarios. The purpose of SFT is to refine the model's performance on the most sensitive aspects of image editing, such as instruction fidelity, local controllability, text accuracy, spatial precision, reference consistency, and difficult long-tail cases that may still be underrepresented in the previous training stages. This stage also allows us to rebalance different data sources and task categories more aggressively, so that the model does not overfit to dominant edit types while maintaining sufficient exposure to specialized tasks. In practice, SFT acts as a targeted adjustment stage that sharpens capability boundaries and improves overall user satisfaction before post-training preference optimization.

\begin{table}[tbhp]
\centering
\caption{Results on GEdit~\cite{liu2025step1x} (Category-wise and Overall Performance). Best results are shown in bold and second-best results are underlined. "PE" denotes the use of the prompt-enhancing technique.}
\label{tab:gedit_benchmark}
\scriptsize
\renewcommand{\arraystretch}{1.2}
\resizebox{0.85\textwidth}{!}{
\begin{tabular}{l|ccc|ccc}
\toprule
\multirow{2}{*}{\textbf{Model}} & \multicolumn{3}{c|}{\textbf{GEdit-Bench-EN}} & \multicolumn{3}{c}{\textbf{GEdit-Bench-CN}} \\
\cmidrule(lr){2-4} \cmidrule(lr){5-7}
& \textbf{G\_SC}$\uparrow$ & \textbf{G\_PQ}$\uparrow$ & \textbf{G\_O}$\uparrow$ 
& \textbf{G\_SC}$\uparrow$ & \textbf{G\_PQ}$\uparrow$ & \textbf{G\_O}$\uparrow$ \\
\midrule

Nano-Banana~\cite{nanopro} 
& 7.396 & {8.454} & 7.291
& 7.540 & {8.424} & 7.399 \\

Seedream4.0~\cite{seedream2025seedream} 
& 8.143 & 8.124 & 7.701 
& 8.159 & 8.074 & 7.692 \\

Nano-Banana-Pro~\cite{nanopro} 
& 8.102 & 8.344 & 7.738 
& 8.135 & 8.306 & 7.799 \\

Seedream4.5~\cite{seedream2025seedream} 
& 8.268 & 8.167 & 7.820
& 8.254 & 8.167 & 7.800 \\

\midrule

FLUX.2 [Dev]~\cite{flux-2-2025} 
& 7.835 & 8.064 & 7.413 
& 7.697 & 8.046 & 7.278 \\

Qwen-Image-Edit-2509~\cite{qwenimage} 
& 7.974 & 7.714 & 7.480 
& 7.988 & 7.679 & 7.467 \\

Step1X-Edit-v1.2~\cite{liu2025step1x} 
& 7.974 & 7.714 & 7.480 
& 7.988 & 7.679 & 7.467 \\

Longcat-Image-Edit~\cite{LongCat-Image}
& 8.128 & 8.177 & 7.748 
& 8.141 & 8.117 & 7.731 \\

Qwen-Image-Edit-2511~\cite{qwenimage} 
& 8.297 & 8.202 & 7.877
& 8.252 & 8.134 & 7.819 \\


FireRed-Image-Edit~\cite{team2026firered}
& 8.363 & 8.245 & \bestsec{7.943} 
& 8.287 & 8.227 & \bestsec{7.887} \\

\rowcolor[HTML]{E1F5FE} \textbf{JoyAI-Image-Edit w/o PE} 
&  {8.829} & 8.120 & {8.276}
& {8.618} & 8.110 & {8.125} \\
\rowcolor[HTML]{E1F5FE} \bestall{JoyAI-Image-Edit w/ PE} 
& {8.806} & {8.273} & \bestall{8.290}
& {8.861} & {8.119} & \bestall{8.208} \\

\bottomrule
\end{tabular}
}
\end{table}

\subsubsection{Post-Training}
In the post-training stage, the model is optimized for higher edit fidelity, better visual quality, and stronger alignment with human preferences. We use DiffusionNFT~\cite{zheng2025diffusionnft} to further enhance consistency, naturalness, and the overall usability of editing results.

\textbf{Rewards}. Reinforcement learning (RL) methods in image editing~\cite{luo2025editscore,li2025uniworld} commonly adopt the \textit{LLM-as-a-Judge} paradigm. This evaluation approach is limited by the inherent capabilities of the Vision-Language Model (VLM) itself, and its assessment of naturalness often fails to align with human preferences. To address these limitations, we utilize {Gemini-3-Flash} and the {HPSv3} model~\cite{ma2025hpsv3} as our reward models. \textbf{(1)} For instruction-following and consistency, we adopt a two-stage scoring pipeline to enable Gemini-3-Flash to provide stable and reliable evaluation scores. Specifically, we first generate textual descriptions of the differences between the reference and the edited images. Then, these descriptions, along with both images, are fed into the VLM to obtain the final scores for instruction-following and consistency. To prevent the model from reward hacking on consistency, inspired by~\cite{zhao2026trust}, we replace simple linear weighting with an instruction-following-prioritized mixing strategy. This design makes instruction-following a necessary condition for receiving a high reward: when the instruction-following score is low, the overall reward remains suppressed regardless of how high the consistency score is. \textbf{(2)} For naturalness: we first use Gemini-3-Flash to pre-generate a target caption that describes the desired edited image, based on the provided reference image and editing instruction. We then compute the HPSv3 score between the target caption and the edited image generated by our model. This score is used to optimize the model for generating more natural and realistic images. Finally, we compute the advantage of each reward~\cite{guo2025deepseek,liu2026gdpo} and then take their weighted sum to obtain the optimality probability.

\textbf{Diffusion Negative-aware FineTuning.} We utilize the aforementioned scoring pipeline to calculate an optimality probability $r \in [0, 1]$ for each sampled edited image $x_0$ based on the reference image $x_r$ and prompt $c$. DiffusionNFT optimizes the training objective:
\begin{equation*}
    \mathcal{L}_{NFT} = \mathbb{E}_{c, \pi^{\text{old}}(x_0|c), t, \epsilon} \left[ r\|v_{\theta}^{+}(x_r, x_t, c, t) - v\|^2_2 + (1 - r)\|v_{\theta}^{-}(x_r, x_t, c, t) - v\|^2_2 \right],
\end{equation*}
\begin{equation*}
        \quad\quad\textit{where} \quad v_\theta^+(x_r, x_t, c, t) \coloneqq (1-\beta) v^\text{old}(x_r, x_t, c, t)+\beta v_\theta(x_r, x_t, c, t), \quad \textit{(Implicit positive policy)}
\end{equation*}
\begin{equation*}
    \;\;\quad\quad\textit{and}\;\; \quad v_\theta^-(x_r, x_t, c, t) \coloneqq (1+\beta) v^\text{old}(x_r, x_t, c, t)-\beta v_\theta(x_r, x_t, c, t). \quad\textit{(Implicit negative policy)}
\end{equation*}

\begin{table}[t]
\centering
\caption{Results on ImgEdit-Bench~\cite{ye2025imgedit} (Category-wise and Overall Performance).
Best results are shown in bold. "PE" denotes the use of the prompt-enhancing technique.}
\label{tab:imgedit_benchmark}
\begingroup
\setlength{\tabcolsep}{2.2pt}
\renewcommand{\arraystretch}{1.05}
\scriptsize
\resizebox{0.98\textwidth}{!}{
\begin{tabular}{l|ccccccccc|c}
\toprule
\textbf{Model} & \textbf{Add} & \textbf{Adjust} & \textbf{Extract} & \textbf{Replace} & \textbf{Remove} & \textbf{Background} & \textbf{Style} & \textbf{Hybrid} & \textbf{Action} & \textbf{Overall}$\uparrow$ \\
\midrule
Nano-Banana~\cite{nanopro}
& {4.62} & 4.41 & 3.68 & 4.34 & 4.39 & 4.40 & 4.18 & 3.72 & {4.83} & 4.29 \\

Seedream4.0~\cite{seedream2025seedream}
& 4.33 & 4.38 & 3.89 & 4.65 & 4.57 & 4.35 & 4.22 & 3.71 & 4.61 & 4.30 \\

Seedream4.5~\cite{seedream2025seedream}
& 4.57 & {4.65} & 2.97 & 4.66 & 4.46 & 4.37 & {4.92} & 3.71 & 4.56 & 4.32 \\

Nano-Banana-Pro~\cite{nanopro}
& 4.44 & 4.62 & 3.42 & 4.60 & 4.63 & 4.32 & {4.97} & 3.64 & 4.69 & 4.37 \\

\midrule
Instruct-Pix2Pix~\cite{brooks2023instructpix2pix}  
& 2.45 & 1.83 & 1.44 & 2.01 & 1.50 & 1.44 & 3.55 & 1.20 & 1.46 & 1.88 \\

MagicBrush~\cite{zhang2023magicbrush}  
& 2.84 & 1.58 & 1.51 & 1.97 & 1.58 & 1.75 & 2.38 & 1.62 & 1.22 & 1.90 \\

AnyEdit~\cite{yu2025anyedit} 
& 3.18 & 2.95 & 1.88 & 2.47 & 2.23 & 2.24 & 2.85 & 1.56 & 2.65 & 2.45 \\

UltraEdit~\cite{zhao2024ultraedit}  
& 3.44 & 2.81 & 2.13 & 2.96 & 1.45 & 2.83 & 3.76 & 1.91 & 2.98 & 2.70 \\

OmniGen~\cite{xiao2025omnigen}  
& 3.47 & 3.04 & 1.71 & 2.94 & 2.43 & 3.21 & 4.19 & 2.24 & 3.38 & 2.96 \\

ICEdit~\cite{zhang2025icedit}  
& 3.58 & 3.39 & 1.73 & 3.15 & 2.93 & 3.08 & 3.84 & 2.04 & 3.68 & 3.05 \\

MindOmni~\cite{xiao2025mindomni}  
& 3.42 & 3.48 & 1.71 & 3.23 & 2.93 & 3.22 & 3.76 & 2.96 & 3.44 & 3.13 \\

BAGEL~\cite{deng2025bagel}  
& 3.56 & 3.31 & 1.70 & 3.30 & 2.62 & 3.24 & 4.49 & 2.38 & 4.17 & 3.20 \\

UniWorld-V1~\cite{lin2025uniworld}  
& 3.82 & 3.64 & 2.27 & 3.47 & 3.24 & 2.99 & 4.21 & 2.96 & 2.74 & 3.26 \\

OmniGen2~\cite{wu2025omnigen2}  
& 3.57 & 3.06 & 1.77 & 3.74 & 3.20 & 3.57 & 4.81 & 2.52 & 4.68 & 3.44 \\

Dreamomini2~\cite{xia2025dreamomni2} 
& 3.93 & 3.09 & 2.11 & 3.95 & 3.64 & 3.75 & 4.38 & 2.90 & 4.04 & 3.53 \\

FLUX.1 Kontext [Dev]~\cite{labs2025flux} 
& 3.99 & 3.88 & 2.19 & 4.27 & 3.13 & 3.98 & 4.51 & 3.23 & 4.18 & 3.71 \\

Step1X-Edit-v1.2~\cite{liu2025step1x}  
& 3.91 & 4.04 & 2.68 & 4.48 & 4.26 & 3.90 & 4.82 & 3.23 & 4.22 & 3.95 \\

Qwen-Image-Edit-2509~\cite{qwenimage}
& 4.34 & 4.27 & 3.42 & 4.73 & 4.36 & 4.37 & 4.91 & 3.56 & 4.80 & 4.31 \\

FLUX.2 [Dev]~\cite{flux-2-2025}
& 4.50 & 4.18 & 3.83 & 4.65 & 4.65 & 4.31 & 4.88 & 3.46 & 4.70 & 4.35 \\

Emu3.5~\cite{cui2025emu35nativemultimodalmodels}
& {4.61} & 4.32 & 3.96 & {4.84} & 4.58 & 4.35 & 4.79 & 3.69 & 4.57 & 4.41 \\

ChronoEdit~\cite{wu2025chronoedit}
& 4.48 & 4.39 & 3.49 & 4.66 & {4.67} & {4.57} & 4.91 & 3.82 & {4.83} & 4.42 \\

LongCat-Image-Edit~\cite{LongCat-Image}
& 4.44 & 4.53 & 3.83 & {4.80} & 4.60 & 4.33 & {4.92} & 3.75 & {4.82} & 4.45 \\

Qwen-Image-Edit-2511~\cite{qwenimage}
& 4.54 & 4.57 & 4.13 & 4.70 & 4.46 & 4.36 & 4.89 & {4.16} & 4.81 & {4.51} \\

FireRed-Image-Edit~\cite{team2026firered}
& 4.55 & {4.66} & 4.34 & 4.75 & 4.58 & {4.45} & {4.97} & {4.07} & 4.71 & \bestsec{4.56} \\

\rowcolor[HTML]{E1F5FE}
\textbf{JoyAI-Image-Edit w/o PE}
& 4.47 & 4.48 & {4.31} & 4.57 & 4.75 & 4.33 & 4.79 & 3.72 & 4.69 & 4.46 \\

\rowcolor[HTML]{E1F5FE}
\textbf{JoyAI-Image-Edit w/ PE}
& 4.63 & 4.52 & {4.32} & 4.71 & {4.76} & 4.53 & 4.88 & 4.09 & 4.69 & \bestall{4.57} \\

\bottomrule
\end{tabular}
}
\endgroup
\end{table}


\begin{table}[tbhp]
\centering
\caption{Comparison between the SFT and RL versions of JoyAI-Image-Edit on GEdit~\cite{liu2025step1x} and ImgEdit-Bench~\cite{ye2025imgedit}.}
\label{tab:sft_vs_rl}
\scriptsize
\renewcommand{\arraystretch}{1.1}
\resizebox{0.9\textwidth}{!}{
\begin{tabular}{l|ccc|ccc|c}
\toprule
\multirow{2}{*}{\textbf{Model}} 
& \multicolumn{3}{c|}{\textbf{GEdit-Bench-EN}} 
& \multicolumn{3}{c|}{\textbf{GEdit-Bench-CN}}  
& \multicolumn{1}{c}{\textbf{ImgEdit-Bench}} \\
\cmidrule(lr){2-4} \cmidrule(lr){5-7} \cmidrule(lr){8-8}
& \textbf{G\_SC}$\uparrow$ & \textbf{G\_PQ}$\uparrow$ & \textbf{G\_O}$\uparrow$
& \textbf{G\_SC}$\uparrow$ & \textbf{G\_PQ}$\uparrow$ & \textbf{G\_O}$\uparrow$
& \textbf{Overall}$\uparrow$ \\
\midrule

JoyAI-Image-Edit (SFT)
& 8.566 & 8.114 & 8.090 
& 8.180 & 7.882 & 7.753 
& 4.40\\

JoyAI-Image-Edit (RL)
& 8.829 & 8.120 & 8.276 
& 8.618 & 8.119 & 8.125 
& 4.46\\

\rowcolor[HTML]{E1F5FE} $\Delta$ (\textit{RL Gain})
& $+0.263$ & $+0.006$ & $+\textbf{0.186}$
& $+0.438$ & $+0.237$ & $+\textbf{0.372}$
& $+\textbf{0.06}$ \\

\bottomrule
\end{tabular}
}
\end{table}

\subsection{Model Performance}
We conduct a comprehensive evaluation of JoyAI-Image-Edit through quantitative benchmarks, human evaluation, and qualitative comparisons. Our evaluation covers both general instruction-based editing and fine-grained spatial editing, with a focus on instruction following, content preservation, perceptual quality, and geometric faithfulness. The results consistently show that JoyAI-Image-Edit maintains strong general editing performance while achieving substantial improvements on spatiall manipulation tasks.

\subsubsection{Benchmark Results}

We evaluate our model on three editing benchmarks, covering both general-purpose instruction-based editing and fine-grained spatial manipulation: GEdit~\cite{liu2025step1x}, ImgEdit~\cite{ye2025imgedit}, and SpatialEdit-Bench~\cite{SpatialEdit-Bench}. 
GEdit and ImgEdit primarily assess general editing quality, including instruction following, semantic consistency, and preservation of irrelevant image content, while SpatialEdit-Bench focuses on geometry-sensitive spatial editing, covering both object-centric manipulation and camera-centric view control. 
This evaluation protocol is aligned with our goal of building a model that not only remains competitive on standard image editing tasks, but also performs faithful geometric transformations under natural language instructions.

\paragraph{Results on GEdit.~\cite{liu2025step1x}}
GEdit~\cite{liu2025step1x} is a representative benchmark for general instruction-based image editing, and mainly evaluates whether a model can follow semantic editing instructions while preserving overall visual coherence. As presented in Table~\ref{tab:gedit_benchmark}, JoyAI-Image-Edit achieves competitive results relative to established open-source and closed-source baselines~\cite{nanopro,seedream2025seedream,qwenimage,uniworld-v2,LongCat-Image}, suggesting that the spatial editing specialization does not degrade its broader editing performance. Specifically, compared with FireRed-Image-Edit~\cite{team2026firered}, our model improves the overall G\_O from 7.943 to 8.290 on GEdit-Bench-EN and from 7.887 to 8.208 on GEdit-Bench-CN. It also achieves the best G\_SC among open-source models on both splits while maintaining strong G\_PQ, suggesting that stronger spatial understanding of images can directly benefit standard editing quality. We further compare the SFT and RL versions of JoyAI-Image-Edit, as shown in Table~\ref{tab:sft_vs_rl}. The RL model consistently outperforms the SFT baseline across all metrics on both the EN and CN splits. In particular, RL improves G\_O by +0.186 on GEdit-Bench-EN and by +0.372 on GEdit-Bench-CN, with especially notable gains in G\_SC and G\_PQ on the CN split. These results suggest that preference optimization further strengthens instruction alignment and perceptual quality, while also narrowing the gap between the EN and CN settings.

\paragraph{Results on ImgEdit.~\cite{ye2025imgedit}}
ImgEdit provides a complementary view of general-purpose editing performance by placing stronger emphasis on instruction adherence, editing quality, and preservation of irrelevant content. As shown in Table~\ref{tab:imgedit_benchmark}, our model also achieves competitive results on this benchmark, further confirming that continued training on spatial editing data does not lead to a noticeable regression in general editing behavior. Notably, JoyAI-Image-Edit w/ PE achieves the best overall score of 4.57 and performs particularly well on more challenging categories such as \textit{Extract}. This suggests that activating stronger spatial capability not only benefits geometry-sensitive edits but also helps maintain a balanced performance across diverse editing tasks. We also compare the SFT and RL versions of JoyAI-Image-Edit on ImgEdit-Bench, as shown in Table~\ref{tab:sft_vs_rl}. The RL version improves the overall score from 4.40 to 4.46, showing that preference optimization brings further gains even on a general-purpose editing benchmark. Together with the GEdit results, this indicates that RL not only benefits spatially challenging edits but also improves overall editing quality, instruction following, and content preservation in standard image editing scenarios.


\begin{table*}[t]
\centering
\caption{Performance comparison of different models on the SpatialEdit-Bench~\cite{SpatialEdit-Bench} benchmark. Higher object editing scores indicate better performance, while lower camera control errors indicate better performance.}
\label{tab:spatialedit-bench}
\scriptsize
\setlength{\tabcolsep}{6pt}
\renewcommand{\arraystretch}{1.1}
\resizebox{0.9\textwidth}{!}{
\begin{tabular}{lcccccc}
\toprule
\multirow{3}{*}{\textbf{Method}} 
& \multicolumn{2}{c}{\textbf{Object}} 
& \multicolumn{2}{c}{\textbf{Camera}} 
& \textbf{Object} 
& \textbf{Camera} \\
& \textbf{Moving} & \textbf{Rotation} & \textbf{Viewpoint} & \textbf{Framing} & \textbf{Overall} & \textbf{Overall} \\
& \textbf{Score $\uparrow$} & \textbf{Score $\uparrow$} & \textbf{Error $\downarrow$} & \textbf{Error $\downarrow$} & \textbf{Score $\uparrow$} & \textbf{Error $\downarrow$} \\
\midrule
\multicolumn{7}{c}{\textit{\textbf{Video World Model}}} \\
Veo3.1~\cite{veo} & N/A    & N/A   
& 1.351 & 0.749 & N/A &1.050\\
ViduQ2-Turbo~\cite{vidu}      & N/A    & N/A   
& 1.022 & 0.771 & N/A    & 0.897 \\
Kling-V2.5~\cite{kling}   & N/A    & N/A     
& 1.051 & 0.733 & N/A &0.892\\
ReCamMaster~\cite{recammaster}   & N/A    & N/A     
& 0.755 & 0.720 & N/A &0.738\\
LingBot-World~\cite{lingbot-world}   & N/A    & N/A     
& 0.696 & 0.701 & N/A &\bestsec{0.699}\\

\midrule
\multicolumn{7}{c}{\textit{\textbf{Closed-Source Image Model}}} \\
Nano-Banana-Pro~\cite{nanopro}        & 0.099 & 0.420 & 0.845 & 0.708 & 0.260 & 0.777 \\
Seedream4~\cite{seedream2025seedream}           & 0.163 & 0.482 & 0.839 & 0.701 & 0.323 & {0.770} \\
\midrule
\multicolumn{7}{c}{\textit{\textbf{Open-Source Image Model}}} \\
QwenImageEdit~\cite{qwenimage}      & 0.311 & 0.531 & 0.922 & 0.692 & 0.421 & 0.807 \\
Edit-R1~\cite{uniworld-v2}             & 0.306 & 0.562 & 0.959 & 0.688 & 0.434 & 0.824 \\
LongCatImage-Edit~\cite{LongCat-Image}  & 0.373 & 0.505 & 0.802 & 0.684 & \bestsec{0.439} & 0.743 \\

\rowcolor[HTML]{E1F5FE}
\textbf{JoyAI-Image-Edit} & {0.652} & {0.646} & {0.290} & {0.568} & \bestall{0.649} & \bestall{0.429} \\
\bottomrule
\end{tabular}
}
\end{table*}


\paragraph{Results on SpatialEdit-Bench.~\cite{SpatialEdit-Bench}}
SpatialEdit-Bench is designed to evaluate fine-grained spatial editing, with a particular focus on whether an edit is geometrically correct rather than merely visually plausible. It covers both object transformation and camera control, making it more diagnostic of true spatial capability than prior general editing benchmarks. Table~\ref{tab:spatialedit-bench} demonstrates that JoyAI-Image-Edit not only exceeds leading image editing approaches~\cite{nanopro,seedream2025seedream,qwenimage,uniworld-v2} but also surpasses recent video world models~\cite{vidu,veo,kling,recammaster,lingbot-world}. Specifically, compared with LongCatImage-Edit~\cite{LongCat-Image}, JoyAI-Image-Edit improves the Moving Score from 0.373 to 0.652, the Rotation Score from 0.505 to 0.646, reduces the Viewpoint Error from 0.802 to 0.290, and reduces the Framing Error from 0.684 to 0.568. This leads to a large gain in Object Overall Score from 0.439 to 0.649 and a major reduction in Camera Overall Error from 0.743 to 0.429. We also compare against representative video models, and our unified image model still achieves clearly better camera-control accuracy, showing that spatial understanding and editing in image-based unified models can be pushed to a new level. Overall, these results show that our approach significantly improves faithful geometric compliance and narrows the gap between semantic plausibility and precise spatial instruction following.

\subsubsection{Human Evaluation}

\begin{figure*}[t]
\begin{center}
   \includegraphics[width=1\linewidth]{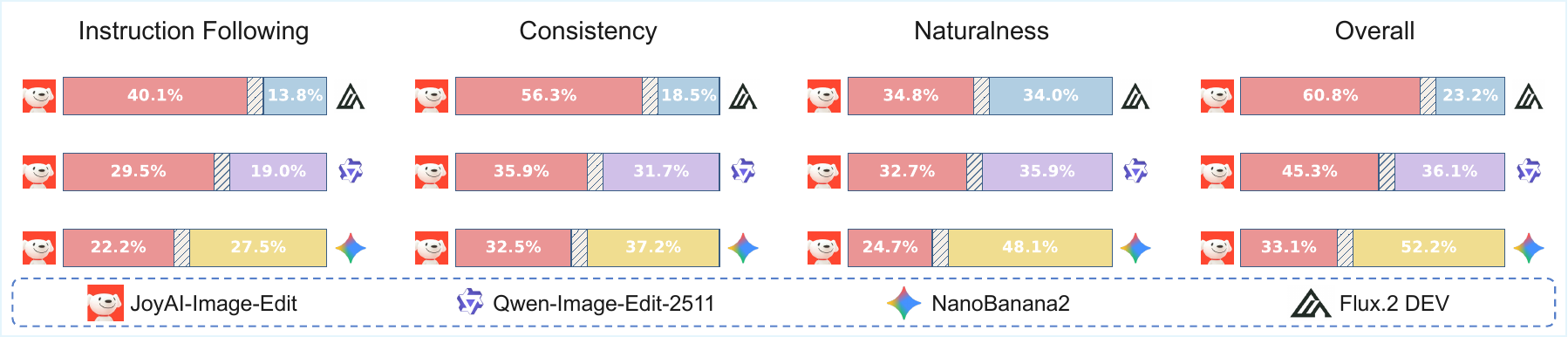}
   \caption{Human evaluation of JoyAI-Image-Edit  against competing models across multiple editing dimensions.}
   \label{fig:abtest}
\end{center}
\end{figure*}

\paragraph{Evaluation Dimensions.}
We further conduct human evaluation to assess the practical editing quality of JoyAI-Image-Edit from three complementary perspectives. {Semantic Following} measures whether the edited image faithfully executes the user instruction, including the requested attribute, object, or spatial transformation, while avoiding under-editing or semantic drift. {Consistency} evaluates how well the model preserves subject identity, scene structure, geometric coherence, and non-target content throughout the edit, which is particularly important for viewpoint change and camera-control scenarios. {Naturalness} measures the perceptual realism of the edited result, including texture continuity, boundary quality, illumination consistency, and the absence of visible artifacts. {Overall} reflects a holistic human preference that jointly considers instruction faithfulness, preservation quality, and visual plausibility. As shown in Figure~\ref{fig:abtest}, each A/B comparison is grouped into four outcomes: JoyAI-Image-Edit is preferred, the competing model is preferred, both outputs are satisfactory, or both outputs are unsatisfactory, with the latter two shown as shaded regions.

\paragraph{AB Test Results.}
As shown in Figure~\ref{fig:abtest}, the human evaluation reveals a clear strength profile of JoyAI-Image-Edit across different competing systems. Against {Qwen-Image-Edit-2511~\cite{qwenimage}}, JoyAI-Image-Edit is preferred on {Semantic Following} (29.5\% vs.\ 19.0\%), {Consistency} (35.9\% vs.\ 31.7\%), and {Overall} (45.3\% vs.\ 36.1\%), while trailing slightly on {Naturalness} (32.7\% vs.\ 35.9\%). This result indicates that JoyAI-Image-Edit more reliably performs the requested transformation and better preserves scene coherence, while perceptual realism remains a relatively closer competition. Against {Nano-Banana-2~\cite{nanopro}}, JoyAI-Image-Edit is less preferred on all four dimensions, including {Semantic Following} (22.2\% vs.\ 27.5\%), {Consistency} (32.5\% vs.\ 37.2\%), {Naturalness} (24.7\% vs.\ 48.1\%), and {Overall} (33.1\% vs.\ 52.2\%). The largest gap appears on {Naturalness}, suggesting that perceptual polish remains the main area for improvement when compared with the strongest baseline in this study.

In contrast, JoyAI-Image-Edit shows a substantial advantage over {Flux.2 [DEV]~\cite{flux-2-2025}} across nearly all dimensions. It is preferred on {Semantic Following} (40.1\% vs.\ 13.8\%), {Consistency} (56.3\% vs.\ 18.5\%), and {Overall} (60.8\% vs.\ 23.2\%), and remains slightly ahead on {Naturalness} (34.8\% vs.\ 34.0\%). The especially large margins on Consistency and Overall suggest that JoyAI-Image-Edit is markedly stronger at balancing faithful edits with preservation of identity, structure, and scene-level coherence. Overall, the human evaluation shows that the main advantage of JoyAI-Image-Edit lies in {instruction faithfulness} and {edit consistency}, especially for structurally constrained or spatially sensitive edits, which is consistent with our geometry-aware training design and the quantitative gains observed on spatial editing benchmarks.

\subsubsection{Qualitative Results}
To comprehensively assess the image editing capability of JoyAI-Image-Edit, we conduct a qualitative evaluation covering both general-purpose editing and spatially grounded editing. For comparison, we benchmark our model against several strong image editing baselines, including QwenImageEdit~\cite{qwenimage}, LongCatImage-Edit~\cite{LongCat-Image}, Nano-Banana-Pro~\cite{nanopro}, Seedream4~\cite{seedream2025seedream}, and GPT Image 1.5~\cite{openai2025gptimage}. As shown in the qualitative comparisons, JoyAI-Image-Edit consistently demonstrates stronger instruction following, better content preservation, and more faithful spatial control, especially on challenging viewpoint and camera manipulation cases.

\begin{figure}[htbp]
\begin{center}
   \includegraphics[width=0.99\linewidth]{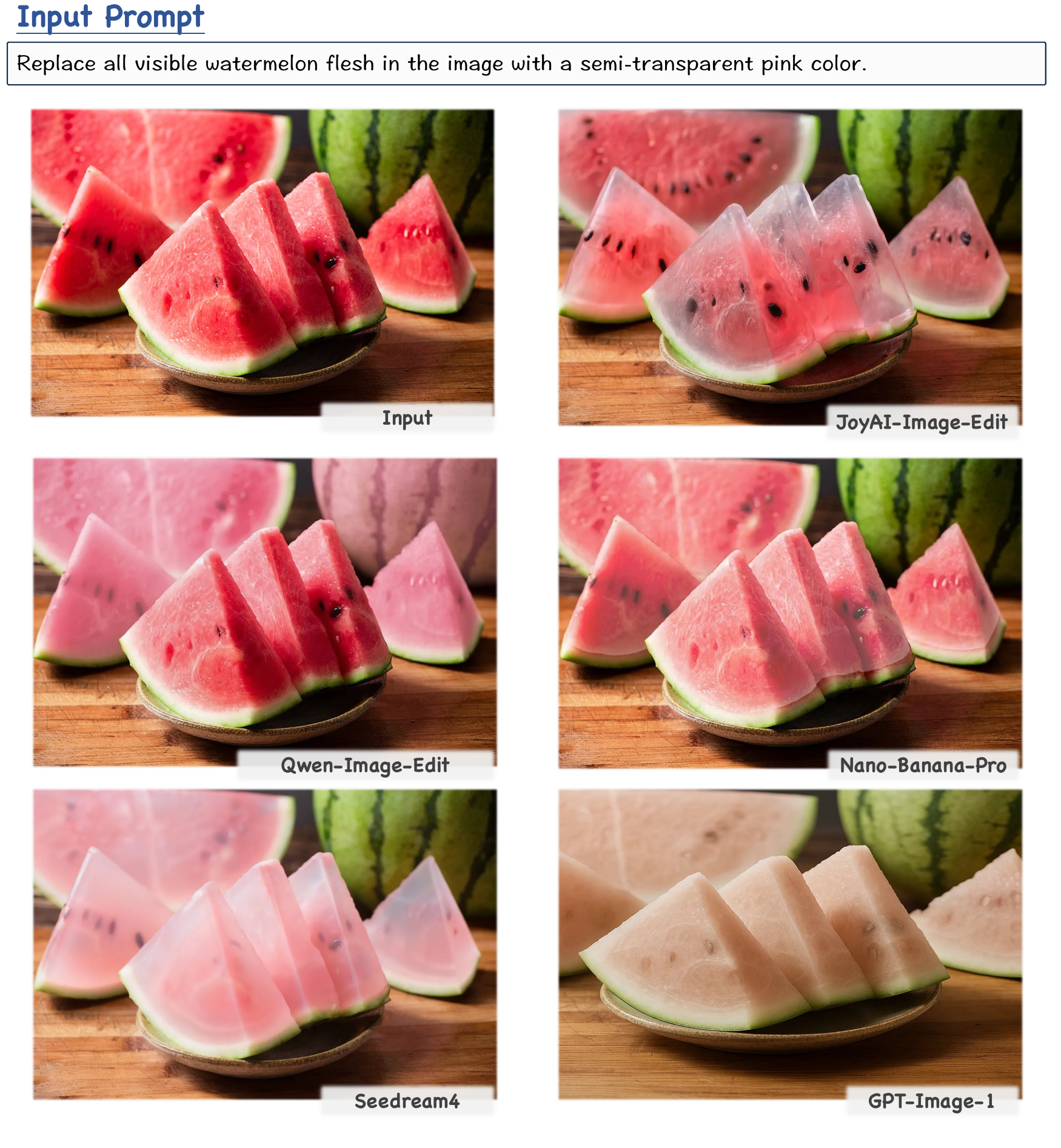}
   \caption{Qualitative comparison of attribute editing. JoyAI-Image better preserves subject identity, scene structure, and lighting while following the instruction, producing more accurate and natural edits than competing methods.
}
   \label{fig:attri-edit}
\end{center}
\end{figure}

\begin{figure}[htbp]
\begin{center}
   \includegraphics[width=0.99\linewidth]{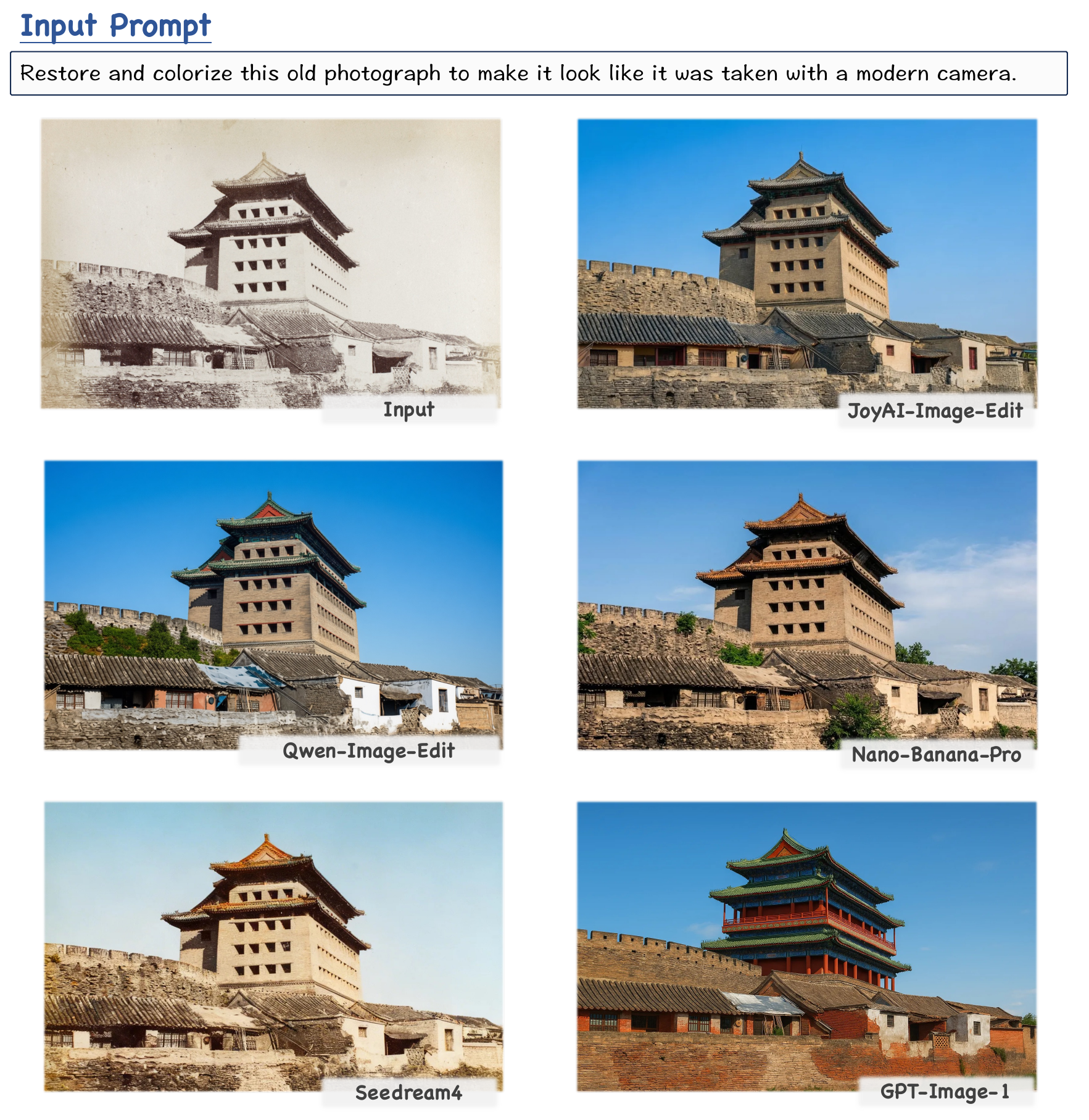}
   \caption{Qualitative comparison of image restoration. JoyAI-Image effectively removes degradation while preserving scene content and visual consistency, producing clearer and more natural restoration results than competing methods.}
   \label{fig:lowlevel-edit}
\end{center}
\end{figure}

\noindent\textbf{General Editing.}
Figure~\ref{fig:attri-edit} and Figure~\ref{fig:lowlevel-edit} present qualitative comparisons across general editing tasks including attribute modification, style transfer, enhancement, and restoration. While several baselines often fail to fully follow instructions or struggle to preserve original structures and textures, JoyAI-Image-Edit demonstrates superior localized precision and content preservation. Whether in multi-subject scenarios or low-level vision tasks, JoyAI-Image-Edit effectively improves image clarity and recovers plausible details without introducing severe artifacts. These examples illustrate that JoyAI-Image-Edit achieves strong general editing performance by maintaining an optimal balance between edit fidelity and structural stability.

\begin{figure}[htbp]
\begin{center}
   \includegraphics[width=0.99\linewidth]{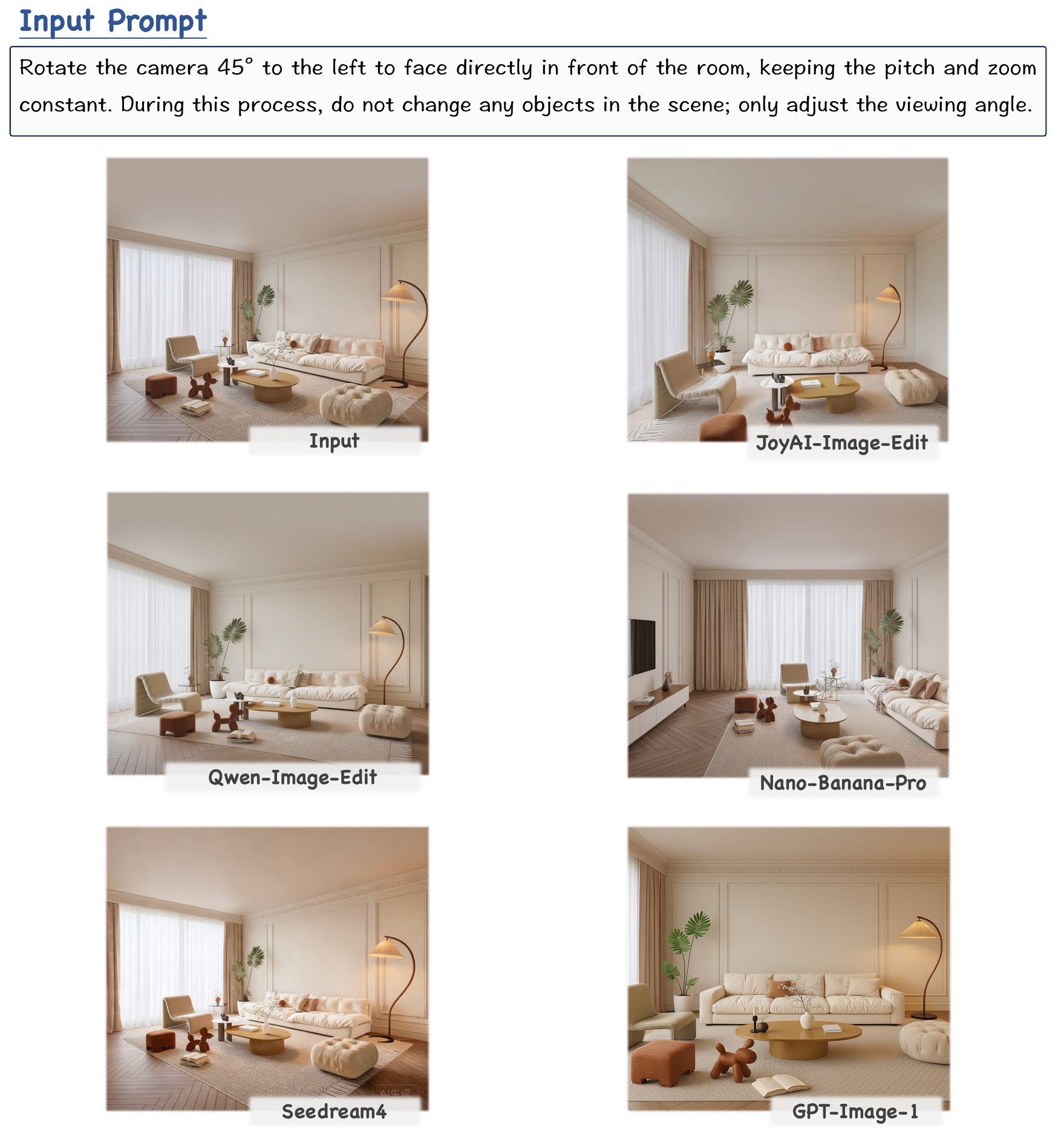}
   \caption{Qualitative comparison of camera control. JoyAI-Image accurately follows camera movement instructions while preserving scene content and visual consistency, producing more precise and natural viewpoint changes than competing methods.}
   \label{fig:se-camera}
\end{center}
\end{figure}

\noindent\textbf{Camera Control.}
Figure~\ref{fig:se-camera} further evaluates JoyAI-Image-Edit on camera control, which is particularly challenging because it requires coordinated updates of global perspective while preserving scene consistency. When moving the camera upward and zooming in on the instrument, JoyAI-Image-Edit best matches the requested viewpoint and focus change without introducing geometric collapse. Similarly, in the indoor scene, JoyAI-Image-Edit rotates the camera to the desired direction while keeping the objects and room layout unchanged, demonstrating stronger 3D consistency and spatial reasoning. These results suggest that image editing models can go beyond appearance manipulation and exhibit early capabilities related to world modeling, as they must implicitly reason about viewpoint change, spatial structure, and scene geometry.

\begin{figure}[htbp]
\begin{center}
   \includegraphics[width=0.99\linewidth]{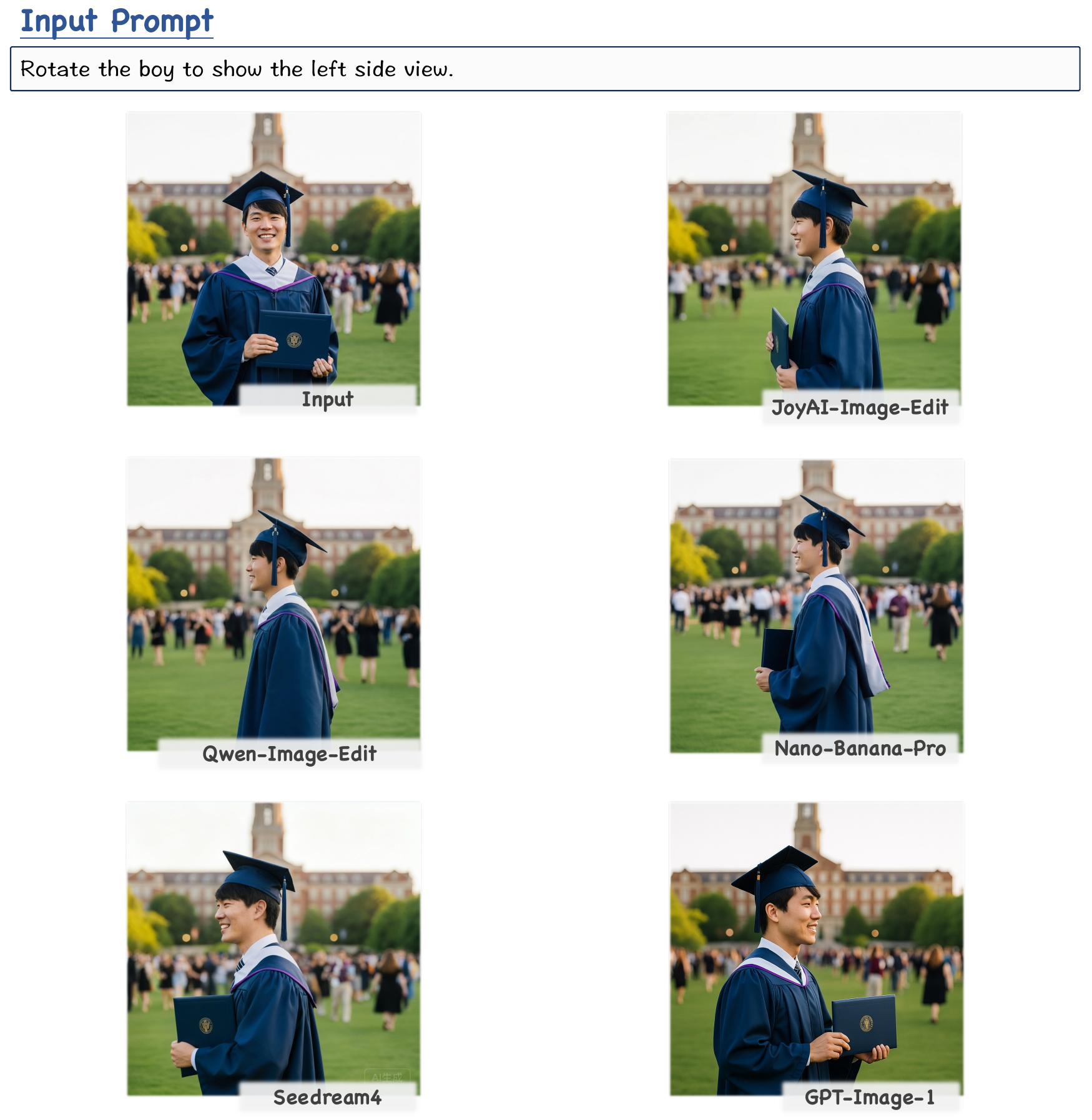}
   \caption{Qualitative comparison of object rotation. JoyAI-Image accurately follows viewpoint instructions for specific objects while preserving scene content and visual consistency, producing more precise and natural results than competing methods.}
   \label{fig:se-object}
\end{center}
\end{figure}

\noindent\textbf{Object Transformation.}
Figure~\ref{fig:se-object} evaluates JoyAI-Image-Edit on object movement and human pose rotation. These cases require accurate target selection, local geometric transformation, and preservation of non-target content. JoyAI-Image-Edit successfully rotates only objects to the requested front-left view while preserving the the background than other methods. These results demonstrate that JoyAI-Image-Edit supports precise local viewpoint manipulation with stronger spatial consistency and less interference to the surrounding scene. Overall, JoyAI-Image-Edit not only performs competitively on general-purpose editing, but also shows clear advantages on spatial editing, where precise geometric compliance and controllable viewpoint manipulation are essential.



\section{Applications}

\subsection{Thinking with Novel Views}

\begin{figure}[htbp]
    \centering
    \includegraphics[width=0.9\textwidth]{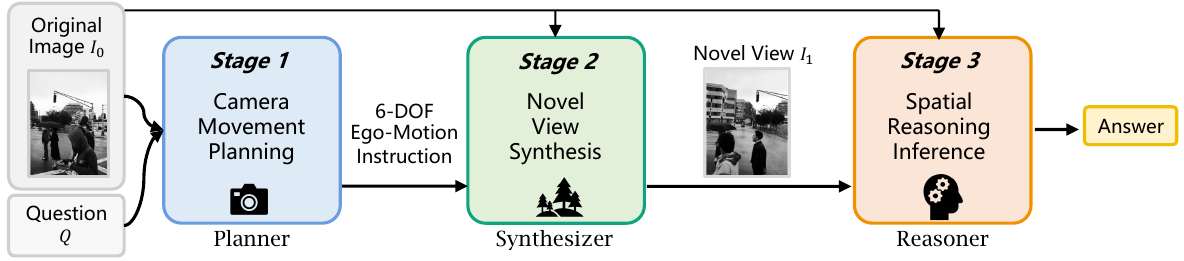}
    \caption{\textbf{The Thinking with Novel Views (TwNV) pipeline.} Given an input image $I_0$ and a spatial question $Q$, the MLLM Planner outputs a 6-DOF camera motion instruction, the Synthesizer renders the target novel view $I_1$, and the MLLM Reasoner infers over $\{I_0, I_1\}$ to answer the question.}
    \label{fig:twnv_pipeline}
\end{figure}

\begin{table}[htbp]
\centering
\small
\caption{Evaluation of the Thinking with Novel Views paradigm.
\textbf{Left:} Comparative analysis of generative Synthesizers using a fixed GPT-5 Reasoner. 
\textbf{Right:} Generalization of JoyAI-Image-Edit across diverse Reasoners. 
``Baseline'' denotes the single-view input without novel-view synthesis. 
``Overall'' represents the sample-count-weighted average across all displayed categories. 
``Rel.'' reports the relative improvement over the corresponding baseline.}
\label{tab:twov_application}
\begin{minipage}[t]{0.48\linewidth}
\centering
\setlength{\tabcolsep}{4pt}
\resizebox{\linewidth}{!}{
\begin{tabular}{lcccc}
\toprule
\textbf{Editor} & \textbf{Orient.} & \textbf{Loc.} &
\textbf{Multi-Obj.} & \textbf{Overall} \\
\midrule
None (w/o NV)  & 63.6 & 80.9 & 60.5 & 68.8 \\
Qwen-Image-Edit  & 61.8 & 77.8 & 61.5 & 67.4 \\
Nano Banana Pro  & 60.9 & 83.0 & 63.6 & \underline{69.5} \\
\rowcolor[HTML]{E1F5FE} JoyAI-Image-Edit & 65.3 & 82.6 & 66.2 & \textbf{71.7} \\
\bottomrule
\end{tabular}
}
\end{minipage}\hfill
\begin{minipage}[t]{0.5\linewidth}
\renewcommand{\arraystretch}{1.08}
\centering
\setlength{\tabcolsep}{3.5pt}
\resizebox{\linewidth}{!}{
\begin{tabular}{lcccc}
\toprule
\textbf{Reasoner} & \textbf{w/o NV} & \textbf{JoyAI-Image-Edit} & $\bm{\Delta}$ & \textbf{Rel.} \\
\midrule
Gemini-3-Flash & 75.5 & \cellcolor[HTML]{E1F5FE}{\textbf{77.2}} & +1.7 & $\uparrow$ 2.3\% \\
GPT-5          & 68.8 & \cellcolor[HTML]{E1F5FE}{\textbf{71.7}} & +2.9 & $\uparrow$ 4.2\% \\
Qwen3-VL-235B  & 58.6 & \cellcolor[HTML]{E1F5FE}{\textbf{61.8}} & +3.2 & $\uparrow$ 5.5\% \\
Qwen3-VL-32B   & 56.2 & \cellcolor[HTML]{E1F5FE}{\textbf{60.6}} & +4.4 & $\uparrow$ 7.8\% \\
\bottomrule
\end{tabular}
}
\end{minipage}
\end{table}

We demonstrate that high-fidelity spatial editing can serve as a powerful catalyst for enhancing spatial reasoning. While conventional Large Multimodal Large Language Models (MLLMs) are often constrained by static inputs or heuristic 2D tool-calling (\textit{e.g.}, cropping or rotation), we transcend these perspective limitations through a generative Thinking with Novel Views (TwNV) paradigm. Our three-stage pipeline (Figure~\ref{fig:twnv_pipeline}), comprising an MLLM \textbf{Planner}, a generative \textbf{Synthesizer}, and an MLLM \textbf{Reasoner}, empowers models to proactively explore and disambiguate complex scenes. Concretely, given an input image and a spatial question, the Planner first predicts a 6-DOF camera motion instruction that specifies how the viewpoint should shift to expose the most informative geometric evidence. The Synthesizer then follows this instruction to generate a novel view, and the Reasoner jointly inspects the original and synthesized images to answer the question. This transition to dynamic view synthesis enables the system to resolve occlusions and geometric uncertainties that are inherently invisible from a single viewpoint.

To rigorously assess spatial reasoning capabilities, we curate a dedicated evaluation suite comprising 695 high-quality samples derived from two primary sources: 575 instances from 3DSRBench~\cite{3dsrbench}, obtained via stratified sampling across all 12 subcategories, and 120 spatially-focused entries from RealWorldQA~\cite{grok15v} (targeting orientation, size, and position). All samples are systematically categorized into three dimensions: Orientation, Location, and Multi-Object Relationship.

Building upon this benchmark, we evaluate the proposed paradigm by employing the frontier MLLM, GPT-5~\cite{ChatGPT}, as both the Planner and Reasoner, while benchmarking various editing models as Synthesizers. Our results (Table~\ref{tab:twov_application}, Left) confirm that leveraging novel-view generation to facilitate spatial reasoning is both feasible and highly effective. Specifically, our JoyAI-Image-Edit model outperforms existing competitors, boosting GPT-5's overall accuracy from 68.8\% to 71.7\% and yielding a substantial 5.7~pp. gain in multi-object relationship tasks. Unlike Nano Banana Pro~\cite{nanopro} or Qwen-Image-Edit~\cite{qwenimage}, which exhibit marginal or even negative gains, JoyAI-Image-Edit provides the rigorous 3D consistency and geometric precision requisite for reliable downstream inference.

Furthermore, we investigate the generalizability of our framework across a diverse spectrum of reasoner capacities, including Gemini-3-Flash~\cite{gemini}, Qwen3-VL-235B~\cite{qwen3vl2025}, and Qwen3-VL-32B~\cite{qwen3vl2025}. As shown in Table~\ref{tab:twov_application} (Right), the TwNV paradigm delivers consistent improvements across all models, with absolute gains ranging from 1.7 to 4.4~pp. Notably, we observe a ``Small-Model Dividend'', where relative improvements are more pronounced in smaller models (\textit{e.g.}, a 7.8\% relative gain for Qwen3-VL-32B compared to 5.5\% for Qwen3-VL-235B~\cite{qwen3vl2025} and 2.3\% for Gemini-3-Flash~\cite{gemini2.5}). This suggests that explicit view synthesis serves as a vital compensatory mechanism for parameter-constrained models. By offloading 3D spatial modeling to an external generative workspace, smaller models can leverage synthesized Chain-of-Thought links to achieve spatial intelligence far exceeding their native capacities.

\begin{figure*}[t]
\centering
\includegraphics[width=.95\textwidth]{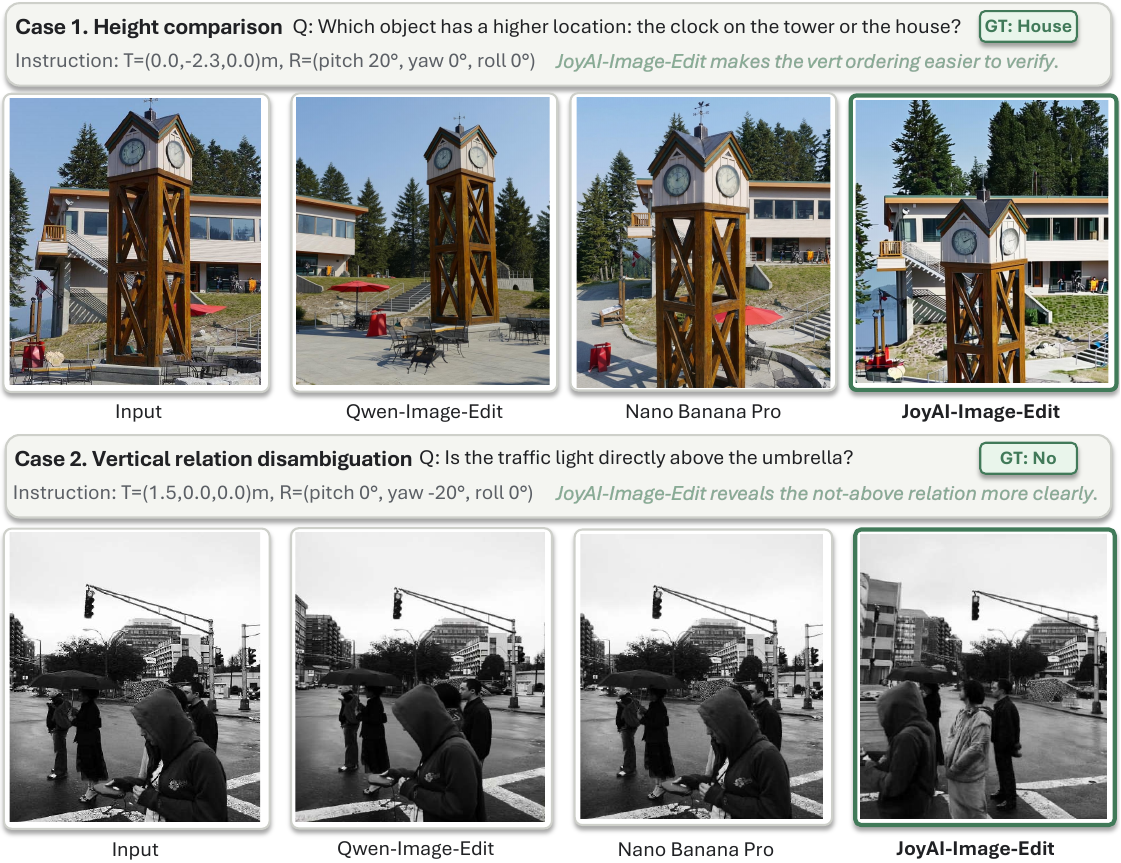}
\caption{Visualization of the Thinking with Novel Views paradigm. We showcase two representative spatial reasoning tasks: height comparison (Top) and vertical relationship (Bottom). For each case, we provide the query, ground truth (GT), and the camera-motion instructions generated by the MLLM Planner. Compared with Qwen-Image-Edit~\cite{qwenimage} and Nano Banana Pro~\cite{nanopro}, JoyAI-Image-Edit synthesizes the most diagnostic viewpoints by faithfully executing camera motions. These high-fidelity novel views effectively disambiguate complex spatial relations, providing clearer visual evidence for downstream reasoning.}
\label{fig:twov_application}
\end{figure*}

Figure~\ref{fig:twov_application} visualizes two representative tasks. Compared with Qwen-Image-Edit~\cite{qwenimage} and Nano Banana Pro~\cite{nanopro}, JoyAI-Image-Edit executes requested camera motions more faithfully, thereby exposing target spatial relations more clearly for downstream reasoning.

\subsection{Reconstruction with Novel Views}
\label{sec:recon_novel_views}

Another useful application of spatial image editing is to expand a single observed image into a set of geometrically meaningful novel views, which can in turn serve as additional input for downstream 3D reconstruction.
This setting is particularly relevant because high-quality reconstruction requires not only photorealistic synthesis, but also faithful preservation of camera geometry, object layout, and cross-view correspondence.
Recent progress in world models and camera-controllable video generation has likewise highlighted 3D reconstruction and world-consistency as an important way to assess whether generated observations are truly spatially coherent rather than merely visually plausible.

\begin{figure}[htbp]
\begin{center}
   \includegraphics[width=0.488\linewidth]{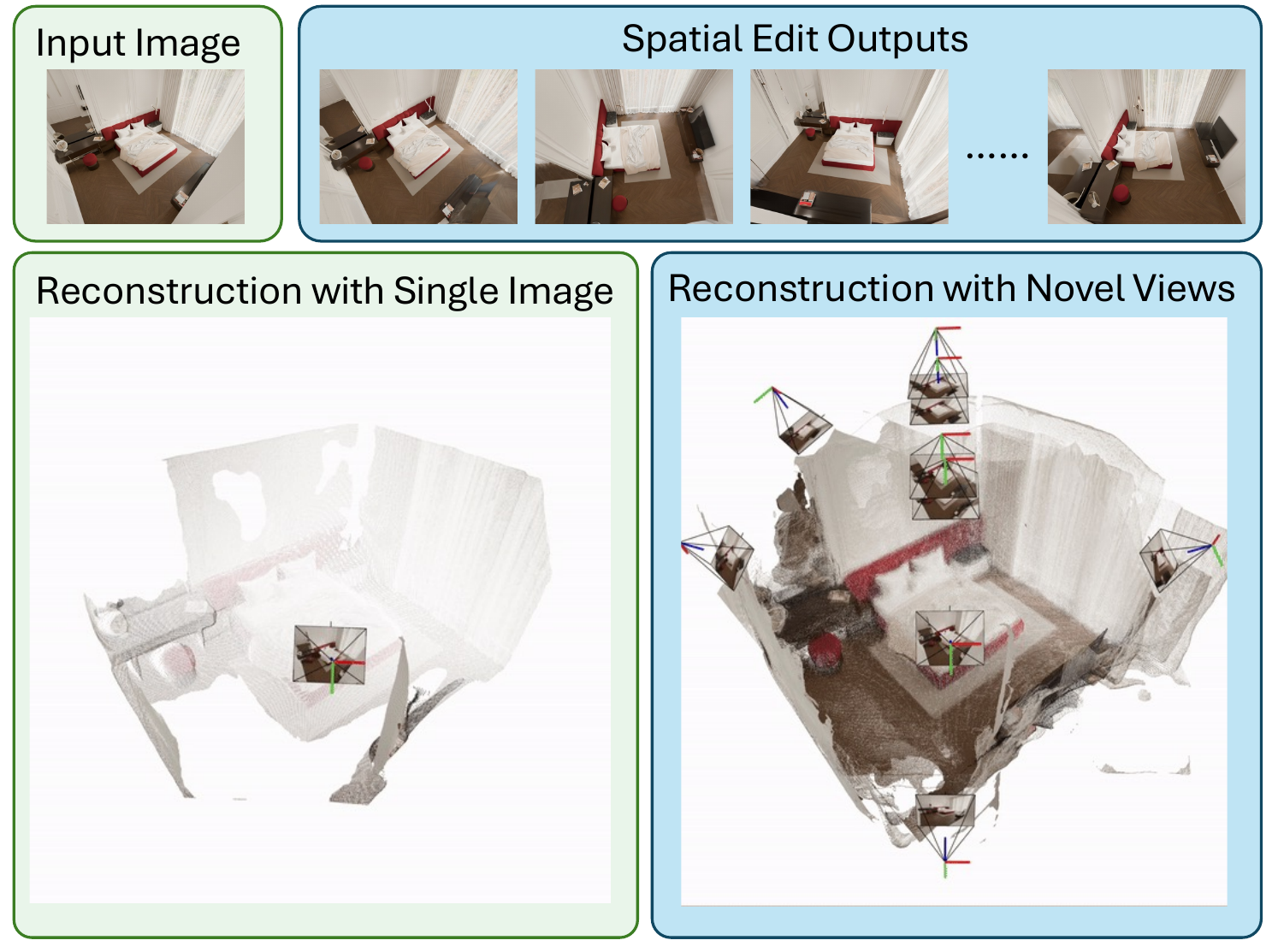}
   \hfill
   \includegraphics[width=0.488\linewidth]{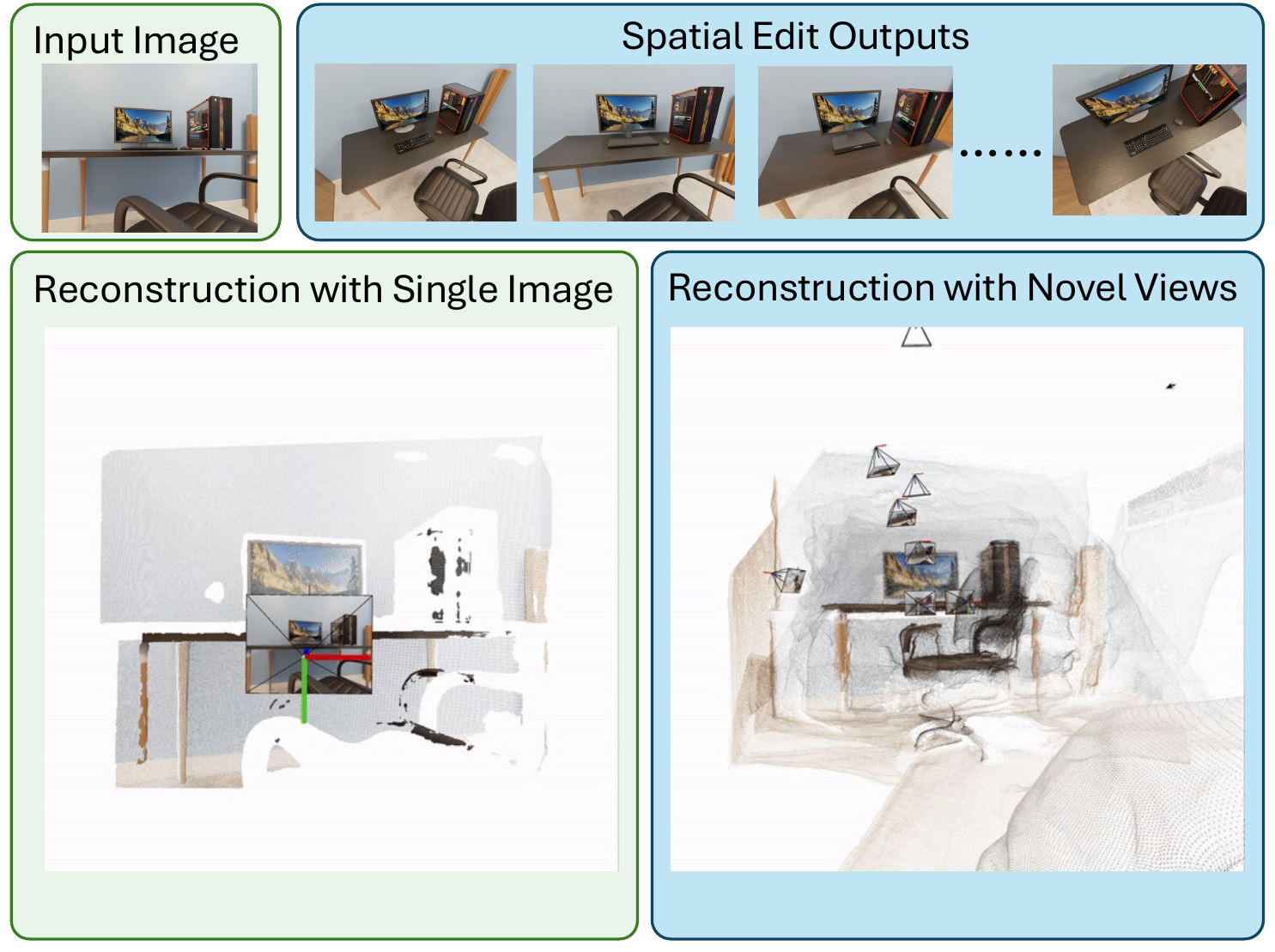}
   \caption{Novel views generated by our spatial model lead to improved 3D reconstruction performance.}
   \label{fig:3d-recon}
\end{center}
\end{figure}

To study this capability, we use our spatial editing model to generate multiple novel views from a single input image by varying the viewpoint while preserving scene structure.
We then feed the generated views into VGGT~\cite{wang2025vggt}, a strong feed-forward 3D reconstruction model, and visualize the resulting point clouds and camera poses.
As shown in Figure~\ref{fig:3d-recon}, reconstruction from the input image alone produces sparse and incomplete geometry, whereas reconstruction with our generated novel views becomes substantially denser and more structurally complete.
The recovered scene layout, dominant surfaces, and object placements are all noticeably improved, indicating that the synthesized views provide complementary geometric evidence that is useful for multi-view reasoning.

More importantly, this experiment also offers an indirect but intuitive validation of our spatial editing ability.
If the generated images were only locally realistic but inconsistent in camera motion, object placement, or scene geometry, they would introduce ambiguity and often degrade reconstruction quality.
Instead, the fact that they consistently improve 3D reconstruction suggests that our model preserves cross-view geometric structure to a meaningful extent.
In other words, the model does not merely generate plausible edits in image space; it produces spatial edits that are sufficiently geometry-consistent to support downstream 3D perception.
This result further demonstrates that our spatial image editing model captures a stronger notion of 3D-aware scene manipulation, which is essential for applications such as embodied world modeling, scene exploration, and controllable visual simulation.


\section{Conclusion}

In this report, we presented JoyAI-Image, a unified multimodal foundation model that brings image understanding, text-to-image generation, and instruction-based image editing into a shared framework centered on a spatially enhanced MLLM and a large-scale MMDiT. By tightly coupling understanding, generation, and editing rather than treating them as isolated capabilities, JoyAI-Image achieves strong performance across broad visual tasks, including spatial understanding, bilingual long-text rendering, controllable editing, and multi-view generation. More importantly, our results show that spatial intelligence can be strengthened as a first-class property of unified visual modeling through data construction, training design, and cross-task interaction.

We view JoyAI-Image as a practical step toward visual systems that can not only perceive and generate, but also reason about structure, transformation, and the geometry of the physical world. Beyond content creation, such capability has broader implications for applications that require grounded visual reasoning under change, including visual-language-action systems, robotics, and world models. We hope this work helps move unified multimodal modeling from broad competence toward genuinely spatially intelligent visual foundation models.
\clearpage
\renewcommand{\thefootnote}{\fnsymbol{footnote}}

\section{Authors}
\label{sec:contributions}

\textbf{Core Contributors}\\[0.5em]
Lin Song\footnotemark[1], Wenbo Li\footnotemark[1], Guoqing Ma\footnotemark[1], Wei Tang, Bo Wang, Yuan Zhang, Yijun Yang, Yicheng Xiao, Jianhui~Liu, Yanbing Zhang, Guohui Zhang,  Wenhu Zhang, Hang Xu, Nan Jiang, Xin Han, Haoze Sun, Maoquan Zhang, Haoyang~Huang\footnotemark[2], Nan Duan

\footnotetext[1]{Equal contribution.}
\footnotetext[2]{Corresponding author: Haoyang Huang<huanghaoyang.ocean@jd.com>.}

\textbf{Contributors}\footnotemark[3]\\[0.5em]
Anson Li, Bi Cheng, Bingning Liu, Boxian Ai, Boyang Li, Chao Xue, Chaocai~Liang, Cheng Zhang, Chenmeijin Liang, Chenyi Li, Chuyang Zhao, Dongjiang Li, Dongyan Yang, Duomi Zhang, Gen Li, Guofeng Chen, Haokun Lin, Haoran Li, He Zhang, Hengshan Ji, Hongcheng Gao, Hu Yu, Huaishao Luo, Haoyu Wu, Jiachen Liu, Jiang Yuan, Jianlong Yuan, JianZhong Shi, Jiaqi Wang, Jiaqi Zhao, Jiaxiu Jiang, Jiayi Deng, Jiazhe Xu, Jie Huang, Jing~Long, Jingdi Chen, Jinghao Zhang, Jiyao Zhang, Jizhi~Li, Jundao Li, Junwu~Xiong, Jiajun Zha, Liang~Lin, Libing Fang, Lichen Ma, Liwei Wang, Lixin Wang, Mingsi Wang, Mingyu Wang, Meihui Wang, Nanhua Lai, Nick, Pan Wang, Peihao Li, Peng~Cao, Qianli Chen, Qiming~Yang, Qingyi Si, Ruofan Lv, Ruize He, Shaonan Wu, Shenghe~Zheng, Shichen Ma, Shiyang Zhou, Shiyi Zhang, Shuai Lu, Siming Fu, Songchun Zhang, Wanting~Xu, Wei Li, Weilin~Jin, Weiyang Jin, Xiaoxiao Huo, Xing Pan, Xinran Qin, Xinyu Lyu, Xionghao Wu, Xuan Yang, Xuanyi~Li, Yan~Li, Yaofeng su, Yaowei Li, Yicheng Gong, Yifan Jiao, Yihang Li, Yijun Liu, Yilang~Sun, Yingzi Han, Yitong Chen, Yuanming Yang, Yubo~Li, Yucheng~Guo, Yuhang Cao, Yujia Liang, Yuming Li, Yuzheng Zhuang, Yue Ma, Yufei Jiao, Zeyue Xue, Zheming Liang, Zhengqi Huang, Zhiliang Zhu, Zhongqi Yang, Ziyu Zhao, Zuopeng~Dong

\footnotetext[3]{Contributors are listed in alphabetical order.}



\begin{figure}[htbp]
\begin{center}
   \includegraphics[width=0.84\linewidth]{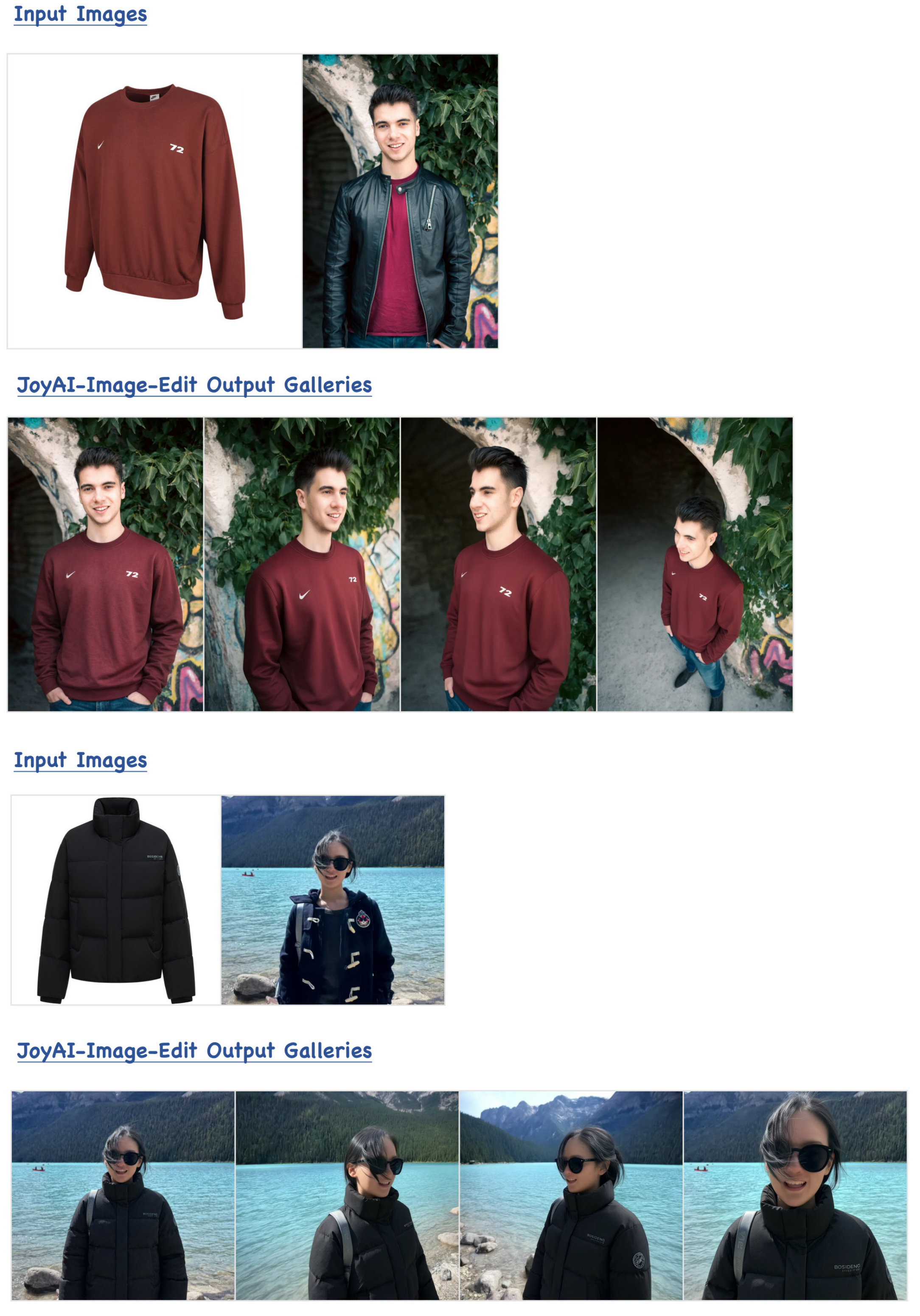}
   \caption{Visual examples of multi-view try-on applications}
   \label{fig:try-on}
\end{center}
\end{figure}

\clearpage
\bibliographystyle{plainnat}
\bibliography{cite}

\begin{thebibliography}{108}
\providecommand{\natexlab}[1]{#1}
\providecommand{\url}[1]{\texttt{#1}}
\expandafter\ifx\csname urlstyle\endcsname\relax
  \providecommand{\doi}[1]{doi: #1}\else
  \providecommand{\doi}{doi: \begingroup \urlstyle{rm}\Url}\fi

\bibitem[Achiam et~al.(2023)Achiam, Adler, Agarwal, Ahmad, Akkaya, Aleman,
  Almeida, Altenschmidt, Altman, Anadkat, et~al.]{gpt4}
Josh Achiam, Steven Adler, Sandhini Agarwal, Lama Ahmad, Ilge Akkaya,
  Florencia~Leoni Aleman, Diogo Almeida, Janko Altenschmidt, Sam Altman,
  Shyamal Anadkat, et~al.
\newblock Gpt-4 technical report.
\newblock \emph{arXiv:2303.08774}, 2023.

\bibitem[Bai et~al.(2025{\natexlab{a}})Bai, Xia, Fu, Wang, Mu, Cao, Liu, Hu,
  Bai, Wan, et~al.]{recammaster}
Jianhong Bai, Menghan Xia, Xiao Fu, Xintao Wang, Lianrui Mu, Jinwen Cao, Zuozhu
  Liu, Haoji Hu, Xiang Bai, Pengfei Wan, et~al.
\newblock Recammaster: Camera-controlled generative rendering from a single
  video.
\newblock In \emph{Proceedings of the IEEE/CVF International Conference on
  Computer Vision}, pages 14834--14844, 2025{\natexlab{a}}.

\bibitem[Bai et~al.(2025{\natexlab{b}})Bai, Cai, Chen, Huang, Li, Lin, Zhu,
  et~al.]{qwen3vl2025}
Shuai Bai, Yuxuan Cai, Xionghui Chen, Qidong Huang, Kaixin Li, Zicheng Lin,
  Keming Zhu, et~al.
\newblock Qwen3-vl technical report.
\newblock \emph{arXiv preprint arXiv:2511.21631}, 2025{\natexlab{b}}.
\newblock URL \url{https://arxiv.org/abs/2511.21631}.

\bibitem[Bai et~al.(2025{\natexlab{c}})Bai, Chen, Liu, Wang, Ge, Song, Dang,
  Wang, Wang, Tang, et~al.]{qwen2.5vl}
Shuai Bai, Keqin Chen, Xuejing Liu, Jialin Wang, Wenbin Ge, Sibo Song, Kai
  Dang, Peng Wang, Shijie Wang, Jun Tang, et~al.
\newblock Qwen2. 5-vl technical report.
\newblock \emph{arXiv:2502.13923}, 2025{\natexlab{c}}.

\bibitem[Baruch et~al.(2021)Baruch, Chen, Dehghan, Dimry, Feigin, Fu, Gebauer,
  Joffe, Kurz, Schwartz, et~al.]{arkitscenes}
Gilad Baruch, Zhuoyuan Chen, Afshin Dehghan, Tal Dimry, Yuri Feigin, Peter Fu,
  Thomas Gebauer, Brandon Joffe, Daniel Kurz, Arik Schwartz, et~al.
\newblock Arkitscenes: A diverse real-world dataset for 3d indoor scene
  understanding using mobile rgb-d data.
\newblock \emph{arXiv:2111.08897}, 2021.

\bibitem[{Black Forest Labs}(2024)]{labs2025flux}
{Black Forest Labs}.
\newblock \emph{FLUX.1 [Dev]}.
\newblock Black Forest Labs, 2024.
\newblock URL \url{https://huggingface.co/black-forest-labs/FLUX.1-dev}.
\newblock Official model card, accessed 2026-03-24.

\bibitem[{Blender Foundation}(2024)]{blender}
{Blender Foundation}.
\newblock Blender, 2024.
\newblock URL \url{https://www.blender.org/}.

\bibitem[Brooks et~al.(2023)Brooks, Holynski, and
  Efros]{brooks2023instructpix2pix}
Tim Brooks, Aleksander Holynski, and Alexei~A Efros.
\newblock Instructpix2pix: Learning to follow image editing instructions.
\newblock In \emph{Proceedings of the IEEE/CVF conference on computer vision
  and pattern recognition}, pages 18392--18402, 2023.

\bibitem[Bruce et~al.(2024)Bruce, Dennis, Edwards, Parker-Holder, Shi, Hughes,
  Lai, Mavalankar, Steigerwald, Apps, et~al.]{bruce2024genie}
Jake Bruce, Michael~D Dennis, Ashley Edwards, Jack Parker-Holder, Yuge Shi,
  Edward Hughes, Matthew Lai, Aditi Mavalankar, Richie Steigerwald, Chris Apps,
  et~al.
\newblock Genie: Generative interactive environments.
\newblock In \emph{Forty-first International Conference on Machine Learning},
  2024.

\bibitem[Cai et~al.(2025)Cai, Chen, Chen, Li, Long, Pan, Qiu, Zhang, Gao, Xu,
  Wang, Yu, Chen, Feng, Gong, Pan, Peng, Tian, Wang, Zhao, Yao, and
  Mei]{cai2025hidream}
Qi~Cai, Jingwen Chen, Yang Chen, Yehao Li, Fuchen Long, Yingwei Pan, Zhaofan
  Qiu, Yiheng Zhang, Fengbin Gao, Peihan Xu, Yimeng Wang, Kai Yu, Wenxuan Chen,
  Ziwei Feng, Zijian Gong, Jianzhuang Pan, Yi~Peng, Rui Tian, Siyu Wang,
  Bo~Zhao, Ting Yao, and Tao Mei.
\newblock Hidream-i1: A high-efficient image generative foundation model with
  sparse diffusion transformer.
\newblock \emph{arXiv preprint arXiv:2505.22705}, 2025.
\newblock URL \url{https://arxiv.org/abs/2505.22705}.

\bibitem[Cao et~al.(2025)Cao, Ma, Li, Li, Shao, Zhu, Zhou, Pu, Wu, Wang,
  et~al.]{cao2025artimuse}
Shuo Cao, Nan Ma, Jiayang Li, Xiaohui Li, Lihao Shao, Kaiwen Zhu, Yu~Zhou,
  Yuandong Pu, Jiarui Wu, Jiaquan Wang, et~al.
\newblock Artimuse: Fine-grained image aesthetics assessment with joint scoring
  and expert-level understanding.
\newblock \emph{arXiv preprint arXiv:2507.14533}, 2025.

\bibitem[Carion et~al.(2025)Carion, Gustafson, Hu, Debnath, Hu, Suris, Ryali,
  Alwala, Khedr, Huang, Lei, Ma, Guo, Kalla, Marks, Greer, Wang, Sun,
  R{\"a}dle, Afouras, Mavroudi, Xu, Wu, Zhou, Momeni, Hazra, Ding, Vaze,
  Porcher, Li, Li, Kamath, Cheng, Doll{\'a}r, Ravi, Saenko, Zhang, and
  Feichtenhofer]{sam3}
Nicolas Carion, Laura Gustafson, Yuan-Ting Hu, Shoubhik Debnath, Ronghang Hu,
  Didac Suris, Chaitanya Ryali, Kalyan~Vasudev Alwala, Haitham Khedr, Andrew
  Huang, Jie Lei, Tengyu Ma, Baishan Guo, Arpit Kalla, Markus Marks, Joseph
  Greer, Meng Wang, Peize Sun, Roman R{\"a}dle, Triantafyllos Afouras,
  Effrosyni Mavroudi, Katherine Xu, Tsung-Han Wu, Yu~Zhou, Liliane Momeni,
  Rishi Hazra, Shuangrui Ding, Sagar Vaze, Francois Porcher, Feng Li, Siyuan
  Li, Aishwarya Kamath, Ho~Kei Cheng, Piotr Doll{\'a}r, Nikhila Ravi, Kate
  Saenko, Pengchuan Zhang, and Christoph Feichtenhofer.
\newblock Sam 3: Segment anything with concepts.
\newblock \emph{arXiv preprint arXiv:2511.16719}, 2025.

\bibitem[Cen et~al.(2025)Cen, Yu, Yuan, Jiang, Huang, Guo, Li, Song, Luo, Wang,
  et~al.]{cen2025worldvla}
Jun Cen, Chaohui Yu, Hangjie Yuan, Yuming Jiang, Siteng Huang, Jiayan Guo, Xin
  Li, Yibing Song, Hao Luo, Fan Wang, et~al.
\newblock Worldvla: Towards autoregressive action world model.
\newblock \emph{arXiv preprint arXiv:2506.21539}, 2025.

\bibitem[Chang et~al.(2017)Chang, Dai, Funkhouser, Halber, Niessner, Savva,
  Song, Zeng, and Zhang]{matterport3d}
Angel Chang, Angela Dai, Thomas Funkhouser, Maciej Halber, Matthias Niessner,
  Manolis Savva, Shuran Song, Andy Zeng, and Yinda Zhang.
\newblock Matterport3d: Learning from rgb-d data in indoor environments.
\newblock \emph{arXiv:1709.06158}, 2017.

\bibitem[Chang et~al.(2025)Chang, Fang, Xing, Wu, Cheng, Wang, Zeng, YU, and
  Chen]{chang2025oneig}
Jingjing Chang, Yixiao Fang, Peng Xing, Shuhan Wu, Wei Cheng, Rui Wang,
  Xianfang Zeng, Gang YU, and Hai-Bao Chen.
\newblock One{IG}-bench: Omni-dimensional nuanced evaluation for image
  generation.
\newblock In \emph{The Thirty-ninth Annual Conference on Neural Information
  Processing Systems Datasets and Benchmarks Track}, 2025.
\newblock URL \url{https://openreview.net/forum?id=S9TQM1Uhpl}.

\bibitem[Chen et~al.(2024{\natexlab{a}})Chen, Huang, Lv, Cui, Chen, and
  Wei]{chen2024textdiffuser}
Jingye Chen, Yupan Huang, Tengchao Lv, Lei Cui, Qifeng Chen, and Furu Wei.
\newblock Textdiffuser-2: Unleashing the power of language models for text
  rendering.
\newblock In \emph{European Conference on Computer Vision}, pages 386--402.
  Springer, 2024{\natexlab{a}}.

\bibitem[Chen et~al.(2025{\natexlab{a}})Chen, Xu, Pan, Hu, Qin, Goldstein,
  Huang, Zhou, Xie, Savarese, Xue, Xiong, and Xu]{chen2025blip3}
Jiuhai Chen, Zhiyang Xu, Xichen Pan, Yushi Hu, Can Qin, Tom Goldstein, Lifu
  Huang, Tianyi Zhou, Saining Xie, Silvio Savarese, Le~Xue, Caiming Xiong, and
  Ran Xu.
\newblock Blip3-o: A family of fully open unified multimodal
  models-architecture, training and dataset.
\newblock \emph{arXiv preprint arXiv:2505.09568}, 2025{\natexlab{a}}.
\newblock URL \url{https://arxiv.org/abs/2505.09568}.

\bibitem[Chen et~al.(2024{\natexlab{b}})Chen, Ge, Xie, Wu, Yao, Ren, Wang, Luo,
  Lu, and Li]{chen2024pixart}
Junsong Chen, Chongjian Ge, Enze Xie, Yue Wu, Lewei Yao, Xiaozhe Ren, Zhongdao
  Wang, Ping Luo, Huchuan Lu, and Zhenguo Li.
\newblock Pixart-$\sigma$: Weak-to-strong training of diffusion transformer for
  4k text-to-image generation.
\newblock In \emph{European Conference on Computer Vision}, pages 74--91.
  Springer, 2024{\natexlab{b}}.

\bibitem[Chen et~al.(2024{\natexlab{c}})Chen, Jincheng, Chongjian, Yao, Xie,
  Wang, Kwok, Luo, Lu, and Li]{chen2023pixart}
Junsong Chen, YU~Jincheng, GE~Chongjian, Lewei Yao, Enze Xie, Zhongdao Wang,
  James Kwok, Ping Luo, Huchuan Lu, and Zhenguo Li.
\newblock Pixart-$\alpha $: Fast training of diffusion transformer for
  photorealistic text-to-image synthesis.
\newblock In \emph{The Twelfth International Conference on Learning
  Representations}, 2024{\natexlab{c}}.

\bibitem[Chen et~al.(2024{\natexlab{d}})Chen, Li, Dong, Zhang, Zang, Chen,
  Duan, Wang, Qiao, Lin, et~al.]{mmstar}
Lin Chen, Jinsong Li, Xiaoyi Dong, Pan Zhang, Yuhang Zang, Zehui Chen, Haodong
  Duan, Jiaqi Wang, Yu~Qiao, Dahua Lin, et~al.
\newblock Are we on the right way for evaluating large vision-language models?
\newblock \emph{arXiv:2403.20330}, 2024{\natexlab{d}}.

\bibitem[Chen et~al.(2025{\natexlab{b}})Chen, Wu, Liu, Pan, Liu, Xie, Yu, and
  Ruan]{chen2025januspro}
Xiaokang Chen, Zhiyu Wu, Xingchao Liu, Zizheng Pan, Wen Liu, Zhenda Xie,
  Xingkai Yu, and Chong Ruan.
\newblock Janus-pro: Unified multimodal understanding and generation with data
  and model scaling.
\newblock \emph{arXiv preprint arXiv:2501.17811}, 2025{\natexlab{b}}.
\newblock URL \url{https://arxiv.org/abs/2501.17811}.

\bibitem[Chen et~al.(2024{\natexlab{e}})Chen, Wang, Cao, Liu, Gao, Cui, Zhu,
  Ye, Tian, Liu, et~al.]{internvl2.5}
Zhe Chen, Weiyun Wang, Yue Cao, Yangzhou Liu, Zhangwei Gao, Erfei Cui, Jinguo
  Zhu, Shenglong Ye, Hao Tian, Zhaoyang Liu, et~al.
\newblock Expanding performance boundaries of open-source multimodal models
  with model, data, and test-time scaling.
\newblock \emph{arXiv:2412.05271}, 2024{\natexlab{e}}.

\bibitem[Comanici et~al.(2025)Comanici, Bieber, Schaekermann, Pasupat,
  Sachdeva, Dhillon, Blistein, Ram, Zhang, Rosen, et~al.]{gemini2.5}
Gheorghe Comanici, Eric Bieber, Mike Schaekermann, Ice Pasupat, Noveen
  Sachdeva, Inderjit Dhillon, Marcel Blistein, Ori Ram, Dan Zhang, Evan Rosen,
  et~al.
\newblock Gemini 2.5: Pushing the frontier with advanced reasoning,
  multimodality, long context, and next generation agentic capabilities.
\newblock \emph{arXiv:2507.06261}, 2025.

\bibitem[Cui et~al.(2026)Cui, Sun, Liang, Gao, Zhang, Liu, Wang, Zhou, Liu,
  Lin, et~al.]{cui2026paddleocr}
Cheng Cui, Ting Sun, Suyin Liang, Tingquan Gao, Zelun Zhang, Jiaxuan Liu,
  Xueqing Wang, Changda Zhou, Hongen Liu, Manhui Lin, et~al.
\newblock Paddleocr-vl-1.5: Towards a multi-task 0.9 b vlm for robust
  in-the-wild document parsing.
\newblock \emph{arXiv preprint arXiv:2601.21957}, 2026.

\bibitem[Cui et~al.(2025)Cui, Chen, Deng, Huang, Li, Liu, Liu, Luo, Wang, Wang,
  Wang, Wang, Zhang, Zhao, Pan, Li, Hao, Ma, Chen, Ao, Huang, Wang, and
  Wang]{cui2025emu35nativemultimodalmodels}
Yufeng Cui, Honghao Chen, Haoge Deng, Xu~Huang, Xinghang Li, Jirong Liu, Yang
  Liu, Zhuoyan Luo, Jinsheng Wang, Wenxuan Wang, Yueze Wang, Chengyuan Wang,
  Fan Zhang, Yingli Zhao, Ting Pan, Xianduo Li, Zecheng Hao, Wenxuan Ma, Zhuo
  Chen, Yulong Ao, Tiejun Huang, Zhongyuan Wang, and Xinlong Wang.
\newblock Emu3.5: Native multimodal models are world learners, 2025.
\newblock URL \url{https://arxiv.org/abs/2510.26583}.

\bibitem[Dai et~al.(2017)Dai, Chang, Savva, Halber, Funkhouser, and
  Nie{\ss}ner]{scannet}
Angela Dai, Angel~X Chang, Manolis Savva, Maciej Halber, Thomas Funkhouser, and
  Matthias Nie{\ss}ner.
\newblock Scannet: Richly-annotated 3d reconstructions of indoor scenes.
\newblock In \emph{CVPR}, 2017.

\bibitem[Deng et~al.(2025{\natexlab{a}})Deng, Zhu, Li, Gou, Li, Wang, Zhong,
  Yu, Nie, Song, Shi, and Fan]{deng2025bagel}
Chaorui Deng, Deyao Zhu, Kunchang Li, Chenhui Gou, Feng Li, Zeyu Wang, Shu
  Zhong, Weihao Yu, Xiaonan Nie, Ziang Song, Guang Shi, and Haoqi Fan.
\newblock Emerging properties in unified multimodal pretraining.
\newblock \emph{arXiv preprint arXiv:2505.14683}, 2025{\natexlab{a}}.

\bibitem[Deng et~al.(2025{\natexlab{b}})Deng, Zhu, Li, Gou, Li, Wang, Zhong,
  Yu, Nie, Song, Shi, and Fan]{deng2025emerging}
Chaorui Deng, Deyao Zhu, Kunchang Li, Chenhui Gou, Feng Li, Zeyu Wang, Shu
  Zhong, Weihao Yu, Xiaonan Nie, Ziang Song, Guang Shi, and Haoqi Fan.
\newblock Emerging properties in unified multimodal pretraining.
\newblock \emph{arXiv preprint arXiv:2505.14683}, 2025{\natexlab{b}}.
\newblock URL \url{https://arxiv.org/abs/2505.14683}.

\bibitem[Duan et~al.(2024)Duan, Yang, Qiao, Fang, Chen, Liu, Dong, Zang, Zhang,
  Wang, et~al.]{duan2024vlmevalkit}
Haodong Duan, Junming Yang, Yuxuan Qiao, Xinyu Fang, Lin Chen, Yuan Liu, Xiaoyi
  Dong, Yuhang Zang, Pan Zhang, Jiaqi Wang, et~al.
\newblock Vlmevalkit: An open-source toolkit for evaluating large
  multi-modality models.
\newblock In \emph{Proceedings of the 32nd ACM international conference on
  multimedia}, pages 11198--11201, 2024.

\bibitem[Esser et~al.(2024)Esser, Kulal, Blattmann, Entezari, M{\"u}ller,
  Saini, Levi, Lorenz, Sauer, Boesel, et~al.]{esser2024scaling}
Patrick Esser, Sumith Kulal, Andreas Blattmann, Rahim Entezari, Jonas
  M{\"u}ller, Harry Saini, Yam Levi, Dominik Lorenz, Axel Sauer, Frederic
  Boesel, et~al.
\newblock Scaling rectified flow transformers for high-resolution image
  synthesis.
\newblock In \emph{ICML}, 2024.

\bibitem[Fu et~al.(2024)Fu, Hu, Li, Feng, Wang, Lin, Roth, Smith, Ma, and
  Krishna]{blink}
Xingyu Fu, Yushi Hu, Bangzheng Li, Yu~Feng, Haoyu Wang, Xudong Lin, Dan Roth,
  Noah~A Smith, Wei-Chiu Ma, and Ranjay Krishna.
\newblock Blink: Multimodal large language models can see but not perceive.
\newblock In \emph{ECCV}, 2024.

\bibitem[Gao et~al.(2025)Gao, Gong, Guo, Hou, Lai, Li, Li, Lian, Liao, Liu,
  Liu, Shi, Sun, Tian, Tian, Wang, Wang, Wang, Wang, Wang, Wu, Wu, Xia, Xiao,
  Zhai, Zhang, Zhang, Zhang, Zhao, Yang, and Huang]{gao2025seedream30}
Yu~Gao, Lixue Gong, Qiushan Guo, Xiaoxia Hou, Zhichao Lai, Fanshi Li, Liang Li,
  Xiaochen Lian, Chao Liao, Liyang Liu, Wei Liu, Yichun Shi, Shiqi Sun,
  Yu~Tian, Zhi Tian, Peng Wang, Rui Wang, Xuanda Wang, Xun Wang, Ye~Wang,
  Guofeng Wu, Jie Wu, Xin Xia, Xuefeng Xiao, Zhonghua Zhai, Xinyu Zhang,
  Qi~Zhang, Yuwei Zhang, Shijia Zhao, Jianchao Yang, and Weilin Huang.
\newblock Seedream 3.0 technical report.
\newblock \emph{arXiv preprint arXiv:2504.11346}, 2025.
\newblock URL \url{https://arxiv.org/abs/2504.11346}.

\bibitem[Geng et~al.(2025)Geng, Wang, Ma, Li, Rao, Gu, Zhong, Lu, Hu, Zhang,
  Linus, Wang, and Jiang]{geng2025xomni}
Zigang Geng, Yibing Wang, Yeyao Ma, Chen Li, Yongming Rao, Shuyang Gu, Zhao
  Zhong, Qinglin Lu, Han Hu, Xiaosong Zhang, Linus, Di~Wang, and Jie Jiang.
\newblock X-omni: Reinforcement learning makes discrete autoregressive image
  generative models great again.
\newblock \emph{CoRR}, abs/2507.22058, 2025.

\bibitem[{Google}(2025)]{google2025gemini3blog}
{Google}.
\newblock A new era of intelligence with gemini 3.
\newblock
  \url{https://blog.google/products-and-platforms/products/gemini/gemini-3/},
  November 2025.
\newblock Google Blog.

\bibitem[Google(2025{\natexlab{a}})]{nanopro}
Google.
\newblock Nano banana pro.
\newblock
  \href{https://storage.googleapis.com/deepmind-media/Model-Cards/Gemini-3-Pro-Image-Model-Card.pdf}{Gemini 3 Pro Image Model Card},
  2025{\natexlab{a}}.

\bibitem[Google(2025{\natexlab{b}})]{veo}
Google.
\newblock Introducing veo 3, our video generation model with expanded creative
  controls – including native audio and extended videos.
\newblock \emph{https://deepmind.google/models/veo/}, 2025{\natexlab{b}}.

\bibitem[Guo et~al.(2025)Guo, Yang, Zhang, Song, Wang, Zhu, Xu, Zhang, Ma, Bi,
  et~al.]{guo2025deepseek}
Daya Guo, Dejian Yang, Haowei Zhang, Junxiao Song, Peiyi Wang, Qihao Zhu,
  Runxin Xu, Ruoyu Zhang, Shirong Ma, Xiao Bi, et~al.
\newblock Deepseek-r1: Incentivizing reasoning capability in llms via
  reinforcement learning.
\newblock \emph{arXiv preprint arXiv:2501.12948}, 2025.

\bibitem[Hu et~al.(2024)Hu, Wang, Fang, Fu, Cheng, and Yu]{hu2024ella}
Xiwei Hu, Rui Wang, Yixiao Fang, Bin Fu, Pei Cheng, and Gang Yu.
\newblock Ella: Equip diffusion models with llm for enhanced semantic
  alignment.
\newblock \emph{arXiv preprint arXiv:2403.05135}, 2024.

\bibitem[Ke et~al.(2021)Ke, Wang, Wang, Milanfar, and Yang]{ke2021musiq}
Junjie Ke, Qifei Wang, Yilin Wang, Peyman Milanfar, and Feng Yang.
\newblock Musiq: Multi-scale image quality transformer.
\newblock In \emph{Proceedings of the IEEE/CVF international conference on
  computer vision}, pages 5148--5157, 2021.

\bibitem[Kim et~al.(2024)Kim, Pertsch, Karamcheti, Xiao, Balakrishna, Nair,
  Rafailov, Foster, Lam, Sanketi, et~al.]{openvla}
Moo~Jin Kim, Karl Pertsch, Siddharth Karamcheti, Ted Xiao, Ashwin Balakrishna,
  Suraj Nair, Rafael Rafailov, Ethan Foster, Grace Lam, Pannag Sanketi, et~al.
\newblock Openvla: An open-source vision-language-action model.
\newblock \emph{arXiv:2406.09246}, 2024.

\bibitem[Kling(2024)]{kling}
Kling.
\newblock Kling.
\newblock \emph{Kling}. Accessed Sept.30, 2024 [Online]
  \url{https://kling.kuaishou.com/en}, 2024.
\newblock URL \url{https://kling.kuaishou.com/en}.

\bibitem[Labs(2024)]{flux2024}
Black~Forest Labs.
\newblock Flux.
\newblock \url{https://github.com/black-forest-labs/flux}, 2024.

\bibitem[Labs(2025)]{flux-2-2025}
Black~Forest Labs.
\newblock {FLUX.2: State-of-the-Art Visual Intelligence}.
\newblock \url{https://bfl.ai/blog/flux-2}, 2025.

\bibitem[Li et~al.(2026)Li, Wang, Hu, Huang, Chen, Ou, Tao, Wan, Qi, and
  Feng]{li2026easier}
Ouxiang Li, Yuan Wang, Xinting Hu, Huijuan Huang, Rui Chen, Jiarong Ou, Xin
  Tao, Pengfei Wan, Xiaojuan Qi, and Fuli Feng.
\newblock Easier painting than thinking: Can text-to-image models set the
  stage, but not direct the play?, 2026.
\newblock URL \url{https://arxiv.org/abs/2509.03516}.

\bibitem[Li et~al.(2025{\natexlab{a}})Li, Liu, Zhang, Lin, Wu, Yuan, Yan, Ye,
  Yu, Niu, et~al.]{li2025uniworld}
Zongjian Li, Zheyuan Liu, Qihui Zhang, Bin Lin, Feize Wu, Shenghai Yuan,
  Zhiyuan Yan, Yang Ye, Wangbo Yu, Yuwei Niu, et~al.
\newblock Uniworld-v2: Reinforce image editing with diffusion negative-aware
  finetuning and mllm implicit feedback.
\newblock \emph{arXiv preprint arXiv:2510.16888}, 2025{\natexlab{a}}.

\bibitem[Li et~al.(2025{\natexlab{b}})Li, Liu, Zhang, Lin, Wu, Yuan, Yan, Ye,
  Yu, Niu, et~al.]{uniworld-v2}
Zongjian Li, Zheyuan Liu, Qihui Zhang, Bin Lin, Feize Wu, Shenghai Yuan,
  Zhiyuan Yan, Yang Ye, Wangbo Yu, Yuwei Niu, et~al.
\newblock Uniworld-v2: Reinforce image editing with diffusion negative-aware
  finetuning and mllm implicit feedback.
\newblock \emph{arXiv preprint arXiv:2510.16888}, 2025{\natexlab{b}}.

\bibitem[Lin et~al.(2025)Lin, Li, Cheng, Niu, Ye, He, Yuan, Yu, Wang, Ge,
  et~al.]{lin2025uniworld}
Bin Lin, Zongjian Li, Xinhua Cheng, Yuwei Niu, Yang Ye, Xianyi He, Shenghai
  Yuan, Wangbo Yu, Shaodong Wang, Yunyang Ge, et~al.
\newblock Uniworld: High-resolution semantic encoders for unified visual
  understanding and generation.
\newblock \emph{arXiv preprint arXiv:2506.03147}, 2025.

\bibitem[Lipman et~al.(2022)Lipman, Chen, Ben-Hamu, Nickel, and
  Le]{lipman2022flow}
Yaron Lipman, Ricky~TQ Chen, Heli Ben-Hamu, Maximilian Nickel, and Matt Le.
\newblock Flow matching for generative modeling.
\newblock \emph{arXiv preprint arXiv:2210.02747}, 2022.

\bibitem[Liu et~al.(2025{\natexlab{a}})Liu, Liu, Liang, Li, Liu, Wang, Wan,
  Zhang, and Ouyang]{liu2025flow}
Jie Liu, Gongye Liu, Jiajun Liang, Yangguang Li, Jiaheng Liu, Xintao Wang,
  Pengfei Wan, Di~Zhang, and Wanli Ouyang.
\newblock Flow-grpo: Training flow matching models via online rl.
\newblock \emph{arXiv preprint arXiv:2505.05470}, 2025{\natexlab{a}}.

\bibitem[Liu et~al.(2026)Liu, Dong, Lu, Diao, Belcak, Liu, Chen, Yin, Wang,
  Cheng, et~al.]{liu2026gdpo}
Shih-Yang Liu, Xin Dong, Ximing Lu, Shizhe Diao, Peter Belcak, Mingjie Liu,
  Min-Hung Chen, Hongxu Yin, Yu-Chiang~Frank Wang, Kwang-Ting Cheng, et~al.
\newblock Gdpo: Group reward-decoupled normalization policy optimization for
  multi-reward rl optimization.
\newblock \emph{arXiv preprint arXiv:2601.05242}, 2026.

\bibitem[Liu et~al.(2025{\natexlab{b}})Liu, Han, Xing, Yin, Wang, Cheng, Liao,
  Wang, Fu, Han, et~al.]{liu2025step1x}
Shiyu Liu, Yucheng Han, Peng Xing, Fukun Yin, Rui Wang, Wei Cheng, Jiaqi Liao,
  Yingming Wang, Honghao Fu, Chunrui Han, et~al.
\newblock Step1x-edit: A practical framework for general image editing.
\newblock \emph{arXiv preprint arXiv:2504.17761}, 2025{\natexlab{b}}.

\bibitem[Liu et~al.(2024{\natexlab{a}})Liu, Duan, Zhang, Li, Zhang, Zhao, Yuan,
  Wang, He, Liu, et~al.]{mmbench}
Yuan Liu, Haodong Duan, Yuanhan Zhang, Bo~Li, Songyang Zhang, Wangbo Zhao, Yike
  Yuan, Jiaqi Wang, Conghui He, Ziwei Liu, et~al.
\newblock Mmbench: Is your multi-modal model an all-around player?
\newblock In \emph{ECCV}, 2024{\natexlab{a}}.

\bibitem[Liu et~al.(2024{\natexlab{b}})Liu, Li, Huang, Yang, Yu, Li, Yin, Liu,
  Jin, and Bai]{ocrbench}
Yuliang Liu, Zhang Li, Mingxin Huang, Biao Yang, Wenwen Yu, Chunyuan Li,
  Xu-Cheng Yin, Cheng-Lin Liu, Lianwen Jin, and Xiang Bai.
\newblock Ocrbench: on the hidden mystery of ocr in large multimodal models.
\newblock \emph{Science China Information Sciences}, 67\penalty0 (12):\penalty0
  220102, 2024{\natexlab{b}}.

\bibitem[Lu et~al.(2023)Lu, Bansal, Xia, Liu, Li, Hajishirzi, Cheng, Chang,
  Galley, and Gao]{lu2023mathvista}
Pan Lu, Hritik Bansal, Tony Xia, Jiacheng Liu, Chunyuan Li, Hannaneh
  Hajishirzi, Hao Cheng, Kai-Wei Chang, Michel Galley, and Jianfeng Gao.
\newblock Mathvista: Evaluating mathematical reasoning of foundation models in
  visual contexts.
\newblock \emph{arXiv preprint arXiv:2310.02255}, 2023.

\bibitem[Luo et~al.(2025)Luo, Wang, Wu, Xiao, Jiang, Lian, Zhang, Liu,
  et~al.]{luo2025editscore}
Xin Luo, Jiahao Wang, Chenyuan Wu, Shitao Xiao, Xiyan Jiang, Defu Lian, Jiajun
  Zhang, Dong Liu, et~al.
\newblock Editscore: Unlocking online rl for image editing via high-fidelity
  reward modeling.
\newblock \emph{arXiv preprint arXiv:2509.23909}, 2025.

\bibitem[Ma et~al.(2025{\natexlab{a}})Ma, Peng, Guo, Chen, Lu, and
  Yang]{ma2025x2i}
Jian Ma, Qirong Peng, Xu~Guo, Chen Chen, Haonan Lu, and Zhenyu Yang.
\newblock X2i: Seamless integration of multimodal understanding into diffusion
  transformer via attention distillation.
\newblock In \emph{Proceedings of the IEEE/CVF International Conference on
  Computer Vision}, pages 16733--16744, 2025{\natexlab{a}}.

\bibitem[Ma et~al.(2024)Ma, Chen, Zhang, Chou, de~Melo, and Yuille]{3dsrbench}
Wufei Ma, Haoyu Chen, Guofeng Zhang, Yu-Cheng Chou, Celso~M de~Melo, and Alan
  Yuille.
\newblock 3dsrbench: A comprehensive 3d spatial reasoning benchmark.
\newblock \emph{arXiv:2412.07825}, 2024.

\bibitem[Ma et~al.(2025{\natexlab{b}})Ma, Wu, Sun, and Li]{ma2025hpsv3}
Yuhang Ma, Xiaoshi Wu, Keqiang Sun, and Hongsheng Li.
\newblock Hpsv3: Towards wide-spectrum human preference score.
\newblock In \emph{Proceedings of the IEEE/CVF International Conference on
  Computer Vision}, pages 15086--15095, 2025{\natexlab{b}}.

\bibitem[Mittal et~al.(2012)Mittal, Soundararajan, and Bovik]{mittal2012making}
Anish Mittal, Rajiv Soundararajan, and Alan~C Bovik.
\newblock Making a “completely blind” image quality analyzer.
\newblock \emph{IEEE Signal processing letters}, 20\penalty0 (3):\penalty0
  209--212, 2012.

\bibitem[OpenAI(2023)]{ChatGPT}
OpenAI.
\newblock Chatgpt.
\newblock \url{https://openai.com/blog/chatgpt/}, 2023.

\bibitem[{OpenAI}(2025)]{openai2025gptimage}
{OpenAI}.
\newblock \emph{GPT Image 1}.
\newblock OpenAI, 2025.
\newblock URL \url{https://developers.openai.com/api/docs/models/gpt-image-1}.
\newblock OpenAI API model documentation, accessed 2026-03-24.

\bibitem[Ouyang et~al.(2025)Ouyang, Liu, Wu, Liu, Zhou, Zhou, Meng, and
  Sun]{spacer}
Kun Ouyang, Yuanxin Liu, Haoning Wu, Yi~Liu, Hao Zhou, Jie Zhou, Fandong Meng,
  and Xu~Sun.
\newblock Spacer: Reinforcing mllms in video spatial reasoning.
\newblock \emph{arXiv:2504.01805}, 2025.

\bibitem[Qian et~al.(2025)Qian, Bocek-Rivele, Song, Tong, Yang, Lu, Hu, and
  Gan]{qian2025picobanana400}
Yusu Qian, Eli Bocek-Rivele, Liangchen Song, Jialing Tong, Yinfei Yang, Jiasen
  Lu, Wenze Hu, and Zhe Gan.
\newblock Pico-banana-400k: A large-scale dataset for text-guided image
  editing, 2025.
\newblock URL \url{https://arxiv.org/abs/2510.19808}.

\bibitem[Qin et~al.(2025)Qin, Wang, Li, Chen, Pei, Li, and Cao]{camedit}
Xinran Qin, Zhixin Wang, Fan Li, Haoyu Chen, RenJing Pei, WenBo Li, and
  XiaoChun Cao.
\newblock Camedit: Continuous camera parameter control for photorealistic image
  editing.
\newblock In \emph{The Thirty-ninth Annual Conference on Neural Information
  Processing Systems}, 2025.

\bibitem[Qu et~al.(2025)Qu, Cheng, Yang, Zhao, Lin, Shi, Li, Wang, Chua, and
  Jiang]{qu2025vincie}
Leigang Qu, Feng Cheng, Ziyan Yang, Qi~Zhao, Shanchuan Lin, Yichun Shi, Yicong
  Li, Wenjie Wang, Tat-Seng Chua, and Lu~Jiang.
\newblock Vincie: Unlocking in-context image editing from video.
\newblock In \emph{The Fourteenth International Conference on Learning
  Representations}, 2025.

\bibitem[Ren et~al.(2024)Ren, Li, Chen, Pei, Shao, Guo, Peng, Song, and
  Zhu]{ultrapixel}
Jingjing Ren, Wenbo Li, Haoyu Chen, Renjing Pei, Bin Shao, Yong Guo, Long Peng,
  Fenglong Song, and Lei Zhu.
\newblock Ultrapixel: Advancing ultra high-resolution image synthesis to new
  peaks.
\newblock \emph{Advances in Neural Information Processing Systems},
  37:\penalty0 111131--111171, 2024.

\bibitem[Roberts et~al.(2021)Roberts, Ramapuram, Ranjan, Kumar, Bautista,
  Paczan, Webb, and Susskind]{hypersim}
Mike Roberts, Jason Ramapuram, Anurag Ranjan, Atulit Kumar, Miguel~Angel
  Bautista, Nathan Paczan, Russ Webb, and Joshua~M Susskind.
\newblock Hypersim: A photorealistic synthetic dataset for holistic indoor
  scene understanding.
\newblock In \emph{ICCV}, 2021.

\bibitem[Rombach et~al.(2022)Rombach, Blattmann, Lorenz, Esser, and
  Ommer]{stablediffusion}
Robin Rombach, Andreas Blattmann, Dominik Lorenz, Patrick Esser, and Bj{\"o}rn
  Ommer.
\newblock High-resolution image synthesis with latent diffusion models.
\newblock In \emph{Proceedings of the IEEE/CVF conference on computer vision
  and pattern recognition}, pages 10684--10695, 2022.

\bibitem[Seedream et~al.(2025)Seedream, Chen, Gao, Gong, Guo, Guo, Guo, Hou,
  Huang, Huang, et~al.]{seedream2025seedream}
Team Seedream, Yunpeng Chen, Yu~Gao, Lixue Gong, Meng Guo, Qiushan Guo, Zhiyao
  Guo, Xiaoxia Hou, Weilin Huang, Yixuan Huang, et~al.
\newblock Seedream 4.0: Toward next-generation multimodal image generation.
\newblock \emph{arXiv preprint arXiv:2509.20427}, 2025.

\bibitem[Team et~al.(2025{\natexlab{a}})Team, Yue, Lin, Song, Wang, Ren, Gu,
  Li, Li, Zhao, Li, Bao, Tian, Zhang, Wang, Zhu, Cici, He, Ye, Shen, Zhang,
  Jiang, Zheng, Song, Luo, Yu, Wang, Tian, Tu, Yan, Huang, Wang, Xu, Song,
  Zhang, Yong, Zhang, Deng, Yang, Ma, Lv, Zhuang, Liu, Deng, Liu, Chen, Yu,
  Liu, Wang, Ma, Wang, Wang, Chen, Zhu, Zhou, Zhou, Fang, Shi, Dong, Xiao, Xu,
  Liu, Xu, Qu, Zhao, Lv, Wang, Zhang, Zhang, Zhang, Ma, Liu, Cai, and
  Xia]{coreteam2025mimovltechnicalreport}
Core Team, Zihao Yue, Zhenru Lin, Yifan Song, Weikun Wang, Shuhuai Ren, Shuhao
  Gu, Shicheng Li, Peidian Li, Liang Zhao, Lei Li, Kainan Bao, Hao Tian, Hailin
  Zhang, Gang Wang, Dawei Zhu, Cici, Chenhong He, Bowen Ye, Bowen Shen, Zihan
  Zhang, Zihan Jiang, Zhixian Zheng, Zhichao Song, Zhenbo Luo, Yue Yu, Yudong
  Wang, Yuanyuan Tian, Yu~Tu, Yihan Yan, Yi~Huang, Xu~Wang, Xinzhe Xu, Xingchen
  Song, Xing Zhang, Xing Yong, Xin Zhang, Xiangwei Deng, Wenyu Yang, Wenhan Ma,
  Weiwei Lv, Weiji Zhuang, Wei Liu, Sirui Deng, Shuo Liu, Shimao Chen, Shihua
  Yu, Shaohui Liu, Shande Wang, Rui Ma, Qiantong Wang, Peng Wang, Nuo Chen,
  Menghang Zhu, Kangyang Zhou, Kang Zhou, Kai Fang, Jun Shi, Jinhao Dong,
  Jiebao Xiao, Jiaming Xu, Huaqiu Liu, Hongshen Xu, Heng Qu, Haochen Zhao,
  Hanglong Lv, Guoan Wang, Duo Zhang, Dong Zhang, Di~Zhang, Chong Ma, Chang
  Liu, Can Cai, and Bingquan Xia.
\newblock Mimo-vl technical report, 2025{\natexlab{a}}.
\newblock URL \url{https://arxiv.org/abs/2506.03569}.

\bibitem[Team et~al.(2023)Team, Anil, Borgeaud, Alayrac, Yu, Soricut,
  Schalkwyk, Dai, Hauth, Millican, et~al.]{gemini}
Gemini Team, Rohan Anil, Sebastian Borgeaud, Jean-Baptiste Alayrac, Jiahui Yu,
  Radu Soricut, Johan Schalkwyk, Andrew~M Dai, Anja Hauth, Katie Millican,
  et~al.
\newblock Gemini: a family of highly capable multimodal models.
\newblock \emph{arXiv:2312.11805}, 2023.

\bibitem[Team(2024)]{kolors}
Kolors Team.
\newblock Kolors: Effective training of diffusion model for photorealistic
  text-to-image synthesis.
\newblock \emph{arXiv preprint}, 2024.

\bibitem[Team et~al.(2025{\natexlab{b}})Team, Ma, Tan, Huang, Wu, He, Gao,
  Xiao, Wei, Ma, Cai, Guan, and Hu]{LongCat-Image}
Meituan~LongCat Team, Hanghang Ma, Haoxian Tan, Jiale Huang, Junqiang Wu,
  Jun-Yan He, Lishuai Gao, Songlin Xiao, Xiaoming Wei, Xiaoqi Ma, Xunliang Cai,
  Yayong Guan, and Jie Hu.
\newblock Longcat-image technical report.
\newblock \emph{arXiv preprint arXiv:2512.07584}, 2025{\natexlab{b}}.

\bibitem[Team et~al.(2026{\natexlab{a}})Team, Gao, Wang, Zeng, Zhu, Cheng, Li,
  Wang, Xu, Ma, et~al.]{lingbot-world}
Robbyant Team, Zelin Gao, Qiuyu Wang, Yanhong Zeng, Jiapeng Zhu, Ka~Leong
  Cheng, Yixuan Li, Hanlin Wang, Yinghao Xu, Shuailei Ma, et~al.
\newblock Advancing open-source world models.
\newblock \emph{arXiv preprint arXiv:2601.20540}, 2026{\natexlab{a}}.

\bibitem[Team et~al.(2026{\natexlab{b}})Team, Qiao, Hui, Li, Wang, Song, Zhang,
  Li, Xiang, Wang, et~al.]{team2026firered}
Super~Intelligence Team, Changhao Qiao, Chao Hui, Chen Li, Cunzheng Wang, Dejia
  Song, Jiale Zhang, Jing Li, Qiang Xiang, Runqi Wang, et~al.
\newblock Firered-image-edit-1.0 techinical report.
\newblock \emph{arXiv preprint arXiv:2602.13344}, 2026{\natexlab{b}}.

\bibitem[Team(2025)]{team2025zimage}
Z-Image Team.
\newblock Z-image: An efficient image generation foundation model with
  single-stream diffusion transformer.
\newblock \emph{arXiv preprint arXiv:2511.22699}, 2025.

\bibitem[Tong et~al.(2024)Tong, Brown, Wu, Woo, IYER, Akula, Yang, Yang,
  Middepogu, Wang, et~al.]{cambrian}
Peter Tong, Ellis Brown, Penghao Wu, Sanghyun Woo, Adithya Jairam~Vedagiri
  IYER, Sai~Charitha Akula, Shusheng Yang, Jihan Yang, Manoj Middepogu, Ziteng
  Wang, et~al.
\newblock Cambrian-1: A fully open, vision-centric exploration of multimodal
  llms.
\newblock \emph{NeurIPS}, 2024.

\bibitem[{Vidu Team}(2024)]{vidu}
{Vidu Team}.
\newblock Vidu: Ai video generator.
\newblock https://www.vidu.cn/, 2024.

\bibitem[Wan et~al.(2025)Wan, Wang, Ai, Wen, Mao, Xie, Chen, Yu, Zhao, Yang,
  et~al.]{wan2025wan}
Team Wan, Ang Wang, Baole Ai, Bin Wen, Chaojie Mao, Chen-Wei Xie, Di~Chen,
  Feiwu Yu, Haiming Zhao, Jianxiao Yang, et~al.
\newblock Wan: Open and advanced large-scale video generative models.
\newblock \emph{arXiv preprint arXiv:2503.20314}, 2025.

\bibitem[Wang et~al.(2023)Wang, Chan, and Loy]{wang2023exploring}
Jianyi Wang, Kelvin~CK Chan, and Chen~Change Loy.
\newblock Exploring clip for assessing the look and feel of images.
\newblock In \emph{Proceedings of the AAAI conference on artificial
  intelligence}, volume~37, pages 2555--2563, 2023.

\bibitem[Wang et~al.(2025{\natexlab{a}})Wang, Chen, Karaev, Vedaldi, Rupprecht,
  and Novotny]{wang2025vggt}
Jianyuan Wang, Minghao Chen, Nikita Karaev, Andrea Vedaldi, Christian
  Rupprecht, and David Novotny.
\newblock Vggt: Visual geometry grounded transformer.
\newblock In \emph{Proceedings of the Computer Vision and Pattern Recognition
  Conference}, pages 5294--5306, 2025{\natexlab{a}}.

\bibitem[Wang et~al.(2024)Wang, Zang, Zhang, Chu, Cao, Sun, Liu, Dong, Wu, Lin,
  Chen, Wang, Meng, Yao, Yang, Wu, Chen, Wu, Jiang, Wu, Chai, Nie, Yan, Wang,
  Zhou, Wang, Huang, Xu, Li, Yuan, Zu, Ha, Gao, and
  Jiao]{wang2024v3detchallenge2024vast}
Jiaqi Wang, Yuhang Zang, Pan Zhang, Tao Chu, Yuhang Cao, Zeyi Sun, Ziyu Liu,
  Xiaoyi Dong, Tong Wu, Dahua Lin, Zeming Chen, Zhi Wang, Lingchen Meng, Wenhao
  Yao, Jianwei Yang, Sihong Wu, Zhineng Chen, Zuxuan Wu, Yu-Gang Jiang, Peixi
  Wu, Bosong Chai, Xuan Nie, Longquan Yan, Zeyu Wang, Qifan Zhou, Boning Wang,
  Jiaqi Huang, Zunnan Xu, Xiu Li, Kehong Yuan, Yanyan Zu, Jiayao Ha, Qiong Gao,
  and Licheng Jiao.
\newblock V3det challenge 2024 on vast vocabulary and open vocabulary object
  detection: Methods and results, 2024.
\newblock URL \url{https://arxiv.org/abs/2406.11739}.

\bibitem[Wang et~al.(2025{\natexlab{b}})Wang, Yang, Zhao, Zhang, Liu, Zhou, and
  Xie]{wang2025gptedit}
Yuhan Wang, Siwei Yang, Bingchen Zhao, Letian Zhang, Qing Liu, Yuyin Zhou, and
  Cihang Xie.
\newblock Gpt-image-edit-1.5 m: A million-scale, gpt-generated image dataset.
\newblock \emph{arXiv preprint arXiv:2507.21033}, 2025{\natexlab{b}}.

\bibitem[Wu et~al.(2025{\natexlab{a}})Wu, Li, Zhou, Lin, Gao, Yan, ming Yin,
  Bai, Xu, Chen, Chen, Tang, Zhang, Wang, Yang, Yu, Cheng, Liu, Li, Zhang,
  Meng, Wei, Ni, Chen, Cao, Peng, Qu, Wu, Wang, Yu, Wen, Feng, Xu, Wang, Zhang,
  Zhu, Wu, Cai, and Liu]{qwenimage}
Chenfei Wu, Jiahao Li, Jingren Zhou, Junyang Lin, Kaiyuan Gao, Kun Yan, Sheng
  ming Yin, Shuai Bai, Xiao Xu, Yilei Chen, Yuxiang Chen, Zecheng Tang, Zekai
  Zhang, Zhengyi Wang, An~Yang, Bowen Yu, Chen Cheng, Dayiheng Liu, Deqing Li,
  Hang Zhang, Hao Meng, Hu~Wei, Jingyuan Ni, Kai Chen, Kuan Cao, Liang Peng,
  Lin Qu, Minggang Wu, Peng Wang, Shuting Yu, Tingkun Wen, Wensen Feng,
  Xiaoxiao Xu, Yi~Wang, Yichang Zhang, Yongqiang Zhu, Yujia Wu, Yuxuan Cai, and
  Zenan Liu.
\newblock Qwen-image technical report.
\newblock \emph{arXiv preprint arXiv:2508.02324}, 2025{\natexlab{a}}.

\bibitem[Wu et~al.(2025{\natexlab{b}})Wu, Li, Zhou, Lin, Gao, Yan, ming Yin,
  Bai, Xu, Chen, Chen, Tang, Zhang, Wang, Yang, Yu, Cheng, Liu, Li, Zhang,
  Meng, Wei, Ni, Chen, Cao, Peng, Qu, Wu, Wang, Yu, Wen, Feng, Xu, Wang, Zhang,
  Zhu, Wu, Cai, and Liu]{wu2025qwenimage}
Chenfei Wu, Jiahao Li, Jingren Zhou, Junyang Lin, Kaiyuan Gao, Kun Yan, Sheng
  ming Yin, Shuai Bai, Xiao Xu, Yilei Chen, Yuxiang Chen, Zecheng Tang, Zekai
  Zhang, Zhengyi Wang, An~Yang, Bowen Yu, Chen Cheng, Dayiheng Liu, Deqing Li,
  Hang Zhang, Hao Meng, Hu~Wei, Jingyuan Ni, Kai Chen, Kuan Cao, Liang Peng,
  Lin Qu, Minggang Wu, Peng Wang, Shuting Yu, Tingkun Wen, Wensen Feng,
  Xiaoxiao Xu, Yi~Wang, Yichang Zhang, Yongqiang Zhu, Yujia Wu, Yuxuan Cai, and
  Zenan Liu.
\newblock Qwen-image technical report.
\newblock \emph{arXiv preprint arXiv:2508.02324}, 2025{\natexlab{b}}.
\newblock URL \url{https://arxiv.org/abs/2508.02324}.

\bibitem[Wu et~al.(2025{\natexlab{c}})Wu, Zheng, Yan, Xiao, Luo, Wang, Li,
  Jiang, Liu, Zhou, et~al.]{wu2025omnigen2}
Chenyuan Wu, Pengfei Zheng, Ruiran Yan, Shitao Xiao, Xin Luo, Yueze Wang, Wanli
  Li, Xiyan Jiang, Yexin Liu, Junjie Zhou, et~al.
\newblock Omnigen2: Exploration to advanced multimodal generation.
\newblock \emph{arXiv preprint arXiv:2506.18871}, 2025{\natexlab{c}}.

\bibitem[Wu et~al.(2025{\natexlab{d}})Wu, Ren, Shen, Cao, He, Lu, Gao, Xie,
  Lan, Alvarez, Gao, Fidler, Wang, and Ling]{wu2025chronoedit}
Jay~Zhangjie Wu, Xuanchi Ren, Tianchang Shen, Tianshi Cao, Kai He, Yifan Lu,
  Ruiyuan Gao, Enze Xie, Shiyi Lan, Jose~M. Alvarez, Jun Gao, Sanja Fidler,
  Zian Wang, and Huan Ling.
\newblock Chronoedit: Towards temporal reasoning for image editing and world
  simulation.
\newblock \emph{arXiv preprint arXiv:2510.04290}, 2025{\natexlab{d}}.

\bibitem[x.ai(2024)]{grok15v}
x.ai.
\newblock Grok-1.5 vision preview, 2024.
\newblock URL \url{https://x.ai/blog/grok-1.5v}.

\bibitem[Xia et~al.(2025)Xia, Peng, Zhang, Huang, Liu, Li, Tan, Wu, Wang, Wang,
  et~al.]{xia2025dreamomni2}
Bin Xia, Bohao Peng, Yuechen Zhang, Junjia Huang, Jiyang Liu, Jingyao Li, Haoru
  Tan, Sitong Wu, Chengyao Wang, Yitong Wang, et~al.
\newblock Dreamomni2: Multimodal instruction-based editing and generation.
\newblock \emph{arXiv preprint arXiv:2510.06679}, 2025.

\bibitem[Xiao et~al.(2025{\natexlab{a}})Xiao, Wang, Zhou, Yuan, Xing, Yan, Li,
  Wang, Huang, and Liu]{xiao2025omnigen}
Shitao Xiao, Yueze Wang, Junjie Zhou, Huaying Yuan, Xingrun Xing, Ruiran Yan,
  Chaofan Li, Shuting Wang, Tiejun Huang, and Zheng Liu.
\newblock Omnigen: Unified image generation.
\newblock In \emph{Proceedings of the Computer Vision and Pattern Recognition
  Conference}, pages 13294--13304, 2025{\natexlab{a}}.

\bibitem[Xiao et~al.(2025{\natexlab{b}})Xiao, Song, Chen, Luo, Chen, Gan,
  Huang, Li, Qi, and Shan]{xiao2025mindomni}
Yicheng Xiao, Lin Song, Yukang Chen, Yingmin Luo, Yuxin Chen, Yukang Gan, Wei
  Huang, Xiu Li, Xiaojuan Qi, and Ying Shan.
\newblock Mindomni: Unleashing reasoning generation in vision language models
  with rgpo.
\newblock \emph{arXiv preprint arXiv:2505.13031}, 2025{\natexlab{b}}.

\bibitem[Xiao et~al.(2026)Xiao, Zhang, Song, Chen, Li, Jiang, Ren, Lin, Huang,
  Huang, Li, Duan, and Qi]{SpatialEdit-Bench}
Yicheng Xiao, Wenhu Zhang, Lin Song, Yukang Chen, Wenbo Li, Nan Jiang, Tianhe
  Ren, Haokun Lin, Wei Huang, Haoyang Huang, Xiu Li, Nan Duan, and Xiaojuan Qi.
\newblock Spatialedit: Benchmarking fine-grained image spatial editing.
\newblock \emph{arXiv preprint arXiv:2604.04911}, 2026.

\bibitem[Yang et~al.(2025{\natexlab{a}})Yang, Li, Yang, Zhang, Hui, Zheng, Yu,
  Gao, Huang, Lv, et~al.]{qwen3}
An~Yang, Anfeng Li, Baosong Yang, Beichen Zhang, Binyuan Hui, Bo~Zheng, Bowen
  Yu, Chang Gao, Chengen Huang, Chenxu Lv, et~al.
\newblock Qwen3 technical report.
\newblock \emph{arXiv:2505.09388}, 2025{\natexlab{a}}.

\bibitem[Yang et~al.(2025{\natexlab{b}})Yang, Yang, Gupta, Han, Fei-Fei, and
  Xie]{vsibench}
Jihan Yang, Shusheng Yang, Anjali~W Gupta, Rilyn Han, Li~Fei-Fei, and Saining
  Xie.
\newblock Thinking in space: How multimodal large language models see,
  remember, and recall spaces.
\newblock In \emph{CVPR}, 2025{\natexlab{b}}.

\bibitem[Yang et~al.(2025{\natexlab{c}})Yang, Zhu, Li, Huang, Yan, Zhou, Liu,
  Li, Li, Wang, et~al.]{vst}
Rui Yang, Ziyu Zhu, Yanwei Li, Jingjia Huang, Shen Yan, Siyuan Zhou, Zhe Liu,
  Xiangtai Li, Shuangye Li, Wenqian Wang, et~al.
\newblock Visual spatial tuning.
\newblock \emph{arXiv preprint arXiv:2511.05491}, 2025{\natexlab{c}}.

\bibitem[Yang et~al.(2025{\natexlab{d}})Yang, Xu, Xie, Yang, Li, Lin, Zhu,
  Chen, Duan, Yue, et~al.]{mmsibench}
Sihan Yang, Runsen Xu, Yiman Xie, Sizhe Yang, Mo~Li, Jingli Lin, Chenming Zhu,
  Xiaochen Chen, Haodong Duan, Xiangyu Yue, et~al.
\newblock Mmsi-bench: A benchmark for multi-image spatial intelligence.
\newblock \emph{arXiv:2505.23764}, 2025{\natexlab{d}}.

\bibitem[Ye et~al.(2025)Ye, He, Li, Lin, Yuan, Yan, Hou, and
  Yuan]{ye2025imgedit}
Yang Ye, Xianyi He, Zongjian Li, Bin Lin, Shenghai Yuan, Zhiyuan Yan, Bohan
  Hou, and Li~Yuan.
\newblock Imgedit: A unified image editing dataset and benchmark.
\newblock \emph{arXiv preprint arXiv:2505.20275}, 2025.

\bibitem[Yeh et~al.(2025)Yeh, Wang, Tong, Cheng, Wang, Chu, Zhai, Chen, Gao,
  and Ma]{yeh2025seeing}
Chun-Hsiao Yeh, Chenyu Wang, Shengbang Tong, Ta-Ying Cheng, Ruoyu Wang, Tianzhe
  Chu, Yuexiang Zhai, Yubei Chen, Shenghua Gao, and Yi~Ma.
\newblock Seeing from another perspective: Evaluating multi-view understanding
  in mllms.
\newblock \emph{arXiv preprint arXiv:2504.15280}, 2025.

\bibitem[Yeshwanth et~al.(2023)Yeshwanth, Liu, Nie{\ss}ner, and Dai]{scannet++}
Chandan Yeshwanth, Yueh-Cheng Liu, Matthias Nie{\ss}ner, and Angela Dai.
\newblock Scannet++: A high-fidelity dataset of 3d indoor scenes.
\newblock In \emph{ICCV}, 2023.

\bibitem[Yu et~al.(2025)Yu, Chow, Yue, Pan, Wu, Wan, Li, Tang, Zhang, and
  Zhuang]{yu2025anyedit}
Qifan Yu, Wei Chow, Zhongqi Yue, Kaihang Pan, Yang Wu, Xiaoyang Wan, Juncheng
  Li, Siliang Tang, Hanwang Zhang, and Yueting Zhuang.
\newblock Anyedit: Mastering unified high-quality image editing for any idea.
\newblock In \emph{Proceedings of the Computer Vision and Pattern Recognition
  Conference}, pages 26125--26135, 2025.

\bibitem[Zhang et~al.(2023)Zhang, Mo, Chen, Sun, and Su]{zhang2023magicbrush}
Kai Zhang, Lingbo Mo, Wenhu Chen, Huan Sun, and Yu~Su.
\newblock Magicbrush: A manually annotated dataset for instruction-guided image
  editing.
\newblock \emph{Advances in Neural Information Processing Systems},
  36:\penalty0 31428--31449, 2023.

\bibitem[Zhang et~al.(2025{\natexlab{a}})Zhang, Sun, Zhou, He, Yin, Wang,
  Sheng, Qiao, Shao, and Liu]{zhang2025bamboo}
Yuanhan Zhang, Qinghong Sun, Yichun Zhou, Zexin He, Zhenfei Yin, Kun Wang,
  Lu~Sheng, Yu~Qiao, Jing Shao, and Ziwei Liu.
\newblock Bamboo: Building mega-scale vision dataset continually with
  human--machine synergy.
\newblock \emph{International Journal of Computer Vision}, 133\penalty0
  (8):\penalty0 5806--5821, 2025{\natexlab{a}}.

\bibitem[Zhang et~al.(2025{\natexlab{b}})Zhang, Xie, Lu, Yang, and
  Yang]{zhang2025icedit}
Zechuan Zhang, Ji~Xie, Yu~Lu, Zongxin Yang, and Yi~Yang.
\newblock In-context edit: Enabling instructional image editing with in-context
  generation in large-scale diffusion transformers.
\newblock In \emph{Advances in Neural Information Processing Systems
  (NeurIPS)}, 2025{\natexlab{b}}.
\newblock arXiv:2504.20690.

\bibitem[Zhao et~al.(2024)Zhao, Ma, Chen, Si, Wu, An, Yu, Zhang, Li, and
  Chang]{zhao2024ultraedit}
Haozhe Zhao, Xiaojian~Shawn Ma, Liang Chen, Shuzheng Si, Rujie Wu, Kaikai An,
  Peiyu Yu, Minjia Zhang, Qing Li, and Baobao Chang.
\newblock Ultraedit: Instruction-based fine-grained image editing at scale.
\newblock \emph{Advances in Neural Information Processing Systems},
  37:\penalty0 3058--3093, 2024.

\bibitem[Zhao et~al.(2026)Zhao, Zhang, Lin, Liang, Duan, Ding, Tian, Zang, Yan,
  and Yang]{zhao2026trust}
Xiangyu Zhao, Peiyuan Zhang, Junming Lin, Tianhao Liang, Yuchen Duan, Shengyuan
  Ding, Changyao Tian, Yuhang Zang, Junchi Yan, and Xue Yang.
\newblock Trust your critic: Robust reward modeling and reinforcement learning
  for faithful image editing and generation.
\newblock \emph{arXiv preprint arXiv:2603.12247}, 2026.

\bibitem[Zheng et~al.(2025)Zheng, Chen, Ye, Wang, Zhang, Jiang, Su, Ermon, Zhu,
  and Liu]{zheng2025diffusionnft}
Kaiwen Zheng, Huayu Chen, Haotian Ye, Haoxiang Wang, Qinsheng Zhang, Kai Jiang,
  Hang Su, Stefano Ermon, Jun Zhu, and Ming-Yu Liu.
\newblock Diffusionnft: Online diffusion reinforcement with forward process.
\newblock \emph{arXiv preprint arXiv:2509.16117}, 2025.

\bibitem[Zhu et~al.(2025)Zhu, Wang, Chen, Liu, Ye, Gu, Tian, Duan, Su, Shao,
  et~al.]{internvl3}
Jinguo Zhu, Weiyun Wang, Zhe Chen, Zhaoyang Liu, Shenglong Ye, Lixin Gu, Hao
  Tian, Yuchen Duan, Weijie Su, Jie Shao, et~al.
\newblock Internvl3: Exploring advanced training and test-time recipes for
  open-source multimodal models.
\newblock \emph{arXiv:2504.10479}, 2025.

\bibitem[Zitkovich et~al.(2023)Zitkovich, Yu, Xu, Xu, Xiao, Xia, Wu, Wohlhart,
  Welker, Wahid, et~al.]{zitkovich2023rt}
Brianna Zitkovich, Tianhe Yu, Sichun Xu, Peng Xu, Ted Xiao, Fei Xia, Jialin Wu,
  Paul Wohlhart, Stefan Welker, Ayzaan Wahid, et~al.
\newblock Rt-2: Vision-language-action models transfer web knowledge to robotic
  control.
\newblock In \emph{Conference on Robot Learning}, pages 2165--2183. PMLR, 2023.

\end{thebibliography}

\end{document}